\documentclass[aps,prd,amsmath,two column,amssymb,showpacs]{revtex4}
\usepackage{amssymb}
\usepackage{txfonts}
\usepackage{mathbbold}
\usepackage{amsfonts}
\usepackage{mathrsfs}
\usepackage{epsfig,bm,dcolumn}
\usepackage{graphicx}
\usepackage{color}
\usepackage{amsmath}

\begin{document}

\title{Pairing fluctuation effects in a strongly coupled color superfluid/superconductor}

\author{Jinyi Pang,$^{1}$ Jincheng Wang,$^{1,3}$ and
Lianyi He$^{2,3}$\footnote{Email address: lianyi@th.physik.uni-frankfurt.de}}
\address{1 Interdisciplinary Center for Theoretical Study and Department of Modern Physics,
University of Science and Technology of China, Hefei 230026, China\\
2 Frankfurt Institute for Advanced Studies, 60438 Frankfurt am Main, Germany\\
3 Institute for Theoretical Physics, J. W. Goethe University, 60438 Frankfurt am
Main, Germany}

\date{\today}

\begin{abstract}
We investigate the effects of pairing fluctuations in fermionic superfluids/superconductors where pairing occurs among three species
(colors) of fermions. Such color superfluids/superconductors can be realized in three-component atomic Fermi gases and in dense quark
matter. The superfluidity/superconductivity is characterized by a three-component order parameter which denotes the pairing among the
three colors of fermions. Because of the SU$(3)$ symmetry of the Hamiltonian, one color does not participate in pairing. This branch of
fermionic excitation is gapless in the naive BCS mean-field description. In this paper, we adopt a pairing fluctuation theory to
investigate the pairing fluctuation effects on the unpaired color in strongly coupled atomic color superfluids and quark color
superconductors. At low temperature, a large pairing gap of the paired colors suppresses the pairing fluctuation effects for the unpaired
color, and the spectral density of the unpaired color shows a single Fermi-liquid peak, which indicates the naive mean-field picture
remains valid. As the temperature is increased, the spectral density of the unpaired color generally exhibits a three-peak structure: The Fermi-liquid peak remains but gets suppressed, and two pseudogaplike peaks appears. At and above the superfluid transition temperature, the Fermi-liquid peak disappears completely and  all three colors exhibit pseudogaplike spectral density. The coexistence of Fermi liquid
and pseudogap behavior is generic for both atomic color superfluids and quark color superconductors.
\end{abstract}


\maketitle

\section{Introduction}

Color superconductivity in dense quark matter, in general, appears due to the attractive interactions in certain diquark
channels~\cite{CSCearly,CSCbegin,CSCreview}. Taking into account only the screened (color) electric interaction which is weakened at
the Debye mass scale $g\mu$ ($g$ is the QCD coupling constant and $\mu$ is the quark chemical potential), the early studies~\cite{CSCearly}
predicted a rather small pairing gap $\Delta\sim 1$MeV for moderate density where $\mu\sim\Lambda_{\text{QCD}}$
($\Lambda_{\text{QCD}}\sim 300$MeV is the QCD energy scale). The breakthrough in this field of research was made in~\cite{CSCbegin} where it was
observed that the pairing gap is about 2 orders of magnitude larger than the previous prediction using the instanton-induced interactions and the phenomenological four-fermion interactions. On the other hand, it was first pointed out by Son that at asymptotic high densities,
the unscreened magnetic interaction dominates~\cite{CSCgap}. This leads to a non-BCS gap $\Delta\sim \mu g^{-5}\exp{\left(-c/g\right)}$~\cite{pQCD} where $c=3\pi^2/\sqrt{2}$, which matches the large magnitude of $\Delta$ at moderate density predicted by the instanton-induced interactions and the phenomenological four-fermion interactions.

Such gaps at moderate density are so large that they may fall outside of the applicability range of the usual BCS-like mean-field
theory. It was estimated that the size of the Cooper pairs or the superconducting coherence length $\xi_c$ becomes comparable to the
averaged interquark distance $d$~\cite{pairsize} at moderate density where $\mu\sim \Lambda_{\text{QCD}}$. This feature is
highly contrasted to the standard BCS superconductivity in metals where $\xi_c\gg d$. Qualitatively, we can examine the ratio of the
superconducting transition temperature $T_c$ to the Fermi energy $E_{\text F}$, $\kappa=T_c/E_{\text F}$~\cite{TC}. We have
$\kappa\sim 10^{-5}$ for ordinary BCS superconductors and $\kappa\sim 10^{-2}$ for high temperature superconductors. For quark
matter at moderate density, taking $E_{\text F}\simeq 400$MeV and $T_c\simeq 50$MeV~\cite{TCQ}, we find that $\kappa$ is even higher,
$\kappa\sim 10^{-1}$, which is close to that for the resonant superfluidity in strongly interacting atomic Fermi gases~\cite{TC}.
This indicates that the color superconducting quark matter at moderate density is in the strongly coupled region or the BCS--Bose-Einstein condensation (BCS-BEC) crossover~\cite{BCSBEC,RBCSBEC,BCSBECQCD}. It is known that the pairing fluctuation effects play important role in
the BCS-BEC crossover~\cite{Levin}. The effects of the pairing fluctuations on the quark spectral properties including possible pseudogap
formation in heated quark matter (above the color superconducting transition temperature) were first elucidated by Kitazawa, Koide, Kunihiro,
and Nemoto~\cite{kitazawa}. One purpose of this paper is to extend these works to the superfluid/superconducting phase, namely, below the
critical temperature.

Many efforts have been made in understanding the single-particle properties and equations of state in the strongly interacting region
of the $s$-wave Fermi superfluids~\cite{Pieri,Hu} where pairing occurs between two components of fermions. Satisfactory agreement with
the experimental data from trapped fermionic atoms has been achieved~\cite{exp1,exp2}. Recently, the single-particle spectral density
in the strongly interacting region has been measured in cold atom experiments. It was found that a pseudogap phase appears above the
critical temperature~\cite{PG1,PG2,PG4}, where the system retains some of the characteristics of the superfluid phase such as a BCS-like
dispersion and a partially gapped density of states but does not exhibit superfluidity. The pseudogap behavior of the single-particle
excitations was also found by quantum Monte Carlo calculations~\cite{PG3}. The agreement between the theoretical predictions and the
experimental data in the equations of state and the single-particle spectral functions indicates that the pairing fluctuation effect
is significant in the strongly interacting region, which defies the conventional BCS-like mean-field theory.

In this paper, we investigate the pairing fluctuation effects in the systems where pairing occurs among three
species (colors) of fermions, in contrast to the ordinary BCS-BEC crossover problem where pairing occurs between two fermion components.
Such color superfluidity/superconductivity can be realized in three-component atomic Fermi gases~\cite{three} and two-flavor dense
quark matter which appears around $\mu\sim 400$MeV in the quark matter phase diagram~\cite{CSCphase}.

\begin{figure}[!htb]
\begin{center}
\includegraphics[width=8cm]{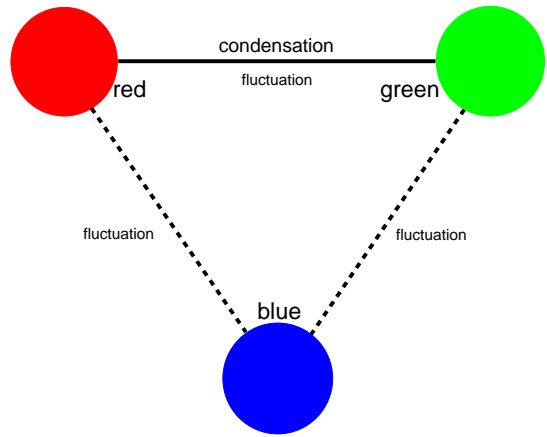}
\caption{(color-online). A schematic plot of the pairing pattern in color
superfluids/superconductors, corresponding to the order parameter gauge
$\mbox{\boldmath{$\Delta$}}=(\Delta,0,0)$. Only the red and
green fermions participate in pairing and their pairs condense.
Even though the blue fermions do not participate in pairing, pairing
fluctuations exist for red-blue and green-blue pairs, in addition to
the red-green pairs.\label{fig1}}
\end{center}
\end{figure}

The color superfluidity/superconductivity is generally characterized by a three-component order parameter $\mbox{\boldmath{$\Delta$}}=(\Delta_1,\Delta_2,\Delta_3)$, where $\Delta_1\sim\langle\psi_{\rm r}\psi_{\rm g}\rangle$,
$\Delta_2\sim\langle\psi_{\rm g}\psi_{\rm b}\rangle$, and $\Delta_3\sim\langle\psi_{\rm r}\psi_{\rm b}\rangle$. Here we use
red (r), green (g), and blue (b) to denote the three fermion species. In QCD, they are the color degrees of freedom for quarks.
If the Hamiltonian of the system has an SU(3) color symmetry, the nonvanishing order parameter breaks the SU(3) symmetry group
down to a subgroup SU(2). However, due to the SU(3) symmetry of the Hamiltonian, the effective potential
${\cal V}(\mbox{\boldmath{$\Delta$}})$ should depend only on the combination
$\mbox{\boldmath{$\Delta$}}\mbox{\boldmath{$\Delta$}}^\dagger=|\Delta_1|^2+|\Delta_2|^2+|\Delta_3|^2$.
Therefore, we can choose a specific gauge for the three-component order parameter, such as $\mbox{\boldmath{$\Delta$}}=(\Delta,0,0)$
without loss of generality. This simple argument shows that there exists a branch of fermion which does not participate in the pairing.
For the gauge $\mbox{\boldmath{$\Delta$}}=(\Delta,0,0)$, only the red and green fermions participate in the pairing, while the blue
fermions do not. A schematic plot of this pairing pattern is shown in Fig.~\ref{fig1}. In the naive BCS-like mean-field description,
the unpaired blue fermions possess a free energy dispersion and are gapless. However, we note that even though the red-blue and
green-blue pairs do not condense, their pairing fluctuations do exist. For weak attraction, the pairing fluctuation effects on the
single-particle excitations can be safely neglected and the BCS-like mean-field description is applicable. In this paper, we are
interested in the strongly interacting systems, where the pairing fluctuation effects become significant. Two systems will be
considered: (i) a resonantly interacting three-component Fermi gas and (ii) a two-flavor quark color superconductor at quark
chemical potential $\mu\sim 400$MeV.

The beyond-mean-field approach for the study of the pairing fluctuation effects adopted in this paper is a generalization of the
$T$-matrix theory for two-component (spin-$\frac{1}{2}$) fermionic systems~\cite{Pieri} to three-component systems and
is also a generalization of the $T$-matrix theory for heated quark matter~\cite{kitazawa} to the low temperature domain.
We will focus on the pairing fluctuation effects on the unpaired fermions in the color superfluids/superconductors, which is absent
in the two-component systems. The paper is organized as follows. In Sec. II, we study the pairing fluctuation effects in atomic
color superfluids, which may be realized in three-component atomic Fermi gases. In Sec. III, the pairing fluctuation effects in
two-flavor color superconductors will be discussed. We summarize in Sec. IV. The natural units $c=\hbar=k_{\rm B}=1$ will be used
throughout the paper.

\section{Color Superfluidity in Atomic Fermi Gases}

In this section we study the pairing fluctuation effects in an atomic color superfluid
\cite{3fermion01,3fermion02,3fermion03,3fermion04,3fermion05,3fermion06,3fermion07,3fermion08,3f09,3f10,3f11}, a cold atom analogue of color superconductivity in dense QCD~\cite{3fermion08}. We consider a dilute Fermi gas composed of three species of fermions with a common mass $m$. Such a system can be
realized in cold atom experiments by trapping the lowest three hyperfine states of the $^6$Li or $^{40}$K atoms in the atomic trap
\cite{three}. We assume that there exist short-range attractive interactions among difference species. The attraction strength can be
tuned from weak to strong by means of the Feshbach resonance. In the dilute limit, the attractive interactions can be modeled by contact
interactions. The Hamiltonian density of the system is given by
\begin{equation}\label{H1}
{\cal H}=\sum_{\alpha=1}^3\psi_\alpha^*\left(-\frac{\nabla^2}{2m}-\mu_\alpha\right)\psi_\alpha
-\sum_{\beta>\alpha=1}^3g_{\alpha\beta}^{\phantom{\dag}}\psi^*_\alpha\psi^*_\beta\psi^{\phantom{\dag}}_\beta\psi^{\phantom{\dag}}_\alpha,
\end{equation}
where $\mu_1,\mu_2,$ and $\mu_3$ are the chemical potentials for the three species, and $g_{12},g_{23},g_{13}$ are the contact
interactions among them.

In the following, we assume that the total particle number is fixed and the chemical potentials become equal, $\mu_1=\mu_2=\mu_3=\mu$.
Further, we assume that the three coupling constants also equal each other, $g_{12}=g_{13}=g_{23}=g$. Then the Hamiltonian (\ref{H1})
has a global SU$(3)$ symmetry. To show this explicitly, we define a three-component fermion field $\psi\equiv(\psi_1,\psi_2,\psi_3)^{\text T}$.
Then the Hamiltonian density (\ref{H1}) can be expressed as \cite{3fermion03}
\begin{equation}\label{H2}
{\cal H}=\psi^\dagger\left(-\frac{\nabla^2}{2m}-\mu\right)\psi+\frac{g}{4}\sum_{{\text a}=2,5,7}
(\psi^\dagger \lambda_{\text a}\psi^*)(\psi^{\text T}\lambda_{\text a}\psi),
\end{equation}
where $\lambda_{\text a}$ are the Gell-Mann matrices in the ``color space" spanned by the three fermion species. The  contact coupling
constant $g$ can be renormalized by introducing the $s$-wave scattering length $a_s$ of the short-range potential via the equation
\begin{equation}\label{NG}
\frac{1}{g(\Lambda)}=-\frac{m}{4\pi a_s}+\sum_{|{\bf k}|<\Lambda}\frac{1}{2\epsilon_{\bf k}}.
\end{equation}
Here $\epsilon_{\bf k}={\bf k}^2/(2m)$ is the free dispersion of the fermions, and the integral over the momentum ${\bf k}$ is associated
with a cutoff $\Lambda$. We can set $\Lambda\rightarrow\infty$ in the physical equations since all UV divergences are removed by Eq. (\ref{NG}).
In general, by tuning the scattering length from small to large negative values, we can realize an evolution from a weakly coupled BCS-like
color superfluid to a strongly coupled color superfluid.

\subsection{BCS mean-field theory}
Because of the attractive interactions among the unlike colors, at low temperature the system is in a color superfluid phase. Different to the
two-component (spin-$\frac{1}{2}$) fermionic systems where the superfluidity is characterized by a single-component order parameter
$\Delta\sim\langle\psi_\uparrow\psi_\downarrow\rangle$, the color superfluidity in the present three-component Fermi gas is
characterized by a three-component order parameter $\mbox{\boldmath{$\Delta$}}=(\Delta_1,\Delta_2,\Delta_3)$, where
\begin{eqnarray}
&&\Delta_1=\frac{g}{2}\langle\psi^{\text T}\lambda_2\psi\rangle=ig\langle\psi_1\psi_2\rangle,\nonumber\\
&&\Delta_2=\frac{g}{2}\langle\psi^{\text T}\lambda_5\psi\rangle=ig\langle\psi_2\psi_3\rangle,\nonumber\\
&&\Delta_3=\frac{g}{2}\langle\psi^{\text T}\lambda_7\psi\rangle=ig\langle\psi_3\psi_1\rangle.
\end{eqnarray}
Since the Hamiltonian (\ref{H2}) has an exact SU$(3)$ symmetry, we can show that the effective potential ${\cal V}(\mbox{\boldmath{$\Delta$}})$
depends only on the combination \cite{3fermion03}
\begin{eqnarray}
\mbox{\boldmath{$\Delta$}}\mbox{\boldmath{$\Delta$}}^\dagger=|\Delta_1|^2+|\Delta_2|^2+|\Delta_3|^2.
\end{eqnarray}
Therefore, different pairing configurations are physically equivalent and we can choose a specific gauge $\Delta_1=\Delta\neq0,
\Delta_2=\Delta_3=0$ without loss of generality. In this gauge, only the red and green fermions participate in the pairing and the red-blue pairs
condense, leaving the blue fermions unpaired. For a general gauge where all three components are nonzero, the unpaired branch is a linear
combination of the three colors. Because of the SU(3) symmetry of the Hamiltonian, different choices of the order parameter lead to the same physical results. For convenience, we use the gauge $\mbox{\boldmath{$\Delta$}}=(\Delta,0,0)$ in the following.

Before going on, we should mention that the BCS-BEC crossover in the present three-component Fermi gases is somewhat different from the
two-component systems due to the possibility of the appearance of a trionic phase, which is a Fermi-liquid state of three-fermion bound states \cite{3fermion02,3f09}. Based on an attractive Fermi Hubbard model, it was shown that there exists a quantum phase transition from the color superfluid phase to the trionic phase when the attraction strength exceeds a critical value \cite{3fermion02}. On the other hand, it was shown that large three-body losses in three-component Fermi gases confined in optical lattices can stabilize the color superfluid phase by suppressing the formation of trions \cite{3fermion06}. Therefore, in this paper we neglect the possibility of the trionic phase and consider the color superfluid phase only.

Now we turn to the general formalism. It is convenient to work with the Nambu-Gor'kov basis $\psi_{\rm{NG}}= (\psi, \psi^*)^{\text T}$.
In the Nambu-Gor'kov representation, the fermion self-energy $\Sigma(K)$ and the dressed fermion propagator ${\cal S}(K)$ are $2\times2$ matrices. They satisfy Dyson's equation
\begin{eqnarray}
&&\left(\begin{array}{cc} {\cal S}_{11}(K)&{\cal S}_{12}(K) \\ {\cal S}_{21}(K)&{\cal S}_{22}(K) \end{array}\right)^{-1}\nonumber\\
&=&\left(\begin{array}{cc} i\omega_n-\xi_{\bf k}&0 \\ 0&i\omega_n+\xi_{\bf k}\end{array}\right)
-\left(\begin{array}{cc} \Sigma_{11}(K)&\Sigma_{12}(K) \\ \Sigma_{21}(K)&\Sigma_{22}(K)\end{array}\right).
\end{eqnarray}
Here and in the following, $K=(i\omega_n,{\bf k})$ with $\omega_n=(2n+1)\pi T$ ($n$ integer) being the fermionic Matsubara frequency, and
$\xi_{\bf k}=\epsilon_{\bf k}-\mu$. From the Green's function relation we have the gap equation
\begin{equation}\label{gapeq}
\Delta=\frac{g}{2}\sum_K\text{Tr}\left[\lambda_2{\cal S}_{12}(K)\right]
\end{equation}
and the number equation
\begin{equation}\label{numeq}
n=\sum_K\text{Tr}\left[{\cal S}_{11}(K)\right].
\end{equation}
Here and in the following the notation $\sum_K=T\sum_n\sum_{\bf k}$
with $\sum_{\bf k}=\int d^3{\bf k}/(2\pi)^3$ is used.

In the BCS mean-field theory, the fermion self-energy is chosen as
\begin{eqnarray}\label{selfbcs}
\Sigma(K)=\Sigma^{\text{BCS}}(K)=\left(\begin{array}{cc} 0&\Delta\lambda_2 \\ \Delta\lambda_2&0 \end{array}\right)
\end{eqnarray}
and hence is momentum independent. Here we have set $\Delta$ to be real without loss of generality. Then the dressed fermion propagator
${\cal S}(K)$ can be evaluated as
\begin{eqnarray}\label{greenbcs}
{\cal S}_{11}^{\text{BCS}}(K)&=&{\cal G}_{\Delta}(K)\lambda_{\text{rg}}+ {\cal G}_{0}(K)\lambda_{\text{b}},\nonumber\\
{\cal S}_{12}^{\text{BCS}}(K)&=&{\cal F}_{\Delta}(K)\lambda_2,\nonumber\\
{\cal S}_{22}^{\text{BCS}}(K)&=&-{\cal S}_{11}^{\text{BCS}}(-K),\nonumber\\
{\cal S}_{21}^{\text{BCS}}(K)&=&{\cal S}_{12}^{\text{BCS}}(K).
\end{eqnarray}
Here we have defined two matrices, $\lambda_{\rm{rg}}=\rm{diag}(1,1,0)$ and $\lambda_{\rm{b}}=\rm{diag}(0,0,1)$, in the color space.
The Green's functions ${\cal G}_\Delta(K), {\cal F}_\Delta(K)$ and ${\cal G}_0(K)$ are analytically given by
\begin{eqnarray}
&&{\cal G}_\Delta(K)=\frac{i\omega_n+\xi_{\bf k}}{(i\omega_n)^2-E_{\bf k}^2},\nonumber\\
&&{\cal F}_{\Delta}(K)=\frac{\Delta}{(i\omega_n)^2-E_{\bf k}^2},\nonumber\\
&&{\cal G}_0(K)=\frac{1}{i\omega_n-\xi_{\bf k}},
\end{eqnarray}
where $E_{\bf k}=\sqrt{\xi_{\bf k}^2+\Delta^2}$ is the BCS-like dispersion.

In the above BCS mean-field description, it is clear that the paired colors possess BCS-like dispersions and obtain an excitation gap $\Delta$,
while the unpaired color has a free dispersion and remains gapless (for $\mu>0$). The physical value of the pairing gap $\Delta$ is determined
by the BCS gap equation
\begin{eqnarray}\label{gapbcs}
-\frac{m}{4\pi a_s}=\sum_{\bf k}\Bigg[\frac{1-2f(E_{\bf k})}{2E_{\bf k}} -\frac{1}{2\epsilon_{\bf k}}\Bigg],
\end{eqnarray}
where $f(E)=1/(e^{E/T}+1)$ is the Fermi-Dirac distribution. However, we note that there exists a serious problem for the naive BCS mean-field
approach. If we consider another system with the couplings, $g_{12}=g$ and $g_{13}=g_{23}=0$, i.e., the blue color is really free, we find no difference in comparison with the present system with $g_{12}=g_{13}=g_{23}=g$. Therefore, the pairing fluctuation effects are likely significant
in the color superfluid, especially when the attractive coupling $g$ is not weak.

\subsection{Fermion self-energy beyond BCS}

\begin{figure}[!htb]
\begin{center}
\includegraphics[width=8.2cm]{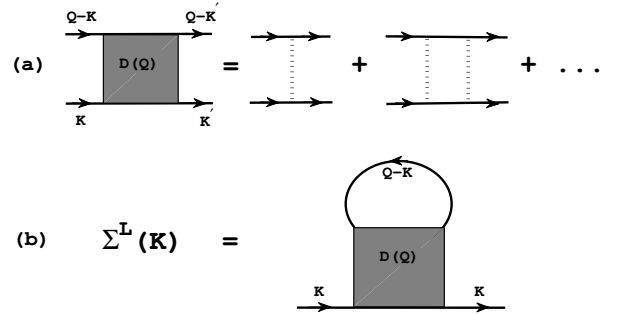}
\caption{The diagrammatic representation of the particle-particle
ladder ${\cal D}(Q)$ and the self-energy $\Sigma^L(K)$. The dashed
lines denote the coupling constant $g$. \label{fig2}}
\end{center}
\end{figure}

Now we take into account the pairing fluctuations and study their effects on the single-particle excitation spectra. To this end, we first
construct the particle-particle ladder ${\cal D}(Q)$ or the ``pair propagator," which is diagrammatically represented in Fig. \ref{fig2}(a).
In the color superfluid phase it is a $2\times2$ matrix,
\begin{equation}\label{ladder}
\left(\begin{array}{cc} {\cal D}_{11}^{\text{ab}}(Q)&{\cal D}_{12}^{\text{ab}}(Q)
\\ {\cal D}_{21}^{\text{ab}}(Q)&{\cal D}_{22}^{\text{ab}}(Q)
\end{array}\right)=\left(\begin{array}{cc} \chi_{11}^{\text{ab}}(Q)&\chi_{12}^{\text{ab}}(Q)
\\ \chi_{21}^{\text{ab}}(Q)&\chi_{22}^{\text{ab}}(Q)
\end{array}\right)^{-1},
\end{equation}
where $\chi(Q)$ is the pair susceptibility. Note that the matrix elements of ${\cal D}(Q)$ and $\chi(Q)$ are also matrices in an adjoint space of the SU(3) group spanned by the indices $\text{a,b}=2,5,7$. The explicit form of the pair susceptibility $\chi(Q)$ is given by
\begin{eqnarray}
\chi_{11}^{\text{ab}}(Q)&=&\frac{\delta_{\text{ab}}}{g}-\frac{1}{2}\sum_K\text{Tr}\left[\lambda_{\text a}{\cal S}_{11}(Q-K)
\lambda_{\text b}{\cal S}_{11}(K)\right],\nonumber\\
\chi_{12}^{\text{ab}}(Q)&=&\frac{1}{2}\sum_K\text{Tr}\left[\lambda_{\text a}{\cal S}_{12}(Q-K)\lambda_{\text b}{\cal S}_{12}(K)\right],
\nonumber\\
\chi_{22}^{{\text {ab}}}(Q)&=&\chi_{11}^{{\text {ab}}}(-Q),\ \ \ \ \ \ \ \ \ \ \ \ \ \ \chi_{21}^{{\text {ab}}}(Q)=\chi_{12}^{{\text {ab}}}(Q).
\end{eqnarray}
Here and in the following, $Q=(i\nu_n,{\bf q})$ with $\omega_n=2n\pi T$ ($n$ integer) being the bosonic Matsubara frequency. The fermion self-energy
beyond the BCS approximation (\ref{selfbcs}) can be considered by taken into account the self-energy $\Sigma^L(K)$ shown in Fig. \ref{fig2}(b).
Now the full fermion self-energy in our consideration is given by
\begin{eqnarray}
\Sigma(K)=\left(\begin{array}{cc} \Sigma_{11}^L(K)&\Sigma_{12}^L(K)\\
\Sigma_{21}^L(K)&\Sigma_{22}^L(K)\end{array}\right)+\Sigma^{\text{BCS}}(K),
\end{eqnarray}
where $\Sigma_{\text{ij}}^L(K)$ can be expressed as
\begin{eqnarray}
\Sigma_{11}^L(K)&=&-\sum_{\rm{a,b}}\sum_Q{\cal D}_{11}^{\text{ab}}(Q)\lambda_{\text a}{\cal S}_{11}(Q-K)\lambda_{\text b},\nonumber\\
\Sigma_{12}^L(K)&=&-\sum_{\rm{a,b}}\sum_Q{\cal D}_{12}^{\text{ab}}(Q)\lambda_{\text a}{\cal S}_{12}(Q-K)\lambda_{\text b},\nonumber\\
\Sigma_{22}^L(K)&=&-\Sigma_{11}^L(-K),\ \ \ \ \ \ \ \Sigma_{21}^L(K)=\Sigma_{12}^L(K).
\end{eqnarray}
We note that for two-component systems, the above $T$-matrix approach is the same as that developed in \cite{Pieri}. Above the superfluid transition
temperature, it also recovers the $T$-matrix approach adopted in \cite{PG4}.

In general, the above equations together with the gap and number equations (\ref{gapeq}) and (\ref{numeq}) form a closed set of integral equations for the superfluid order parameter $\Delta$, the chemical potential $\mu$, and the dressed fermion propagator ${\cal S}(K)$. However, to keep the
Goldstone's theorem and recover the correct pair excitation spectrum at strong coupling, we adopt the following prescriptions suggested in \cite{Pieri}: (i) In evaluating the pair susceptibility $\chi(Q)$ and the self-energy $\Sigma^L(K)$, we use the the fermion propagator
${\cal S}(K)$ of its BCS mean-field form (\ref{greenbcs}); (ii) the off-diagonal fermion propagator ${\cal S}_{12}(K)$ in the gap equation (\ref{gapeq}) is also replaced by its BCS mean-field form, while in the number equation (\ref{numeq}) we take into account the pairing fluctuation effects on the diagonal fermion propagator ${\cal S}_{11}(K)$. Therefore, the gap equation (\ref{gapeq}) for the superfluid order parameter $\Delta$ still takes its BCS form (\ref{gapbcs}), which ensures a gapless pair excitation spectrum, i.e., the Goldstone's theorem. The beyond-mean-field corrections for the pairing gap $\Delta$ and chemical potential $\mu$ are reflected in the full number equation (\ref{numeq}). The pairing fluctuation effects on the single-particle excitation spectra are included in the self-energy $\Sigma^L(K)$.

After some simple matrix algebras, we find that the pair susceptibility $\chi(Q)$ and the particle-particle ladder ${\cal D}(Q)$ are both diagonal
in the adjoint space, i.e.,
\begin{eqnarray}
\chi^{\text{ab}}_{\text{ij}}(Q)=\chi^{\text a}_{\text{ij}}(Q)\delta_{\text{ab}},\ \ \ \ \ \ \ \
{\cal D}^{\text{ab}}_{\text{ij}}(Q)={\cal D}^{\text a}_{\text{ij}}(Q)\delta_{\text{ab}}.
\end{eqnarray}
For the ${\text a}=2$ sector which represents the red-green pairing, the pair susceptibility $\chi^{2}(Q)$ can be evaluated as
\begin{eqnarray}
\chi_{11}^{2}(Q)&=&-\frac{m}{4\pi a_s}-\sum_{\bf k}\left[T\sum_n{\cal G}_{\Delta}(K){\cal G}_{\Delta}(Q-K)
-\frac{1}{2\epsilon_{\bf k}}\right],\nonumber\\
\chi_{12}^{2}(Q)&=&\sum_K{\cal F}_{\Delta}(K){\cal F}_{\Delta}(Q-K).
\end{eqnarray}
The corresponding particle-particle ladder ${\cal D}^{2}(Q)$ reads
\begin{eqnarray}
{\cal D}_{11}^{2}(Q)&=&\frac{\chi_{11}^{2}(-Q)}{\chi_{\text{11}}^{2}(Q)\chi_{\text{11}}^{2}(-Q)-[\chi_{\text{12}}^{2}(Q)]^2},\nonumber\\
{\cal D}_{12}^{2}(Q)&=&\frac{\chi_{12}^{2}(Q)}{\chi_{\text{11}}^{2}(Q)\chi_{\text{11}}^{2}(-Q)-[\chi_{\text{12}}^{2}(Q)]^2}.
\end{eqnarray}
Using the BCS gap equation (\ref{gapbcs}), we can show that
\begin{eqnarray}
\chi_{\text{11}}^{2}(0,{\bf 0})=\chi_{\text{12}}^{2}(0,{\bf 0}).
\end{eqnarray}
Therefore, the pair excitation from the $\rm{a}=2$ sector is gapless, corresponding to one broken generator $\lambda_{\rm{rg}}$. We note that
this Goldstone mode is essentially the same as the Anderson-Bogoliubov mode in the conventional two-component fermionic superfluids \cite{two}.
It possesses a linear dispersion $\omega({\bf q})\propto |{\bf q}|$ in the low momentum and frequency limit. In the strong coupling limit, it recovers the Bogoliubov excitation spectrum for a weakly interacting Bose gas \cite{two,Pieri}.

The sectors ${\text a}=5$ and ${\text a}=7$ represent the red-blue and green-blue pairings, respectively. They are degenerate, corresponding
to the residual SU(2) symmetry group. Since these pairs do not condense according to our order parameter gauge, the off-diagonal components
vanish. We have
\begin{eqnarray}
\chi_{12}^{5}(Q)=\chi_{12}^{7}(Q)=0,\ \ \ \ \ \ \ {\cal D}_{12}^{5}(Q)={\cal D}_{12}^{7}(Q)=0.
\end{eqnarray}
The nonvanishing diagonal component of the pair susceptibility can be evaluated as
\begin{eqnarray}
\chi_{11}^{\text a}(Q)=-\frac{m}{4\pi a_s}-\sum_{\bf k}\left[T\sum_n{\cal G}_{\Delta}(K){\cal G}_{0}(Q-K)-\frac{1}{2\epsilon_{\bf k}}\right].
\end{eqnarray}
The particle-particle ladder is also diagonal and is given by
\begin{eqnarray}
{\cal D}_{11}^{\text a}(Q)&=&\frac{1}{\chi_{11}^{\text a}(Q)},\ \ \
{\text a}=5,7.
\end{eqnarray}
Using the BCS gap equation (\ref{gapbcs}), we find that
\begin{eqnarray}
\chi_{\text{11}}^{\rm a}(0,{\bf 0})=0,\ \ \ {\text a}=5,7,
\end{eqnarray}
which indicates some additional Goldstone modes corresponding to the broken generators $\lambda_4,\lambda_5,\lambda_6,\lambda_7$ of the SU(3)
group \cite{3fermion03}. They possess a quadratic dispersion $\omega({\bf q})\sim {\bf q}^2$ at low momentum and frequency due to the asymmetry between the paired and unpaired colors \cite{3fermion03}.

Using the above matrix structure of the particle-particle ladder ${\cal D}(Q)$, we find that the diagonal component of the self-energy $\Sigma_{11}^L(K)$ takes the form
\begin{eqnarray}
\Sigma_{11}^L(K)&=&\Sigma_{\text{rg}}(K)\lambda_{\text{rg}}+\Sigma_{\text b}(K)\lambda_{\text b},
\end{eqnarray}
where $\Sigma_{\text{rg}}(K)$ and $\Sigma_{\text{b}}(K)$ correspond to the beyond-BCS self-energies for the paired and unpaired colors,
respectively. Their explicit forms are given by
\begin{eqnarray}\label{self1}
\Sigma_{\text{rg}}(K)&=&-\sum_Q\left[{\cal D}_{11}^2(Q){\cal G}_{\Delta}(Q-K)+{\cal D}_{11}^5(Q){\cal G}_{0}(Q-K)\right],\nonumber\\
\Sigma_{\text{b}}(K)&=&-2\sum_Q{\cal D}_{11}^5(Q){\cal G}_{\Delta}(Q-K).
\end{eqnarray}
These results manifest the schematic plot in Fig. \ref{fig1}: Each color couples to the other two colors via the corresponding particle-particle ladders. The off-diagonal component $\Sigma_{12}^L(K)$ can be evaluated as
\begin{eqnarray}
\Sigma_{12}^L(K)=\sum_Q{\cal D}_{12}^{2}(Q){\cal F}_{\Delta}(Q-K)\lambda_2.
\end{eqnarray}
We can neglect this off-diagonal contribution since it is generally much smaller than the BCS self-energy $\Sigma^{\rm BCS}$ \cite{Pieri}.
However, this approximation has no effect on the spectrum of the unpaired color we are interested in in the following.

It is worth comparing the self-energy $\Sigma_{\text{rg}}(K)$ with that in the two-component (spin-$\frac{1}{2}$) system. In the two-component
system, the self-energy $\Sigma_{\text{rg}}(K)$ reads \cite{Pieri}
\begin{eqnarray}
\Sigma_{\text{rg}}(K)=-\sum_Q{\cal D}_{11}^2(Q){\cal G}_{\Delta}(Q-K).
\end{eqnarray}
The additional contribution $-\sum_Q{\cal D}_{11}^5(Q){\cal G}_{0}(Q-K)$ is absent in two-component systems. On the other hand, in the
normal phase $T\geq T_c$, the SU(3) symmetry is restored and the three color becomes degenerate. For the three-component system we have
\begin{eqnarray}
\Sigma_{\text{rg}}(K)=\Sigma_{\rm b}(K)=-2\sum_Q{\cal D}_0(Q){\cal G}_0(Q-K),
\end{eqnarray}
where ${\cal D}_0(Q)$ is the common particle-particle ladder in the normal state with $\Delta=0$. However, for the two-component system,
the self-energy in the normal phase is \cite{Pieri,PG4}
\begin{eqnarray}
\Sigma_{\text{rg}}(K)=-\sum_Q{\cal D}_0(Q){\cal G}_0(Q-K).
\end{eqnarray}
Therefore, the self-energy induced by the pairing fluctuation effects in the normal phase is enhanced by a factor of 2 in comparison with
the two-component system.
\begin{figure*}
\begin{center}
\includegraphics[width=8cm]{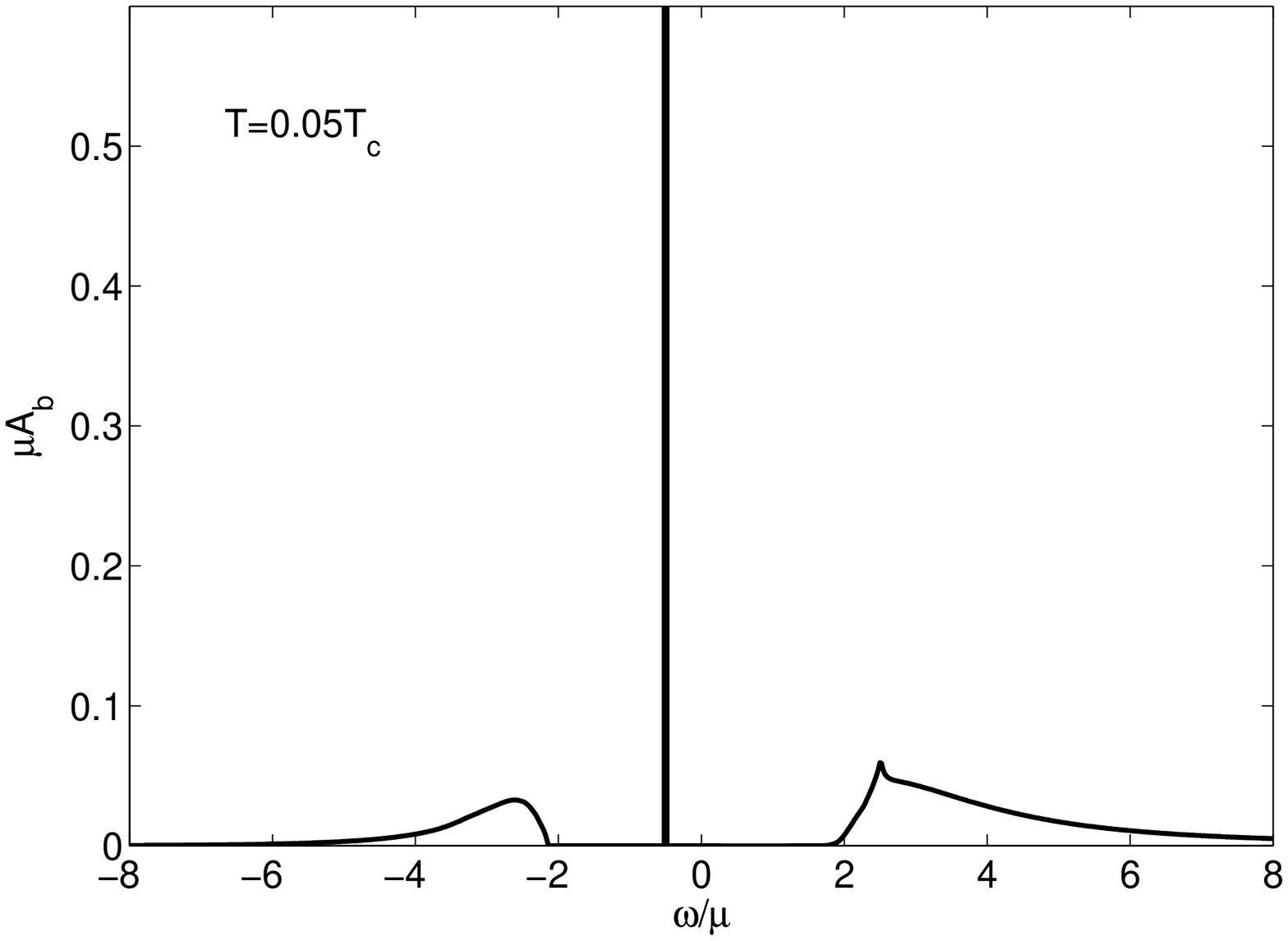}
\includegraphics[width=8cm]{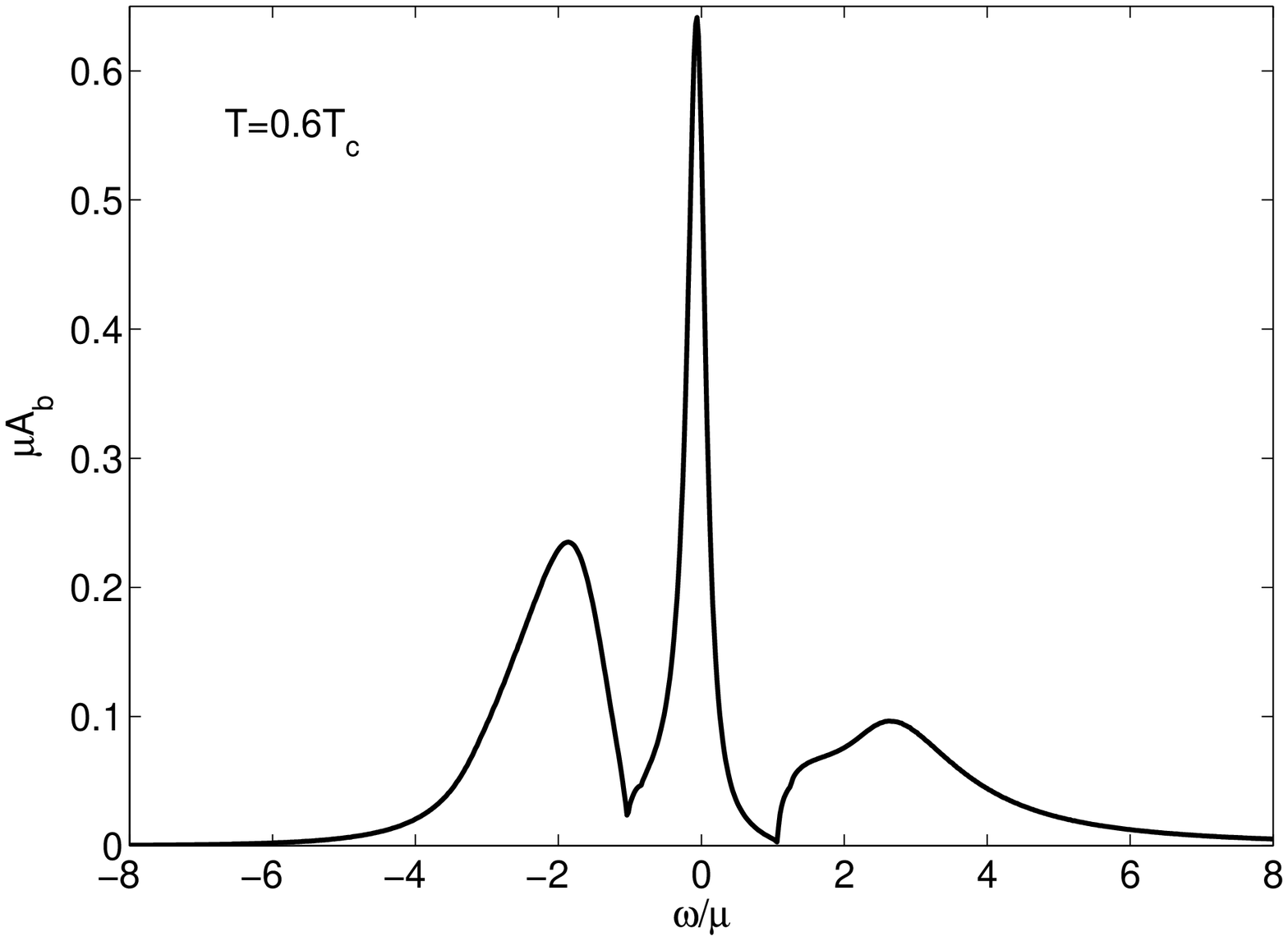}
\includegraphics[width=8cm]{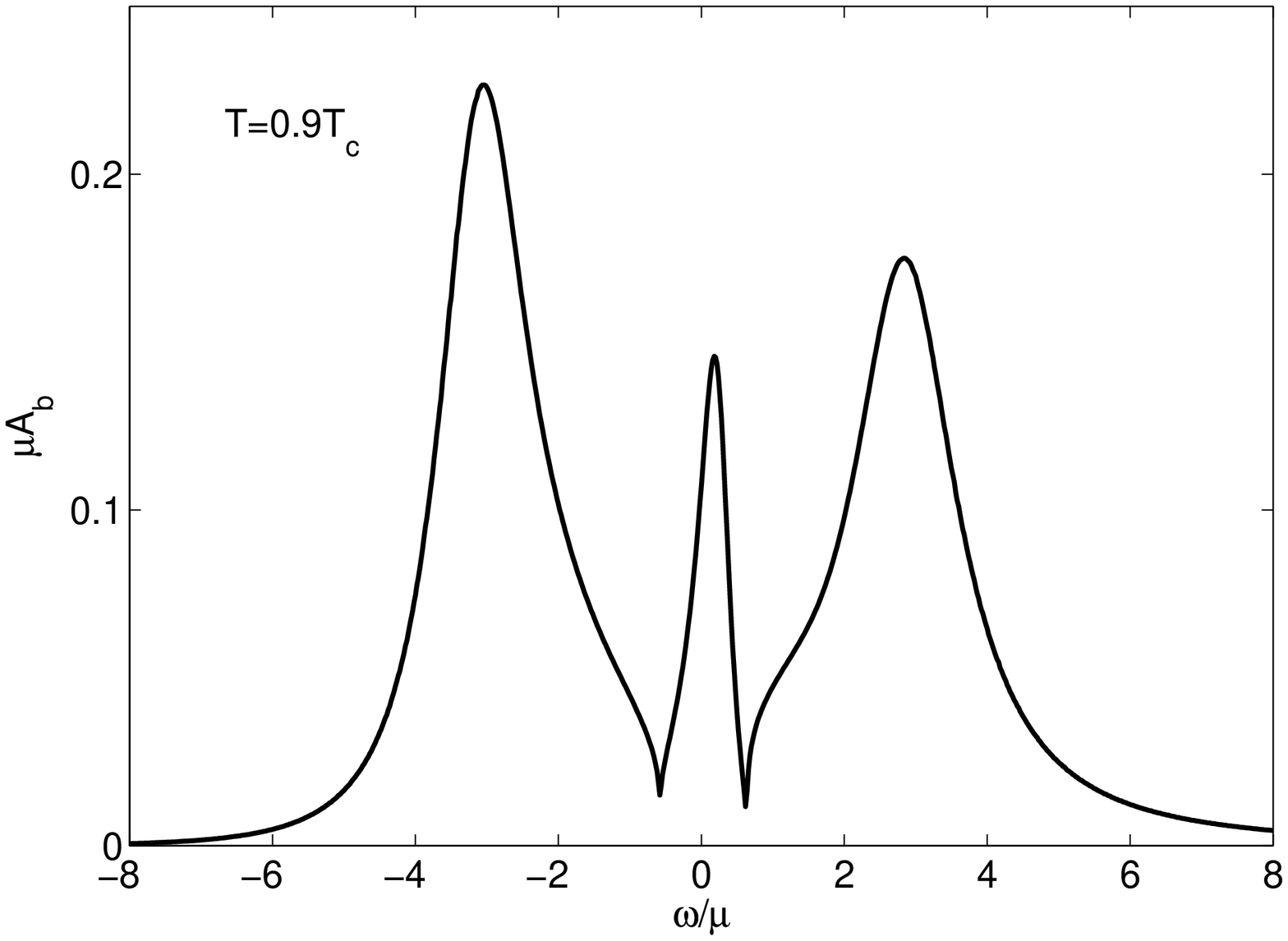}
\includegraphics[width=8cm]{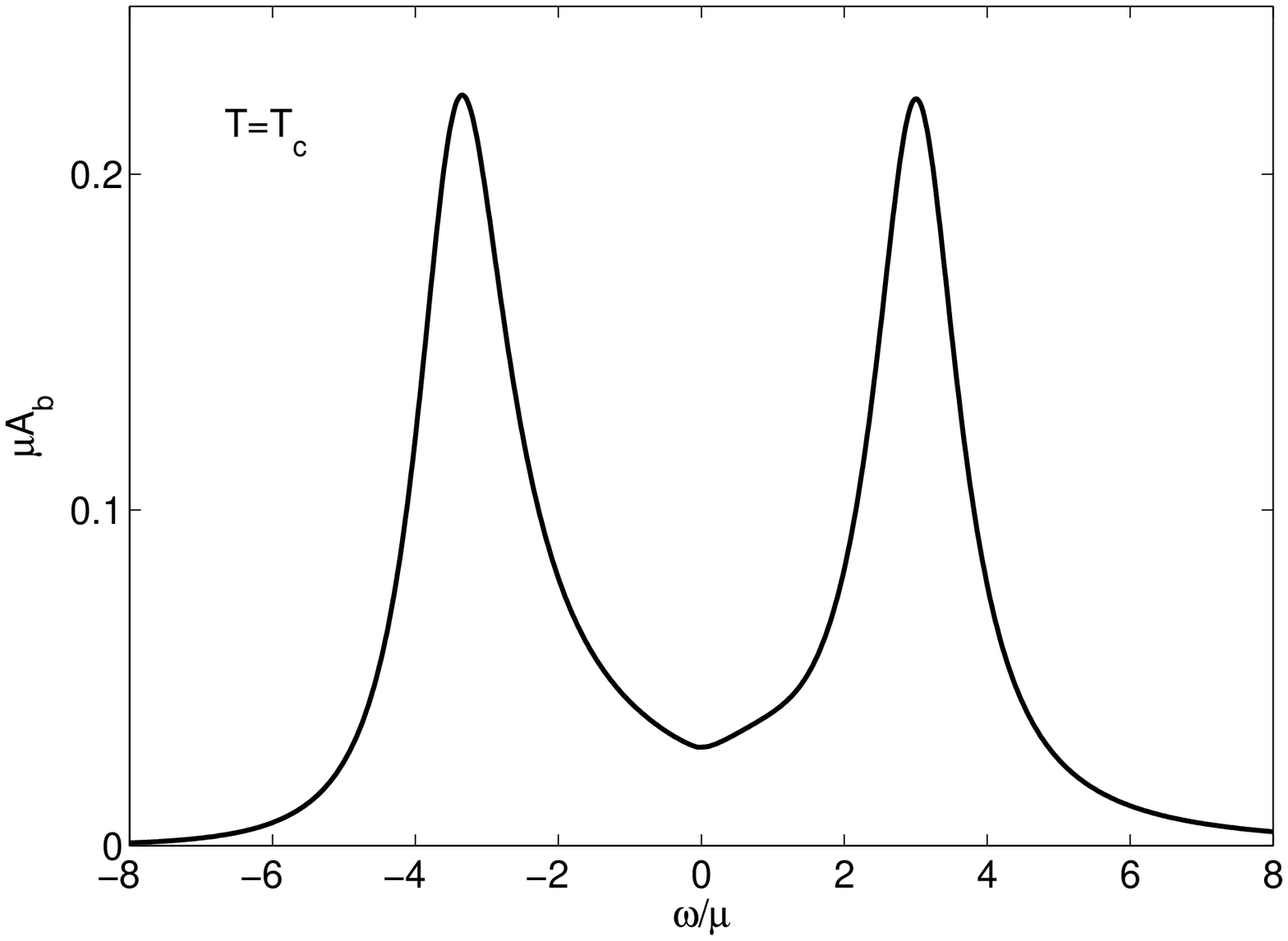}
\caption{The spectral density ${\cal A}_{\rm b}(\omega,k)$ of the unpaired blue color for $k=k_\mu=\sqrt{2m\mu}$ at various temperatures. \label{fig3}}
\end{center}
\end{figure*}

\subsection{Fermion spectral density}
To study the pairing fluctuation effects on the fermionic excitations, we investigate the single-particle spectral density function
${\cal A}(\omega,{\bf k})$. It can be obtained from the dressed fermion Green's function ${\cal S}_{11}(K)$. Ensured by the residual SU(2)
symmetry, the paired and unpaired colors decouple even though the pairing fluctuation effects are taken into account. We have
\begin{eqnarray}
{\cal S}_{11}(K)&=&{\cal G}_{\text{rg}}(K)\lambda_{\text{rg}}+{\cal G}_{\text{b}}(K)\lambda_{\text{b}},
\end{eqnarray}
where ${\cal G}_{\text{rg}}(K)$ and ${\cal G}_{\text{b}}(K)$ are the dressed propagators for the paired and unpaired colors, which can be
expressed as
\begin{eqnarray}
{\cal G}_{\text{rg}}(i\omega_n,{\bf k})=\frac{1}{i\omega_n-\xi_{\bf k}-\Sigma_{\text{rg}}(i\omega_n,{\bf k})
-\frac{\Delta^2}{i\omega_n+\xi_{\bf k}+\Sigma_{\text{rg}}(-i\omega_n,{\bf k})}}
\end{eqnarray}
and
\begin{eqnarray}
{\cal G}_{\text{b}}(i\omega_n,{\bf k})=\frac{1}{i\omega_n-\xi_{\bf
k}-\Sigma_{\text{b}}(i\omega_n,{\bf k})},
\end{eqnarray}
respectively. We emphasize that the expression for ${\cal G}_{\text{b}}(i\omega_n,{\bf k})$ holds even though the off-diagonal self-energy
$\Sigma_{12}^L(K)$ is taken into account. In the BCS mean-field approximation the pairing fluctuation effects are absent and we have
${\cal G}_{\text{rg}}(K)={\cal G}_\Delta(K)$ and ${\cal G}_{\text{b}}(K)={\cal G}_0(K)$.

The spectral functions ${\cal A}_{\alpha}(\omega,{\bf k})$ for the paired and unpaired colors are defined as
\begin{eqnarray}
{\cal A}_\alpha(\omega,{\bf k}) =-\frac{1}{\pi}\text{Im}{\cal G}_{\alpha}^{\rm R}(\omega,{\bf k}),\ \ \alpha={\text{rg}},{\text{b}}.
\end{eqnarray}
Here and in the following, the retarded Green's functions are denoted by the subscript R, i.e., $X^{\rm R}(\omega)\equiv X(\omega+i\epsilon)$
where $\epsilon=0^+$. In the BCS mean-field approximation we have
\begin{eqnarray}
{\cal A}_{\rm{rg}}^{\rm{BCS}}(\omega,{\bf k})=u_{\bf k}^2\delta(\omega-E_{\bf k})+\upsilon_{\bf k}^2\delta(\omega+E_{\bf k})
\end{eqnarray}
and
\begin{eqnarray}
{\cal A}_{\rm{b}}^{\rm{BCS}}(\omega,{\bf k})=\delta(\omega-\xi_{\bf k}).
\end{eqnarray}
Here $u_{\bf k}^2=(1/2)(1+\xi_{\bf k}/E_{\bf k})$ and $\upsilon_{\bf k}^2=(1/2)(1-\xi_{\bf k}/E_{\bf k})$ are the BCS distribution functions.
Since the paired colors obtain a large pairing gap $\sim\Delta$ at strong coupling, their spectral density function ${\cal A}_{\rm{rg}}
(\omega,{\bf k})$ remains gaplike even though the pairing fluctuation effects are taken into account. While there exists an additional
contribution in the self-energy $\Sigma_{\rm{rg}}(K)$ which is absent in the two-component systems, we expect that the qualitative feature of
the spectral density ${\cal A}_{\rm{rg}}(\omega,{\bf k})$ is similar to the results for the two-component systems \cite{Pieri}. In general,
pairing fluctuation and temperature effects lead to broadening of the gaplike peaks.

\begin{figure*}
\begin{center}
\includegraphics[width=8cm]{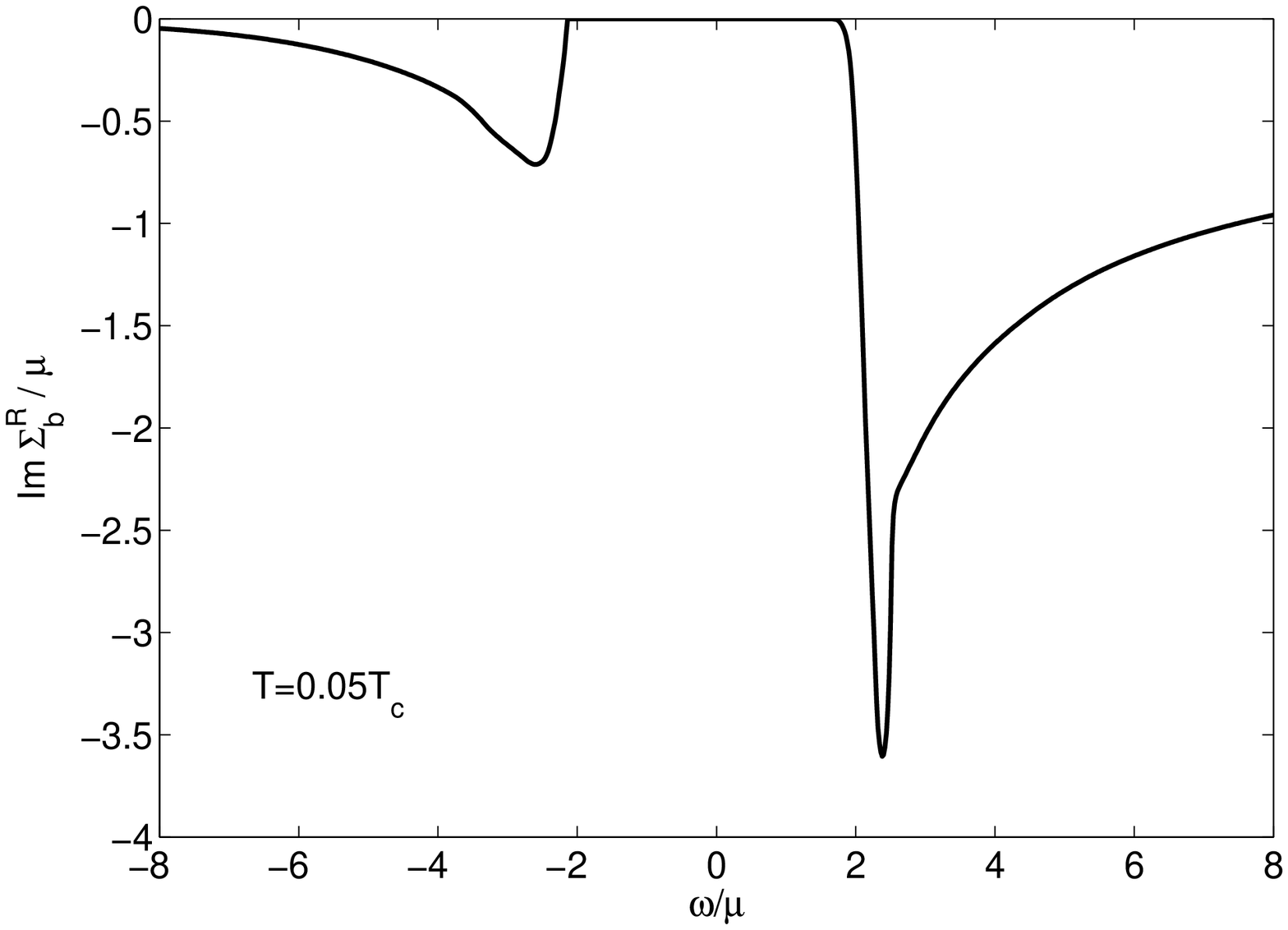}
\includegraphics[width=8cm]{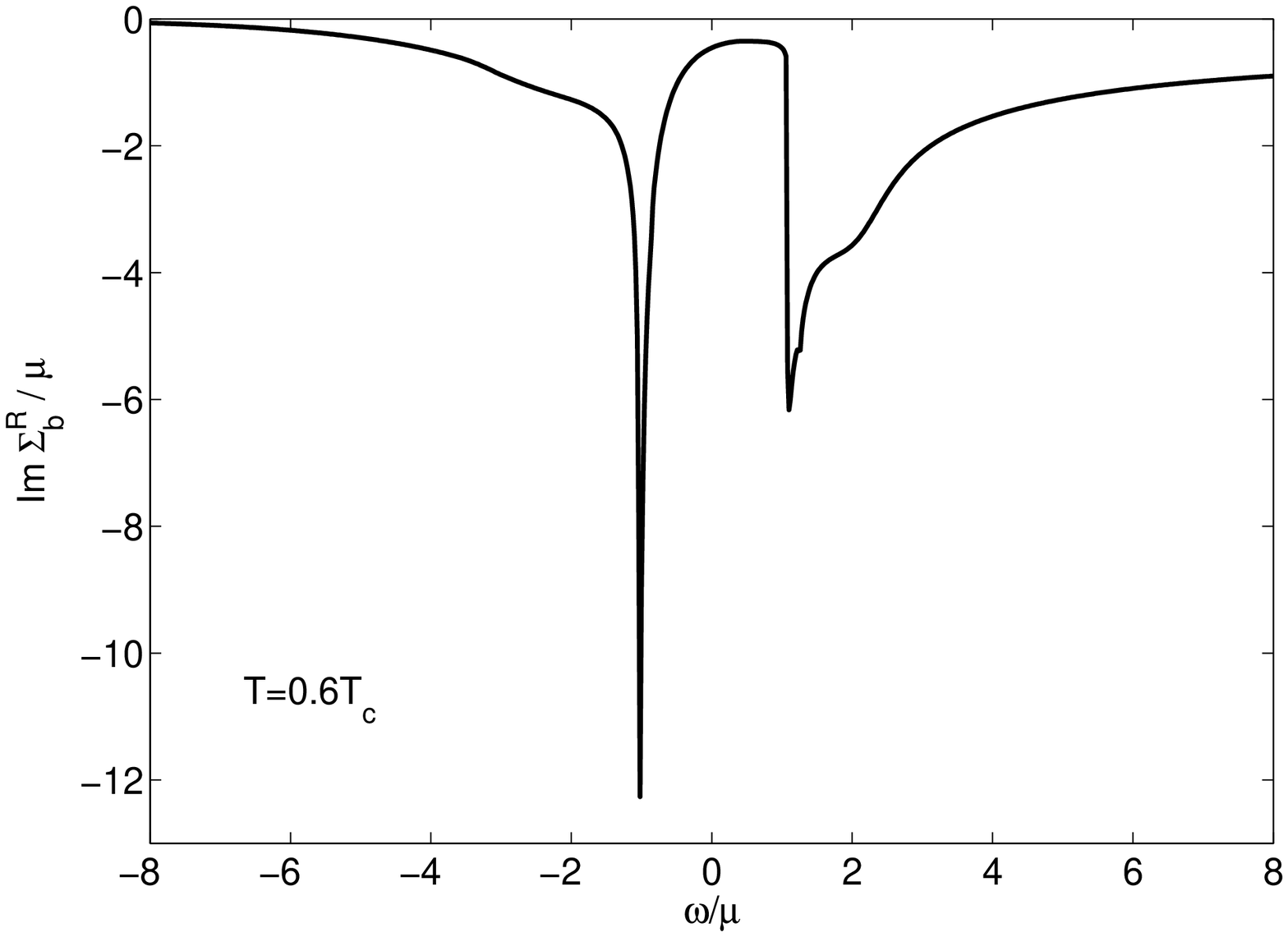}
\includegraphics[width=8cm]{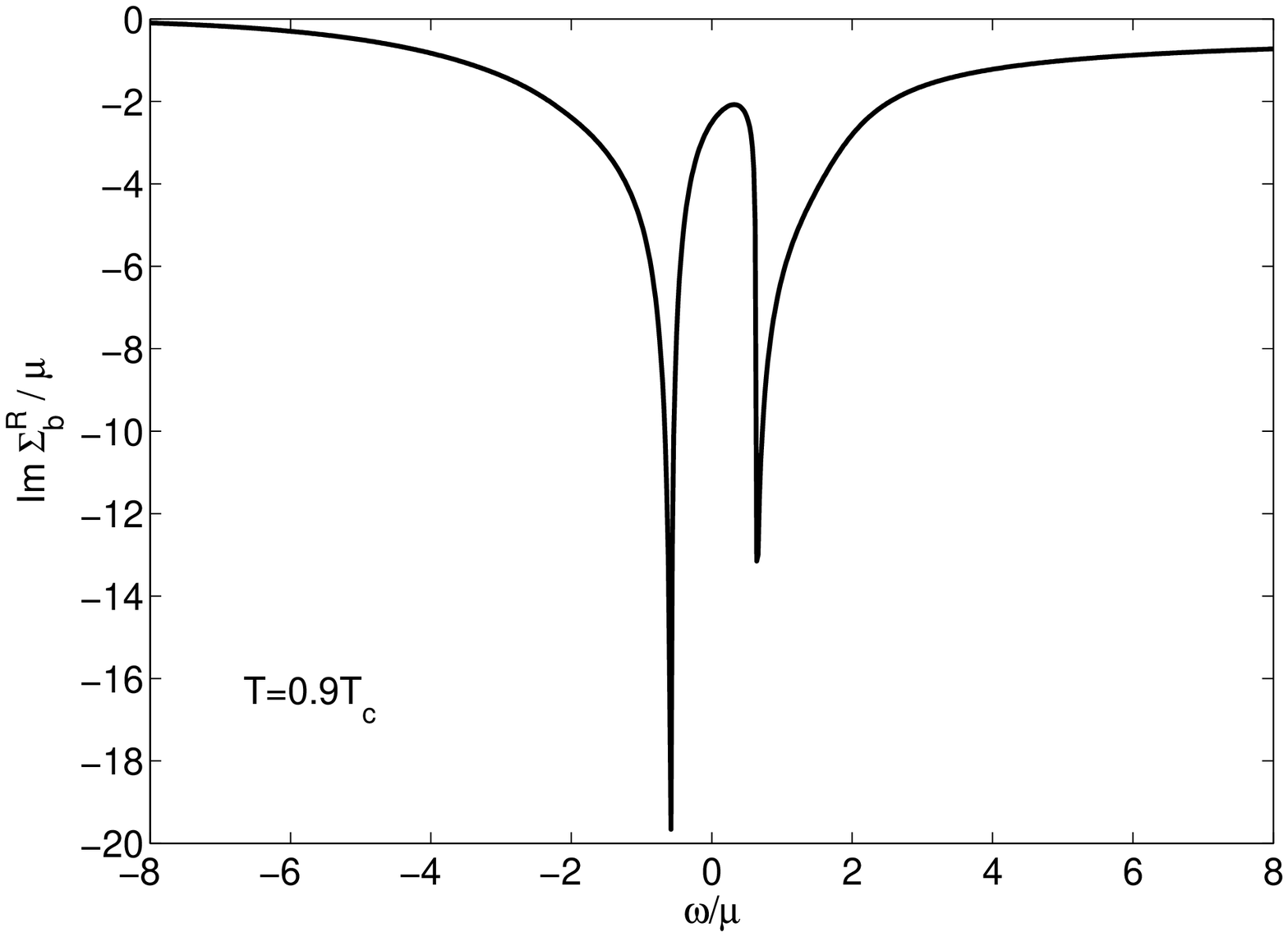}
\includegraphics[width=8cm]{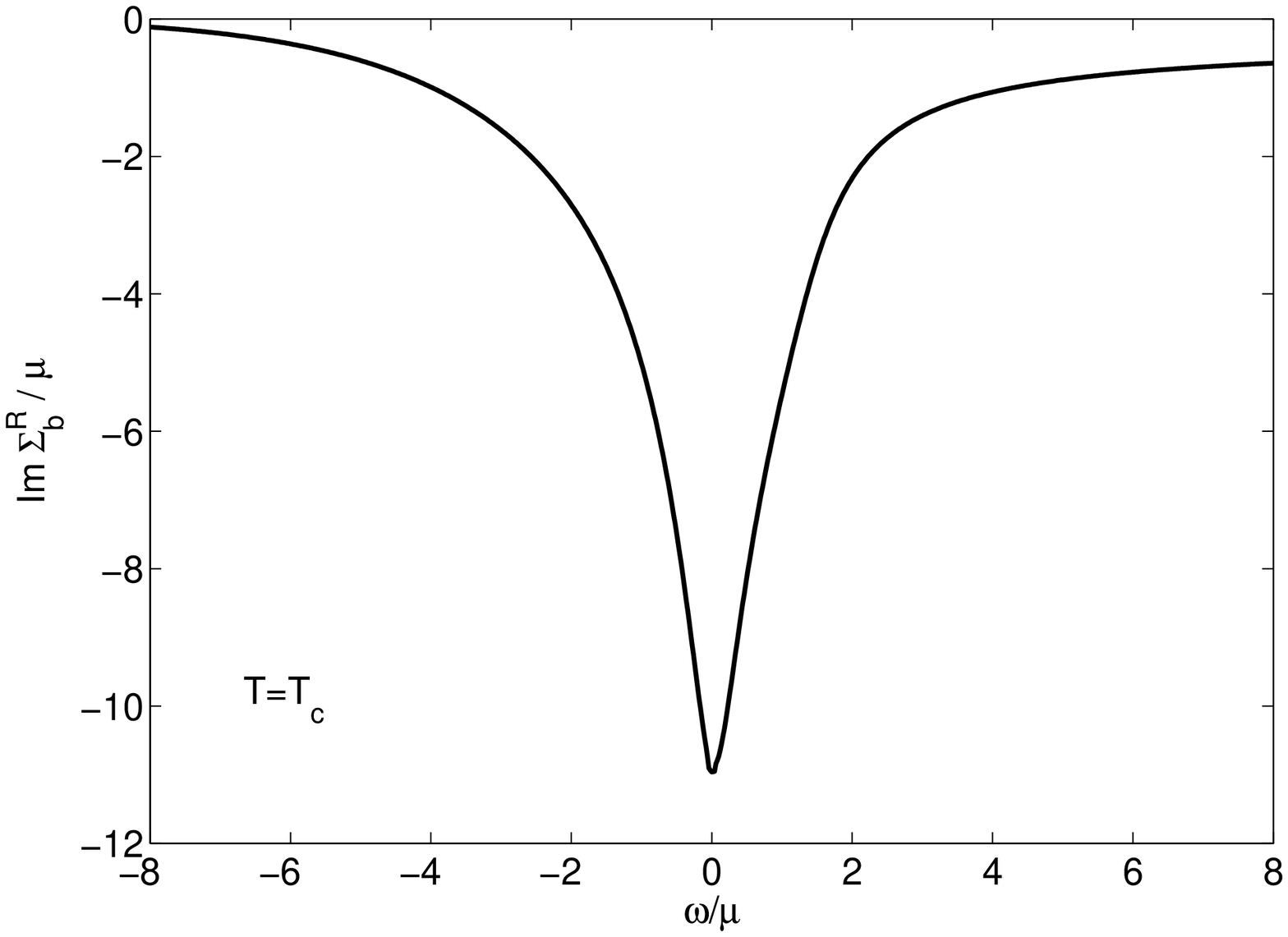}
\caption{The imaginary part of the retarded self-energy $\Sigma_{\rm{b}}^{\rm R}(\omega,k)$ of the unpaired blue color for $k=k_\mu=\sqrt{2m\mu}$ at various temperatures. \label{fig4}}
\end{center}
\end{figure*}

\begin{figure*}
\begin{center}
\includegraphics[width=8cm]{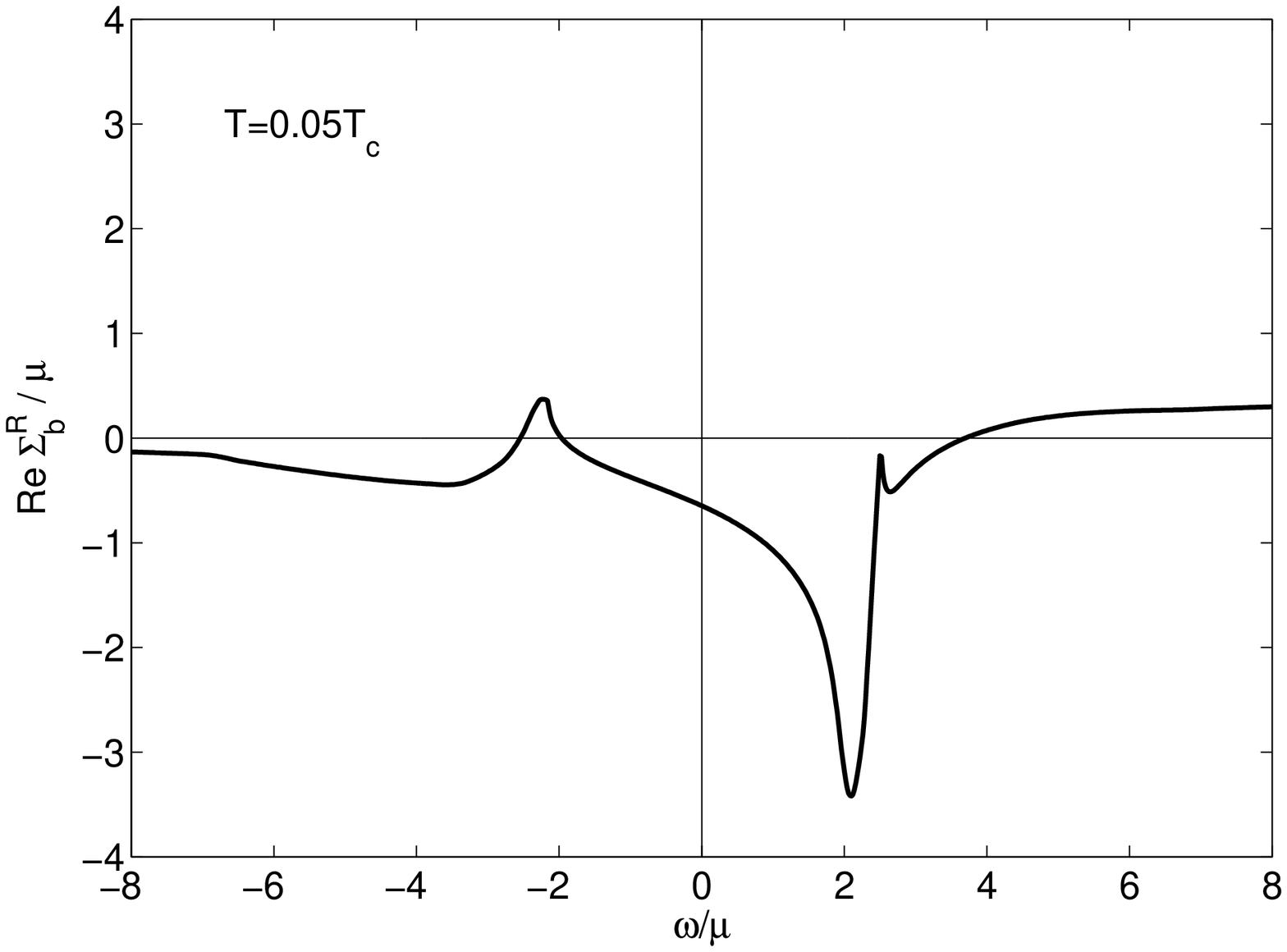}
\includegraphics[width=8cm]{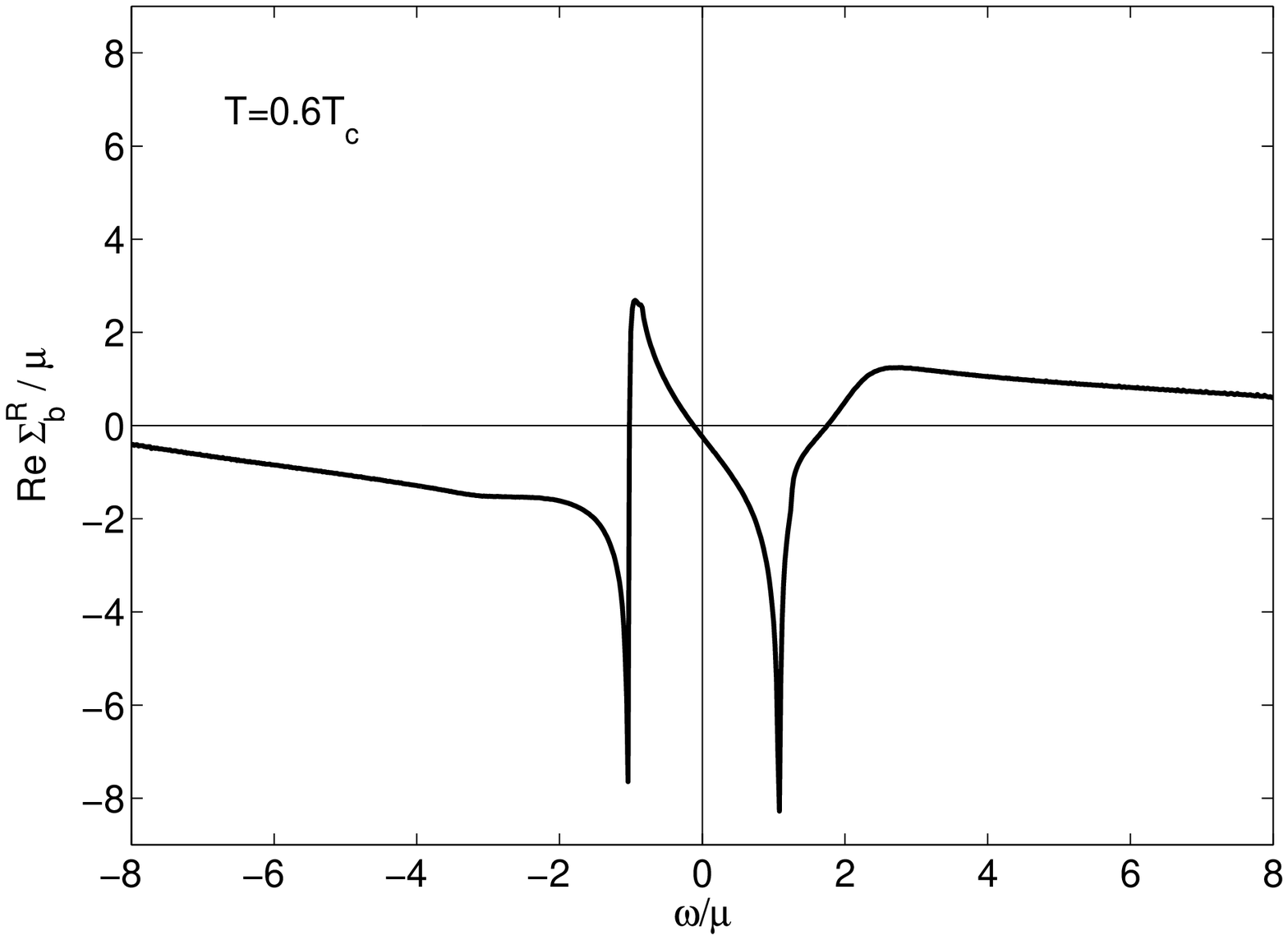}
\includegraphics[width=8cm]{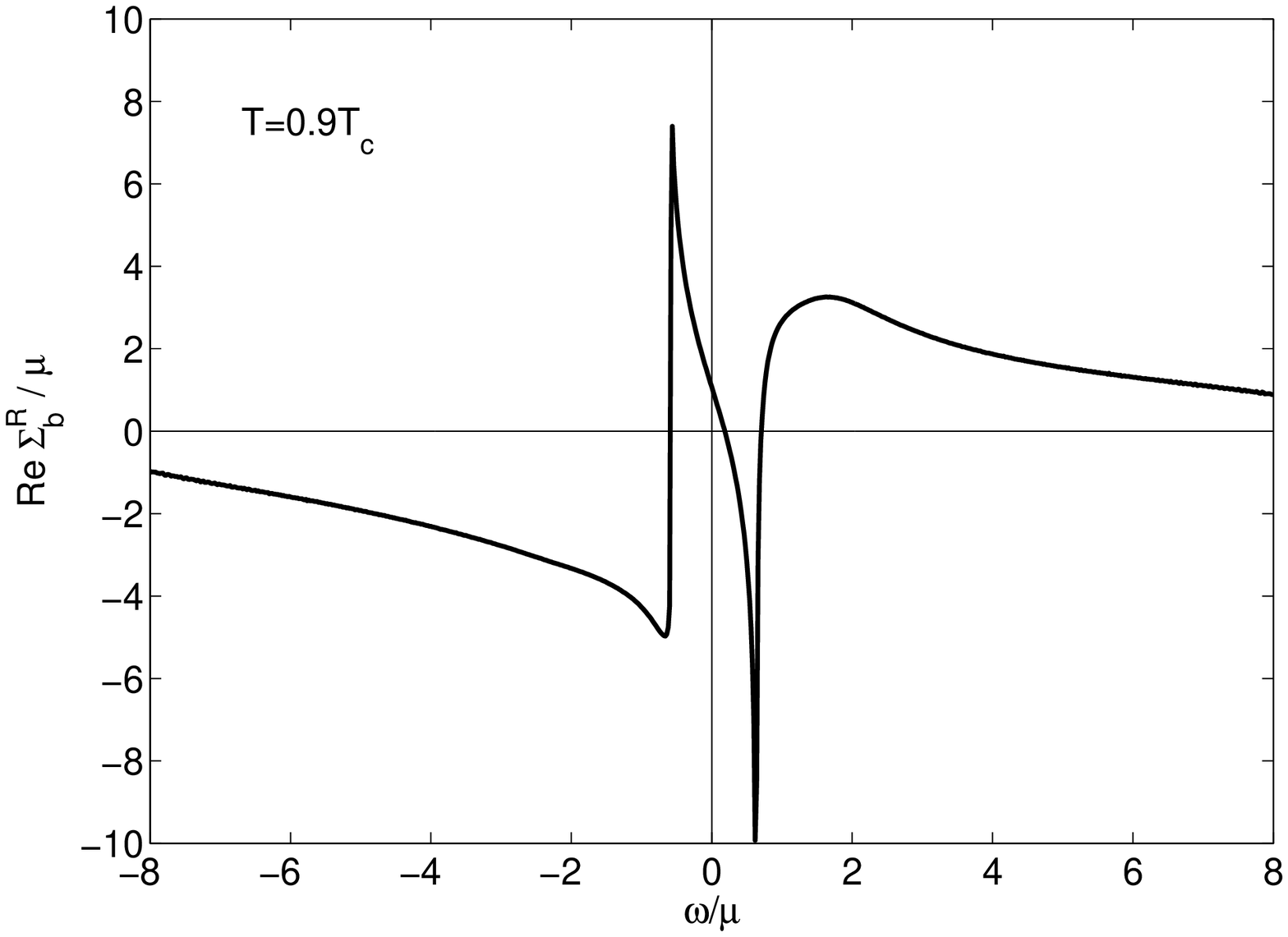}
\includegraphics[width=8cm]{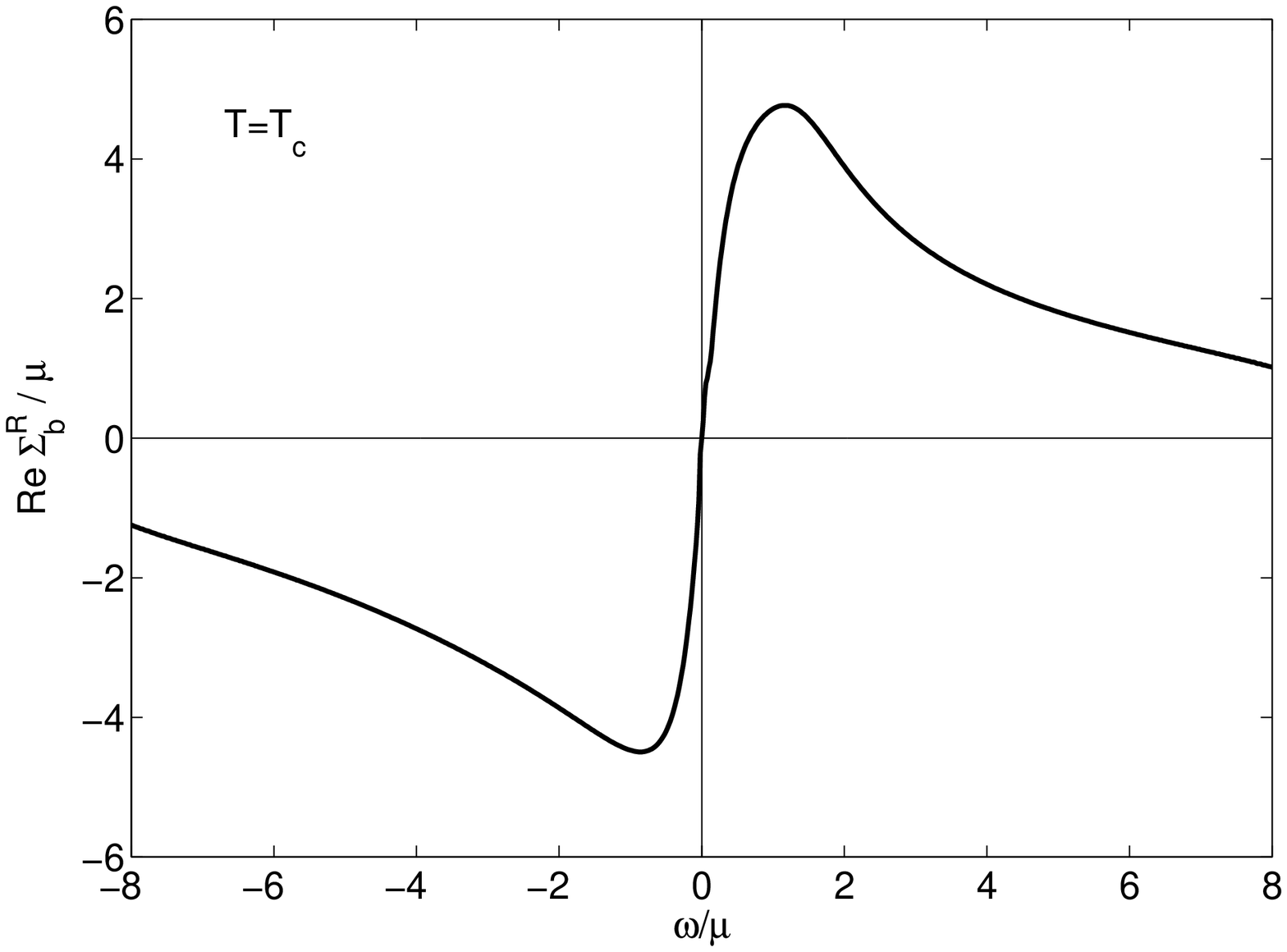}
\caption{The real part of the retarded self-energy $\Sigma_{\rm{b}}^{\rm R}(\omega,k)$ of the unpaired blue color for $k=k_\mu=\sqrt{2m\mu}$ at various temperatures. \label{fig5}}
\end{center}
\end{figure*}

In the following, we will focus on the spectrum of the unpaired blue color, which is unique in the present three-component Fermi system. It is interesting to study how the pairing fluctuation effects influence the spectrum of the unpaired color which takes a free dispersion in the naive
BCS mean-field description. The spectral density function ${\cal A}_{\rm{b}}(\omega,{\bf k})$ can be expressed as
\begin{eqnarray}\label{sp1}
{\cal A}_{\rm{b}}(\omega,{\bf k})=-\frac{1}{\pi}\frac{\rm{Im}\Sigma_{\rm{b}}^{\rm R}(\omega,{\bf k})}
{[\omega-\xi_{\bf k}-\rm{Re}\Sigma_{\rm{b}}^{\rm R}(\omega,{\bf k})]^2+[\rm{Im}\Sigma_{\rm{b}}^{\rm R}(\omega,{\bf k})]^2}.
\end{eqnarray}
The real and imaginary parts of the retarded self-energy $\Sigma_{\rm{b}}^{\rm R}(\omega,{\bf k})$ are related by the dispersion relation
\begin{eqnarray}\label{sp2}
{\rm Re}\Sigma_{\rm{b}}^{\rm R}(\omega,{\bf k})={\rm P}\int_{-\infty}^{+\infty}\frac{d\omega^\prime}{\pi}
\frac{{\rm Im}\Sigma_{\rm b}^{\rm R}(\omega^\prime,{\bf k})}{\omega^\prime-\omega}.
\end{eqnarray}
The imaginary part is explicitly given by
\begin{eqnarray}\label{sp3}
{\rm Im}\Sigma_{\rm b}^{\rm R}(\omega,{\bf k})&=&2\sum_{\bf q}\Big\{u_{{\bf q}-{\bf k}}^2
{\rm{Im}}{\cal D}_{11}^{5\rm{R}}(\omega+E_{{\bf q}-{\bf k}},{\bf q})\nonumber\\
&\times&[f(E_{{\bf q}-{\bf k}})+b(\omega+E_{{\bf q}-{\bf k}})]\nonumber\\
&+&\upsilon_{{\bf q}-{\bf k}}^2{\rm{Im}}{\cal D}_{11}^{5\rm{R}}(\omega-E_{{\bf q}-{\bf k}},{\bf q})\nonumber\\
&\times&[f(-E_{{\bf q}-{\bf k}})+b(\omega-E_{{\bf q}-{\bf k}})]\Big\},
\end{eqnarray}
where $b(E)=1/(e^{E/T}-1)$ is the Bose-Einstein distribution. The imaginary part of ${\cal D}_{11}^{5\text{R}}(\omega,{\bf q})$ is given by
\begin{eqnarray}
\text{Im}{\cal D}_{11}^{5\text{R}}(\omega,{\bf q})=-\frac{\text{Im}\chi_{11}^{5\rm R}(\omega,{\bf q})}
{[\text{Re}\chi_{11}^{5\rm R}(\omega,{\bf q})]^2+[\text{Im}\chi_{11}^{5\rm R}(\omega,{\bf q})]^2},
\end{eqnarray}
where the pair susceptibility $\chi_{11}^5(\omega,{\bf q})$ reads
\begin{eqnarray}\label{sp4}
\chi_{11}^5(\omega,{\bf q})&=&-\frac{m}{4\pi a_s}-\sum_{\bf k}\Bigg[\frac{1-f(E_{\bf k})-f(\xi_{{\bf q}-{\bf k}})}
{E_{\bf k}+\xi_{{\bf q}-{\bf k}}-\omega}u_{\bf k}^2\nonumber\\
&-&\frac{f(E_{\bf k})-f(\xi_{{\bf q}-{\bf k}})}{E_{\bf k}-\xi_{{\bf q}-{\bf k}}+\omega} \upsilon_{\bf k}^2
-\frac{1}{2\epsilon_{\bf k}}\Bigg].
\end{eqnarray}

Now we turn to the numerical results for the spectral density ${\cal A}_{\rm b}(\omega,k)$. The system is characterized by a single dimensionless coupling parameter $1/(k_{\rm F}a_s)$ where $k_{\rm F}$ is defined via the total density as $n=k_{\rm F}^3/(2\pi^2)$. To achieve strong coupling,
we consider the resonant interaction with $a_s\rightarrow\infty$ ($1/a_s=0$). In this case, all quantities can be expressed via two parameters:
the chemical potential $\mu$ and the effective Fermi momentum $k_\mu=\sqrt{2m\mu}$. This also simplifies the numerical procedures. In this case,
we do not need to solve the number equation (\ref{numeq}) to obtain the chemical potential $\mu$ in units of the Fermi energy
$E_{\rm F}=k_{\rm F}^2/(2m)$. The order parameter $\Delta$ in units of $\mu$ as a function of $T/\mu$ is solved from the BCS gap equation (\ref{gapbcs}). Then we can calculate the spectral density ${\cal A}_{\rm b}(\omega,k)$ in units of $1/\mu$ through the Eqs. (\ref{sp1}),
(\ref{sp2}), and  (\ref{sp3}).

The numerical results of the spectral density ${\cal A}_{\rm b}(\omega)$ for fermion momentum $k=k_\mu$ are shown in Fig. \ref{fig3}. At very low temperature $T\rightarrow0$, we find that there exists a very sharp Fermi-liquid peak around $\omega=0$. The spectral weight of the continuum part
is very small. Therefore, the Fermi-liquid picture of the unpaired color remains valid even though the pairing fluctuation effects are taken into account. This can be understood by the behavior of the imaginary part of the self-energy $\Sigma_{\rm{b}}^{\rm R}(\omega)$ shown in Fig. \ref{fig4}. Because of the formation of a large pairing gap $\Delta$ for the paired colors, the imaginary part vanishes in a wide regime around $\omega=0$,
which can be seen from the properties of the Fermi-Dirac and Bose-Einstein distribution functions in Eq. (\ref{sp3}). The real part of the self-energy $\Sigma_{\rm{b}}^{\rm R}(\omega)$ shown in Fig. \ref{fig5} leads to a correction to the effective Fermi surface, which is shown by the fact that the Fermi-liquid peak is no longer located precisely at $\omega=0$.

However, as the temperature is increased, the pairing gap $\Delta$ is reduced and the contribution from the imaginary part of the self-energy $\Sigma_{\rm{b}}^{\rm R}(\omega)$ becomes more and more important. When the temperature is high enough but below the critical one $T_c$, the imaginary part ${\rm Im}\Sigma_{\rm{b}}^{\rm R}(\omega)$ has two main effects: (i) It becomes finite in the regime around $\omega=0$ and hence
the Fermi-liquid peak broadens; (ii) it has two sharp minima which breach the Fermi-liquid peak and the other two gaplike peaks. Physically,
the imaginary part of the self-energy $\Sigma_{\rm b}$ corresponds to the decay processes of the fermions of the blue color~\cite{kitazawa}.
According to the Feynman diagram for the $T$-matrix approximation shown in Fig. \ref{fig2}, the decay process of the blue fermion can be described
as ${\rm b}\rightarrow {\rm h}+(\rm{fb})$ where ${\rm b}$ denotes the blue fermion, $f$ denotes the quasifermion which is a mixture of red and green colors, ${\rm h}$ is the hole excitation corresponding to the quasifermion, and $(\rm{fb})$ is the collective mode with propagator
${\cal D}_{11}^{5,7}$. Since the quasifermion ${\rm f}$ is fully gapped, at zero temperature the decay process is strictly suppressed around $\omega=0$. As the temperature increases, the imaginary part of the self-energy becomes nonzero but is still suppressed by the pairing gap around $\omega=0$. On the other hand, the imaginary part of the self-energy should vanish for $\omega\rightarrow\pm\infty$. Therefore, for $T<T_c$, there arises two sharp peaks for the imaginary part of the self-energy at finite frequency. The decay process of the blue fermion is most enhanced at these peak frequencies.

As the temperature increases, the spectral weight of the Fermi-liquid peak becomes smaller and smaller and two gaplike peaks appear
around the Fermi-liquid peak. These two gaplike peaks can be referred to as the ``pseudogap peaks", since they are induced by the pairing
fluctuation effects rather than the pairing gap or superfluid order parameter $\Delta$. When the temperature reaches $T_c$, the imaginary part
${\rm Im}\Sigma_{\rm{b}}^{\rm R}(\omega)$ undergoes a characteristic change: The two sharp minima combine to a smooth minimum at $\omega=0$.
Therefore, the Fermi-liquid peak disappears completely at $T=T_c$ and only the two pseudogap peaks remain. Note that at $T=T_c$ the SU(3)
symmetry is restored and therefore the spectral density shown in Fig. \ref{fig3} is also true for the red and green colors. This indicates that
there exists a pseudogap phase at $T>T_c$, similar to that observed in two-component systems~\cite{PG1,PG2,PG4}.

\begin{figure*}
\begin{center}
\includegraphics[width=8cm]{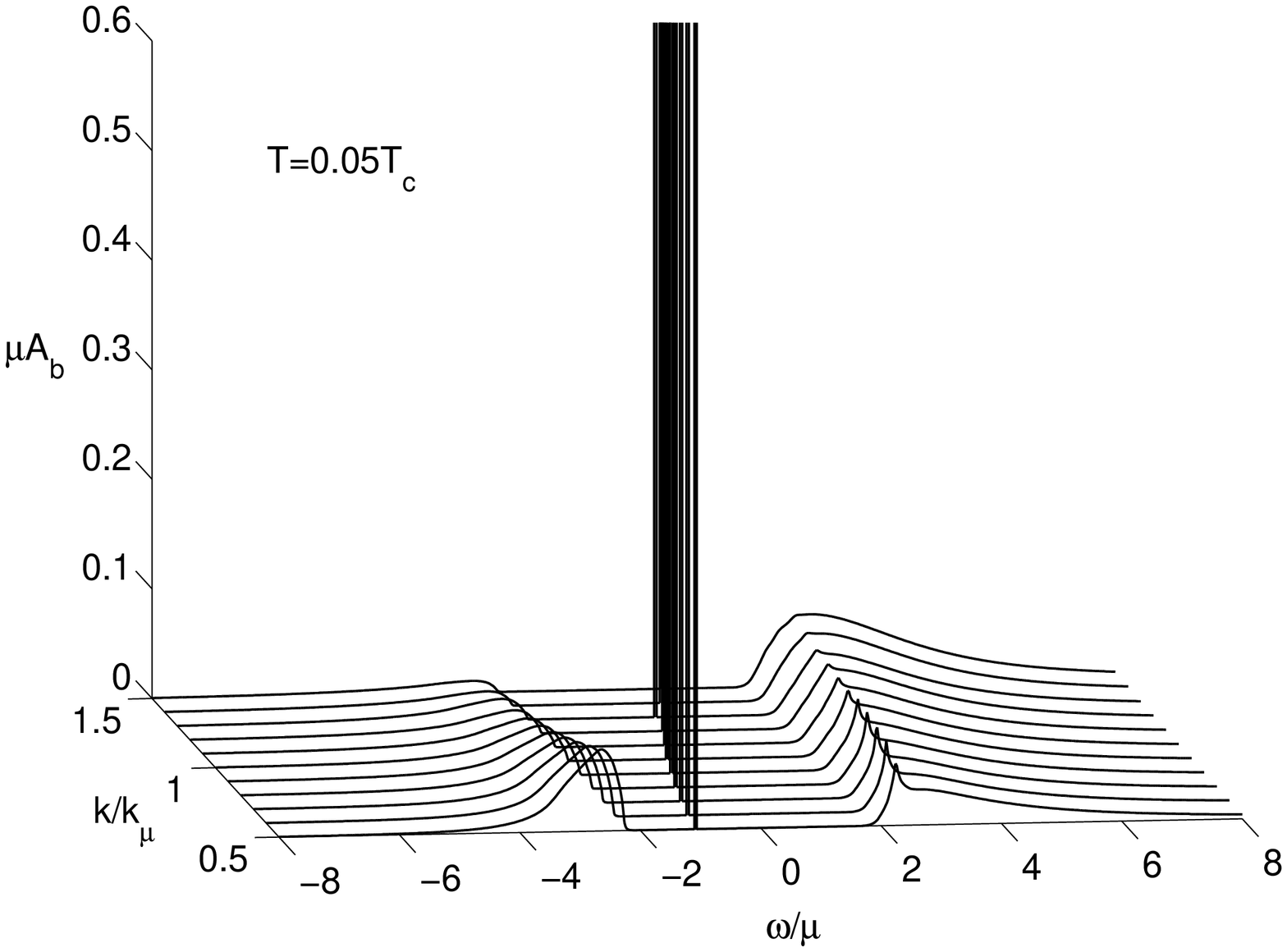}
\includegraphics[width=8cm]{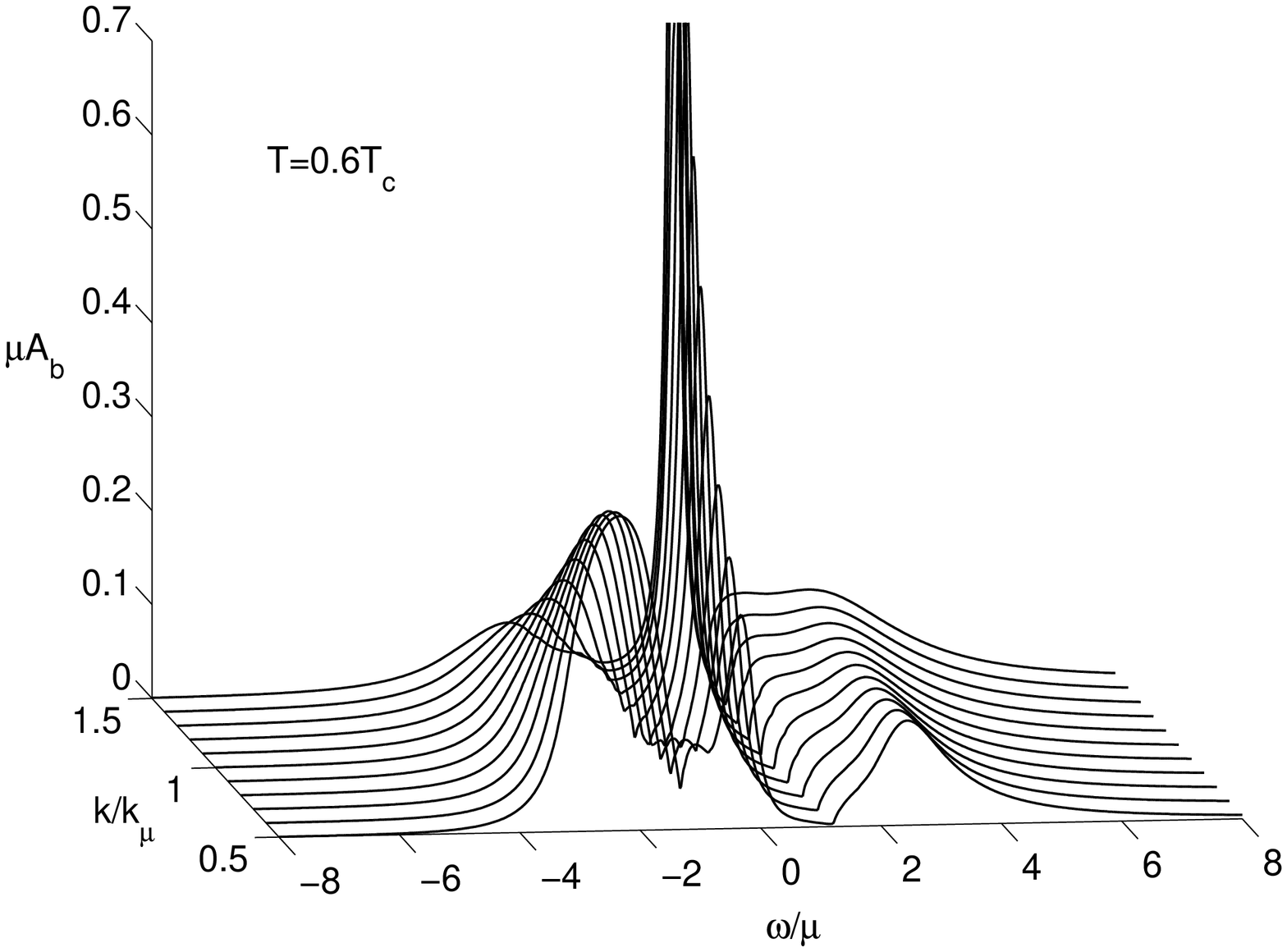}
\includegraphics[width=8cm]{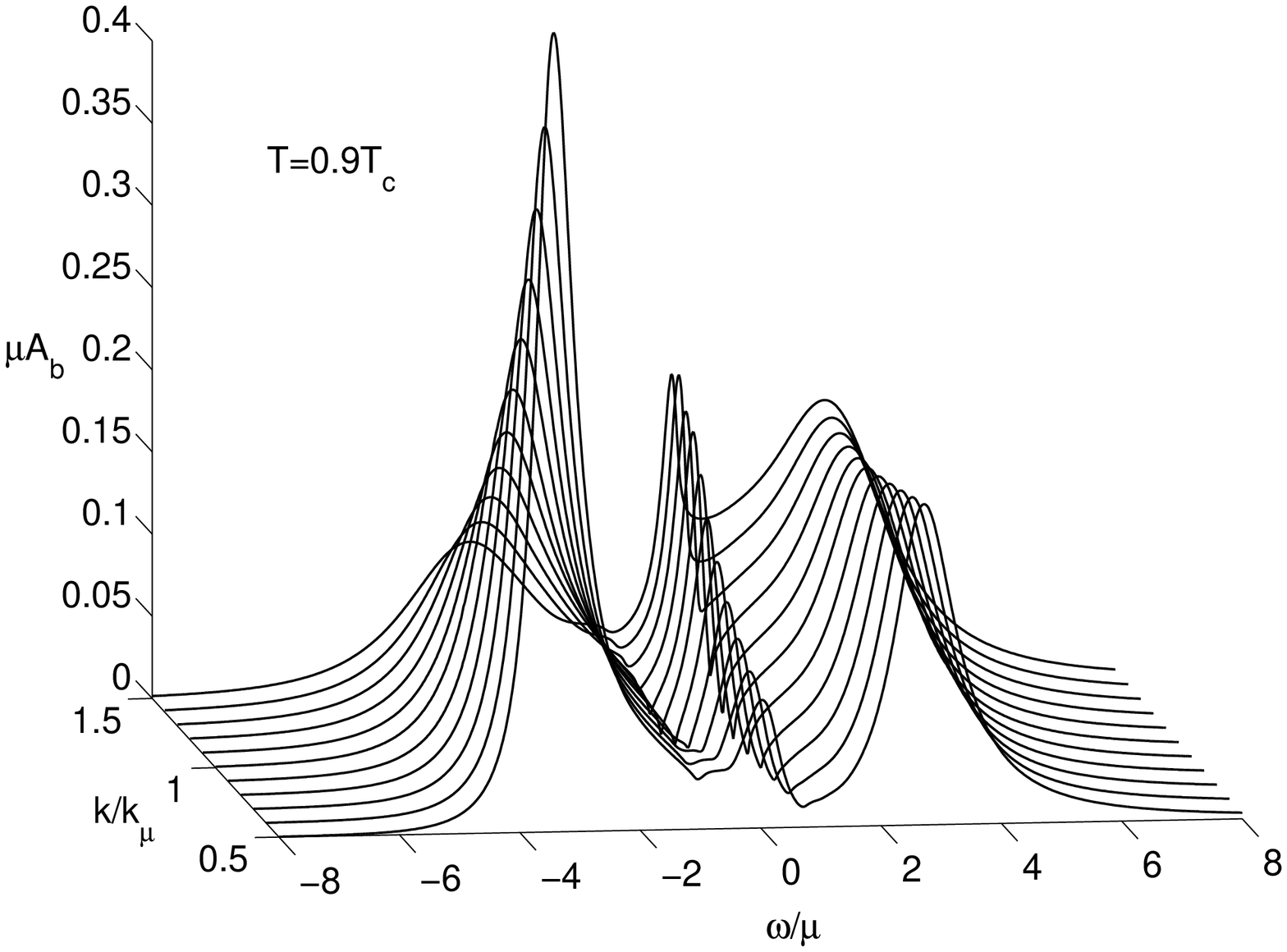}
\includegraphics[width=8cm]{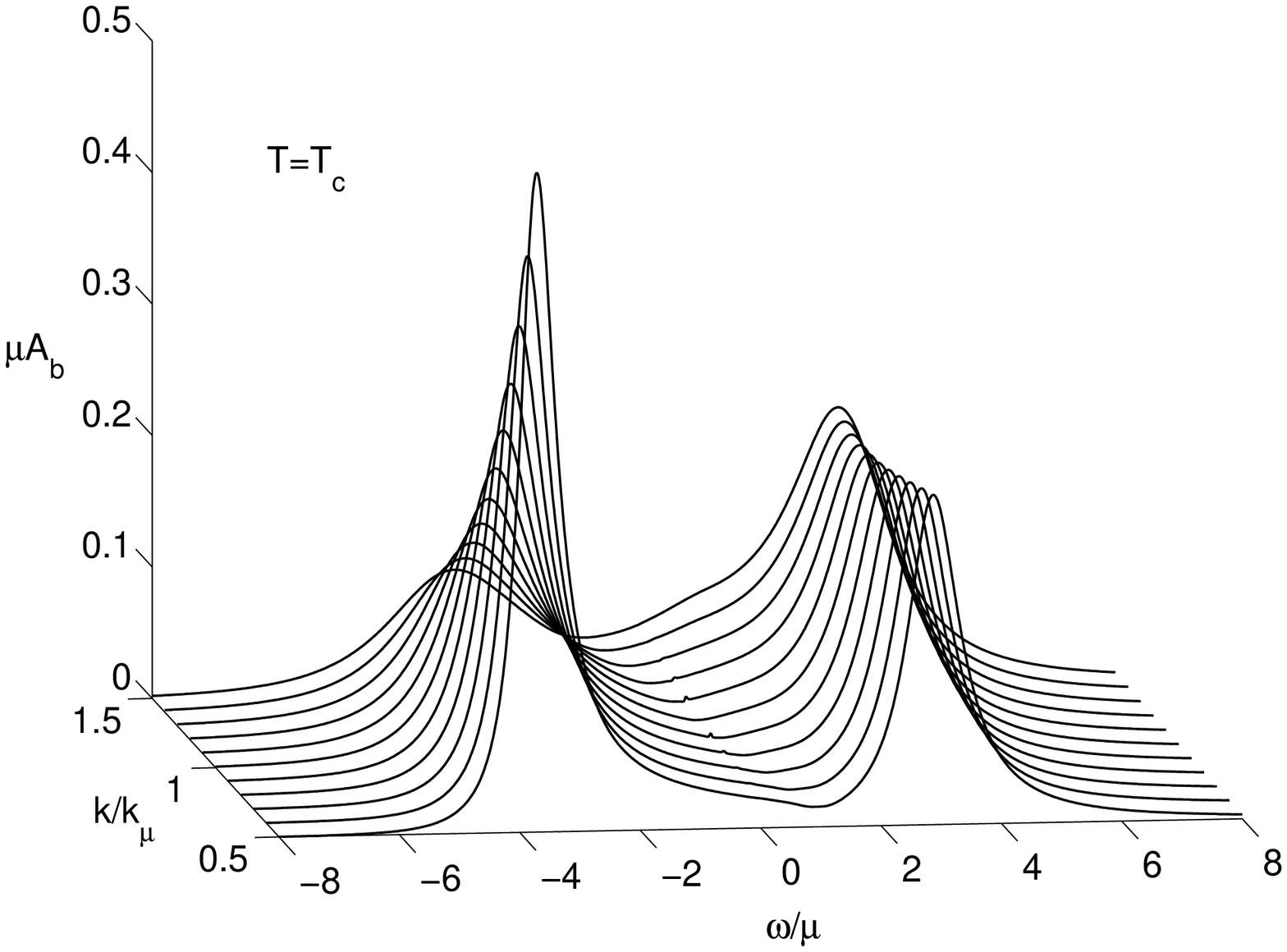}
\caption{The spectral density ${\cal A}_{\rm b}(\omega,k)$ of the unpaired blue color for different momenta $k$ at various temperatures. \label{fig6}}
\end{center}
\end{figure*}

The above discussions show clearly how the Fermi-liquid behavior of the unpaired color at low temperature evolves to the pseudogap behavior at and above $T_c$ due to the pairing fluctuation effects. In a wide temperature regime below $T_c$, we find that the Fermi-liquid peak and the pseudogap peaks coexist. It is intuitive to understand this coexistence through a naive approximation for the fermion self-energy employed in the pseudogap theory of the BCS-BEC crossover \cite{Levin,qijin}. Since the pair propagator ${\cal D}_{11}^5(Q)$ is peaked around $Q=0$, we can approximate the fermion self-energy $\Sigma_{\text{b}}(K)$ as
\begin{eqnarray}\label{pgapp}
\Sigma_{\text{b}}(K)\simeq-\Delta_{\rm{pg}}^2{\cal G}_{\Delta}(-K),
\end{eqnarray}
where the pseudogap energy $\Delta_{\rm{pg}}$ is defined as
\begin{eqnarray}
\Delta_{\rm{pg}}^2=2\sum_Q{\cal D}_{11}^5(Q).
\end{eqnarray}
Under this approximation, the propagator ${\cal G}_{\rm b}(K)$ can be analytically evaluated as
\begin{eqnarray}
{\cal G}_{\rm b}(K)=\frac{(i\omega_n)^2-\Delta^2}{(i\omega_n-\xi_{\bf k})[(i\omega_n)^2-\xi_{\bf k}^2-\Delta^2-\Delta_{\rm{pg}}^2]}.
\end{eqnarray}
Therefore, for $k=k_\mu$, the spectral density ${\cal A}_{\rm b}(\omega)$ reads
\begin{eqnarray}
{\cal A}_{\rm b}(\omega)&=&\frac{\Delta^2}{\Delta^2+\Delta_{\rm{pg}}^2}\delta(\omega)
+\frac{1}{2}\frac{\Delta_{\rm{pg}}^2}{\Delta^2+\Delta_{\rm{pg}}^2}\nonumber\\
&\times&\left[\delta(\omega-\sqrt{\Delta^2+\Delta_{\rm{pg}}^2})+\delta(\omega+\sqrt{\Delta^2+\Delta_{\rm{pg}}^2})\right].
\end{eqnarray}
While the broadening effects of the peaks are neglected in this naive approximation, this analytical expression clearly shows a
three-peak structure. The spectral weights of the Fermi-liquid peak and the pseudogap peaks are given by
\begin{eqnarray}\label{weight}
Z_{\rm{fl}}=\frac{\Delta^2}{\Delta^2+\Delta_{\rm{pg}}^2},\ \ \ \ Z_{\rm{pg}}=\frac{1}{2}\frac{\Delta_{\rm{pg}}^2}{\Delta^2+\Delta_{\rm{pg}}^2}.
\end{eqnarray}
Further, the pseudogap energy can be expressed as
\begin{eqnarray}
\Delta_{\rm{pg}}^2=2\sum_{\bf q}\int_{-\infty}^\infty\frac{d\omega}{\pi}b(\omega){\rm Im}{\cal D}_{11}^{5\rm R}(\omega,{\bf q}).
\end{eqnarray}
As a result of the Bose-Einstein distribution function $b(\omega)$, we find that $\Delta_{\rm{pg}}\rightarrow0$ for $T\rightarrow0$. In general,
the pseudogap energy increases with the increased temperature. Therefore, from Eq. (\ref{weight}), we see clearly that the spectral weight of the
pseudogap peak can be neglected at low temperature and the Fermi-liquid peak disappears at $T=T_c$ where $\Delta$ vanishes.

So far we have only studied the spectral properties of the blue fermion at the ``Fermi surface" $k=k_\mu$. It is instructive to show the spectral
density for momenta $k$ away from $k=k_\mu$. The results for the spectral density ${\cal A}_{\rm b}(\omega,k)$ at different momenta $k\neq k_\mu$
are shown in Fig. \ref{fig6}. For this resonantly interacting Fermi system, we find that the qualitative feature discussed above remains for momenta
$k$ around $k=k_\mu$. The change is that the pseudogap peaks become more asymmetric as the momenta goes away from $k=k_\mu$.

Finally, we present a brief discussion about the case that the SU(3) symmetry is explicitly broken. Let us consider the case that the
attractions among the three colors are not equal, i.e., $g_{13}=g_{23}=g^\prime<g_{12}=g$. In this case, the pairing of red and green color is favored. Since the pair excitations corresponding to the particle-particle ladders ${\cal D}^5(Q)$ and ${\cal D}^7(Q)$ (they are still degenerate since $g_{13}=g_{23}$) are no longer gapless, the pairing fluctuation effects become weaker and weaker as the coupling strength $g^\prime$
decreases. For $g^\prime\rightarrow0$, the particle-particle ladders ${\cal D}^5(Q)$ and ${\cal D}^7(Q)$ vanish and hence the pairing fluctuation
induced self-energy $\Sigma_{\rm b}(Q)$ vanishes. In this case, the Fermi-liquid behavior persists at any temperature, as we expect.

\section{Color Superconductivity in two-flavor quark matter}

In this section we study quark color superconductors. The perturbative-QCD calculation for color superconductivity \cite{CSCgap,pQCD} is
applicable for quark chemical potentials $\mu\gg\Lambda_{\rm{QCD}}$. However, here we are interested in the moderate density regime of
quark matter where the quark chemical potential $\mu\sim400$MeV, and the perturbative approach is not applicable. Therefore, we adopt a phenomenological four-fermion interaction model of QCD \cite{CSCbegin,CSCreview}. Moreover, since the strange quark degree of freedom is
not activated at $\mu\sim400$MeV, we consider only the two light quark flavors ($u$ and $d$). The Lagrangian density of our four-fermion
interaction model is given by
\begin{eqnarray}
{\cal L}=\bar{q}\left(i\partial\!\!\!/+\mu\gamma_0\right)q+{\cal L}_{\text{int}},
\end{eqnarray}
where the interaction part is modeled by a QCD-motivated contact current-current interaction
\begin{eqnarray}\label{njlmodel}
{\cal L}_{\text{int}}=-G_{\text c}\sum_{{\text a}=1}^8(\bar{q} \gamma_\mu \lambda_{\text a}q)
(\bar{q} \gamma^\mu \lambda_{\text a}q).
\end{eqnarray}
Here $q$ denotes the quark field containing two flavors and three colors, and $G_{\rm c}$ is a phenomenological coupling constant
which should be, in principle, determined by the vacuum phenomenology of QCD. In real QCD, the ultraviolet modes decouple because of
asymptotic freedom, but in the present four-fermion interaction model we have to add this feature by hand, through a UV momentum
cutoff $\Lambda$ in the quark momentum integrals. The model therefore has two parameters: the coupling constant $G_{\text c}$ and
the three momentum cutoff $\Lambda$.

\subsection{BCS mean-field theory}

The contact current-current interaction (\ref{njlmodel}) includes all mesonlike and diquarklike interaction channels, which can be obtained
via the Fierz transformation. Since we are interested in the temperature regime $T\sim O(10$MeV), it is sufficient to consider only the $J^P=0^+$ diquark pairing channel. The pairing gap and critical temperature of other diquark pairing channels are much smaller than that of the $J^P=0^+$ pairing channel \cite{Alford}. Therefore, we have
\begin{eqnarray}
{\cal L}_{\text{int}}=\frac{G}{4}\sum_{{\text a}=2,5,7}\left(\bar{q} i\gamma_5\tau_2\lambda_{\text a}C\bar{q}^{\rm T}\right)
\left(q^{\text T}C i\gamma_5\tau_2\lambda_{\text a}q\right)+\cdots,
\end{eqnarray}
where $C=i\gamma_0\gamma_2$ is the charge conjugate matrix and $\tau_{\text i}~({\text i}=1,2,3)$ are the Pauli matrices in the flavor space.
The new coupling constant $G>0$ denotes the attraction in the $J^P=0^+$ diquark pairing channel.

Because of the attraction among the unlike colors, at low temperature the quark matter is a color superconductor. This is due to the fact that some gluons obtain Meissner masses (color Meissner effect) through the Anderson-Higgs mechanism \cite{CMeissner}. The color superconductivity in the two-flavor quark matter is also characterized by a three-component order parameter $\mbox{\boldmath{$\Delta$}}=(\Delta_1,\Delta_2,\Delta_3)$,
where
\begin{eqnarray}
&&\Delta_1=\frac{G}{2}\langle q^{\text T}C i\gamma_5\tau_2\lambda_2q\rangle,\nonumber\\
&&\Delta_2=\frac{G}{2}\langle q^{\text T}C i\gamma_5\tau_2\lambda_5q\rangle,\nonumber\\
&&\Delta_3=\frac{G}{2}\langle q^{\text T}C i\gamma_5\tau_2\lambda_7q\rangle.
\end{eqnarray}
Because of the color SU$(3)$ symmetry of the QCD Lagrangian (and hence the QCD-motivated model), the effective potential ${\cal V}(\mbox{\boldmath{$\Delta$}})$ depends only on the combination $\mbox{\boldmath{$\Delta$}}\mbox{\boldmath{$\Delta$}}^\dagger=|\Delta_1|^2+|\Delta_2|^2+|\Delta_3|^2$.
Therefore, like the atomic color superfluidity, we can choose a specific gauge $\Delta_1=\Delta\neq0, \Delta_2=\Delta_3=0$ without loss of generality. In this gauge, only the red and green quarks participate in the pairing and the red-green diquark pairs condense, leaving the blue
quarks unpaired. In the following calculations we will adopt this gauge.

The Nambu-Gor'kov basis for the present case can be defined as $\Psi= (q, C\bar{q}^{\text{T}})^{\rm T}$. In the Nambu-Gor'kov representation,
the quark self-energy $\Sigma(K)$ and the dressed quark propagator satisfy Dyson's equation
\begin{eqnarray}
&&\left(\begin{array}{cc} {\cal S}_{11}(K)&{\cal S}_{12}(K) \\ {\cal S}_{21}(K)&{\cal S}_{22}(K) \end{array}\right)^{-1}\nonumber\\
&=&\left(\begin{array}{cc} K\!\!\!/+\mu\gamma^0&0 \\ 0&K\!\!\!/-\mu\gamma^0 \end{array}\right)
-\left(\begin{array}{cc} \Sigma_{11}(K)&\Sigma_{12}(K) \\ \Sigma_{21}(K)&\Sigma_{22}(K) \end{array}\right).
\end{eqnarray}
Again, from the Green's function relation we obtain the gap equation
\begin{equation}
\Delta=\frac{G}{2}\sum_K\text{Tr}\left[i\gamma_5\tau_2\lambda_2{\cal S}_{12}(K)\right]
\end{equation}
and the baryon number density
\begin{equation}
n_{\text B}=\frac{1}{3}\sum_K\text{Tr}\left[\gamma_0{\cal S}_{11}(K)\right].
\end{equation}

In the BCS mean-field theory, the quark self-energy is chosen as
\begin{eqnarray}
\Sigma(K)=\Sigma^{\text{BCS}}(K)=\left(\begin{array}{cc}0&i\gamma_5\Delta\tau_2\lambda_2
\\ i\gamma_5\Delta\tau_2\lambda_2&0\end{array}\right).
\end{eqnarray}
Again, we set $\Delta$ to be real without loss of generality. In the BCS approximation, the dressed quark propagator can be explicitly evaluated as
\begin{eqnarray}
{\cal S}_{11}^{\text{BCS}}(K)&=&{\cal G}_{\Delta}(K)\tau_0\lambda_{\text{rg}}+{\cal G}_{0}(K)\tau_0\lambda_{\text b},\nonumber\\
{\cal S}_{12}^{\text{BCS}}(K)&=&{\cal F}_{\Delta}(K)\tau_2\lambda_2,\nonumber\\
{\cal S}_{22}^{\text{BCS}}(K)&=&\gamma_5{\cal S}_{11}^{\text{BCS}}(-K)\gamma_5,\nonumber\\
{\cal S}_{21}^{\text{BCS}}(K)&=&\gamma_5{\cal S}_{12}^{\text{BCS}}(-K)\gamma_5,
\end{eqnarray}
where $\tau_0$ is the identity matrix in the flavor space. Here the Green's functions ${\cal G}_\Delta(K), {\cal F}_\Delta(K)$, and $ {\cal G}_0(K)$
are given by
\begin{eqnarray}
{\cal G}_{\Delta}(K)&=& {i\omega_n+\xi_{\bf k}^-\over(i\omega_n)^2-(E_{\bf k}^-)^2}\Lambda_{\bf k}^-\gamma_0
+ {i\omega_n-\xi_{\bf k}^+\over (i\omega_n)^2-(E_{\bf k}^+)^2}\Lambda_{\bf k}^+\gamma_0 ,\nonumber\\
{\cal F}_{\Delta}(K)&=& {i\Delta\over (i\omega_n)^2-(E_{\bf k}^-)^2}\Lambda_{\bf k}^-\gamma_5
+ {i\Delta\over (i\omega_n)^2-(E_{\bf k}^+)^2}\Lambda_{\bf k}^+\gamma_5,\nonumber\\
{\cal G}_{0}(K)&=&{1\over i\omega_n-\xi_{\bf k}^-}\Lambda_{\bf k}^-\gamma_0
+ {1\over i\omega_n+\xi_{\bf k}^+}\Lambda_{\bf k}^+\gamma_0,
\end{eqnarray}
where $E_{\bf k}^\pm=\sqrt{(\xi_{\bf k}^\pm)^2+\Delta^2}$ are quasiparticle dispersions, $\xi_{\bf k}^\pm=|{\bf k}|\pm\mu$, and
$\Lambda_{\bf k}^{\mp}=(1\pm{\gamma_0\mbox{\boldmath{$\gamma$}}\cdot\hat{{\bf k}}})/2$ ($\hat{\bf k}\equiv{\bf k}/|{\bf k}|$) are the energy projection operators for massless Dirac fermions. The
physical value of the pairing gap $\Delta$ is determined by the BCS
gap equation
\begin{eqnarray}\label{gapq}
\frac{1}{G}=4\sum_{\bf k}\Bigg[\frac{1-2f(E_{\bf k}^-)}{2E_{\bf k}^-}+\frac{1-2f(E_{\bf k}^+)}{2E_{\bf k}^+} \Bigg].
\end{eqnarray}

The new feature here is the presence of antiquark excitations in this relativistic quark system. They generally have an excitation gap
$\geq\mu$. At high density and low temperature, they are therefore irrelevant degrees of freedom. Regardless of these antiparticle
excitations, we find similar conclusion in comparison with the atomic color superfluidity: The paired quarks obtain an excitation gap
$\Delta$, while the unpaired blue quarks are gapless and possess a free dispersion in the BCS mean-field description. Therefore,
the same problem we addressed in Sec. II also appears in the present color superconductor.

\subsection{Quark self-energy beyond BCS}
The following considerations are highly parallel to the studies in Sec. II.  To establish the quark self-energy beyond the BCS mean-field
approximation, we first construct the particle-particle ladder or the ``diquark propagator" ${\cal D}(Q)$. In the color superconducting phase,
it takes the same form as (\ref{ladder}), i.e.,
\begin{equation}
\left(\begin{array}{cc} {\cal D}_{11}^{\text{ab}}(Q)&{\cal D}_{12}^{\text{ab}}(Q)
\\ {\cal D}_{21}^{\text{ab}}(Q)&{\cal D}_{22}^{\text{ab}}(Q)\end{array}\right)
=\left(\begin{array}{cc} \chi_{11}^{\text{ab}}(Q)&\chi_{12}^{\text{ab}}(Q)
\\ \chi_{21}^{\text{ab}}(Q)&\chi_{22}^{\text{ab}}(Q)\end{array}\right)^{-1}.
\end{equation}
The diquark pair susceptibility $\chi(Q)$ here is given by
\begin{eqnarray}
\chi_{11}^{{\text{ab}}}(Q)&=&\frac{\delta_{\text{ab}}}{G}-\frac{1}{2}\sum_K\text{Tr}\left[\tau_2\lambda_{\text a}
{\cal S}_{11}(Q-K)\tau_2\lambda_{\text b}{\cal S}_{11}(K)\right],\nonumber\\
\chi_{12}^{{\text{ab}}}(Q)&=&-\frac{1}{2}\sum_K\text{Tr}\left[\tau_2\lambda_{\text a}{\cal S}_{12}(Q-K)
\tau_2\lambda_{\text b}{\cal S}_{12}(K)\right],\nonumber\\
\chi_{22}^{{\text {ab}}}(Q)&=&\chi_{11}^{{\text {ab}}}(-Q),\ \ \ \ \ \ \ \
\chi_{21}^{{\text {ab}}}(Q)=\chi_{12}^{{\text {ab}}}(Q).
\end{eqnarray}
Note that the trace here should be taken simultaneously in the color, flavor, and the Dirac spin space. With the diquark propagator ${\cal D}(Q)$, the full quark self-energy is given by
\begin{eqnarray}
\Sigma(K)
=\left(\begin{array}{cc} \Sigma_{11}^L(K)&\Sigma_{12}^L(K)\\
\Sigma_{21}^L(K)&\Sigma_{22}^L(K)\end{array}\right)+\Sigma^{\text{BCS}}(K),
\end{eqnarray}
where the beyond-mean-field contribution $\Sigma^L(K)$ reads
\begin{eqnarray}
\Sigma_{11}^L(K)&=&-\sum_{\text{a},\text{b}}\sum_Q{\cal
D}_{11}^{{\text {ab}}}(Q)\tau_2\lambda_{\text a}{\cal
S}_{11}(Q-K)\tau_2\lambda_{\text b},\nonumber\\
\Sigma_{12}^L(K)&=&-\sum_{\text{a},\text{b}}\sum_Q{\cal
D}_{12}^{{\text {ab}}}(Q)\tau_2\lambda_{\text a}{\cal
S}_{12}(Q-K)\tau_2\lambda_{\text b},\nonumber\\
\Sigma_{22}^L(K)&=&\gamma_5\Sigma_{11}^L(-K)\gamma_5,\ \ \
\Sigma_{21}^L(K)=\gamma_5\Sigma_{12}^L(-K)\gamma_5.
\end{eqnarray}
Because of the same reason we addressed in Sec. II, we use the quark propagator of its BCS form ${\cal S}^{\text{BCS}}(K)$ in evaluating the
pair susceptibility $\chi(Q)$ and the self-energy $\Sigma^L(K)$. The pairing gap $\Delta$ is determined by the BCS gap equation (\ref{gapq})
and the Goldstone's theorem holds.

Because of the same symmetry structure as the atomic system in Sec. II, we find again that $\chi(Q)$ and ${\cal D}(Q)$ are diagonal in the
adjoint space, i.e., $\chi^{{\text {ab}}}_{\text{ij}}=\chi^{\text a}_{\text{ij}}\delta_{{\text {ab}}}$ and
${\cal D}^{{\text{ab}}}_{\text{ij}}={\cal D}^{\text a}_{\text{ij}}\delta_{{\text{ab}}}$. For the ${\text a}=2$ sector, we have
\begin{eqnarray}
\chi_{11}^{2}(Q)&=&\frac{1}{G}-2\sum_K\text{Tr}\left[{\cal G}_{\Delta}(K){\cal G}_{\Delta}(Q-K)\right],\nonumber\\
\chi_{12}^{2}(Q)&=&-2\sum_K\text{Tr}\left[{\cal F}_{\Delta}(K){\cal F}_{\Delta}(Q-K)\right]
\end{eqnarray}
and
\begin{eqnarray}
{\cal D}_{11}^{2}(Q)&=&\frac{\chi_{11}^{2}(-Q)}{\chi_{\text{11}}^{2}(Q)\chi_{\text{11}}^{2}(-Q)-[\chi_{\text{12}}^{2}(Q)]^2},\nonumber\\
{\cal D}_{12}^{2}(Q)&=&\frac{\chi_{12}^{2}(Q)}{\chi_{\text{11}}^{2}(Q)\chi_{\text{11}}^{2}(-Q)-[\chi_{\text{12}}^{2}(Q)]^2}.
\end{eqnarray}
The ${\text a}=5$ and ${\text a}=7$ sectors are degenerate due to the residue SU(2) symmetry group, and $\chi_{12}^{5}(Q)=\chi_{12}^{7}(Q)=0$
since the blue quarks do not participate in pairing. We have
\begin{eqnarray}
&&\chi_{11}^{\rm a}(Q)=[{\cal D}_{11}^{\rm a}(Q)]^{-1}\nonumber\\
&=&\frac{1}{G}-\sum_K\text{Tr}\left[{\cal G}_{\Delta}(K){\cal G}_{0}(Q-K)+{\cal G}_{0}(K){\cal G}_{\Delta}(Q-K)\right]
\end{eqnarray}
for ${\rm a}=5,7$. Using the BCS gap equation (\ref{gapq}), we find that ${\cal D}^{\text a}(Q)$ diverges at $Q=0$ for ${\text a}=2,5,7$,
corresponding to the spontaneous breaking of the color SU(3) symmetry \cite{diquark}.

\begin{figure*}
\begin{center}
\includegraphics[width=8cm]{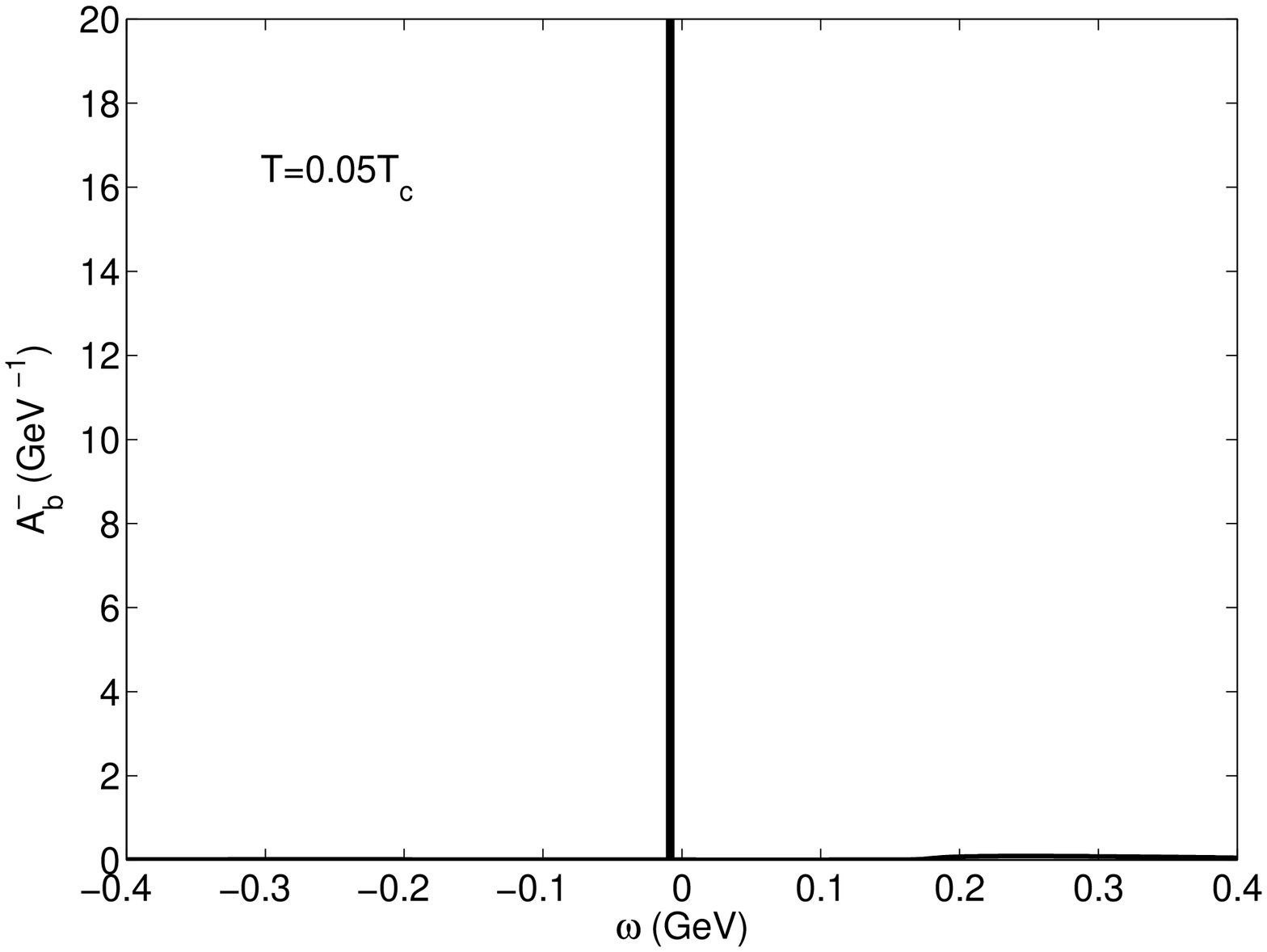}
\includegraphics[width=8cm]{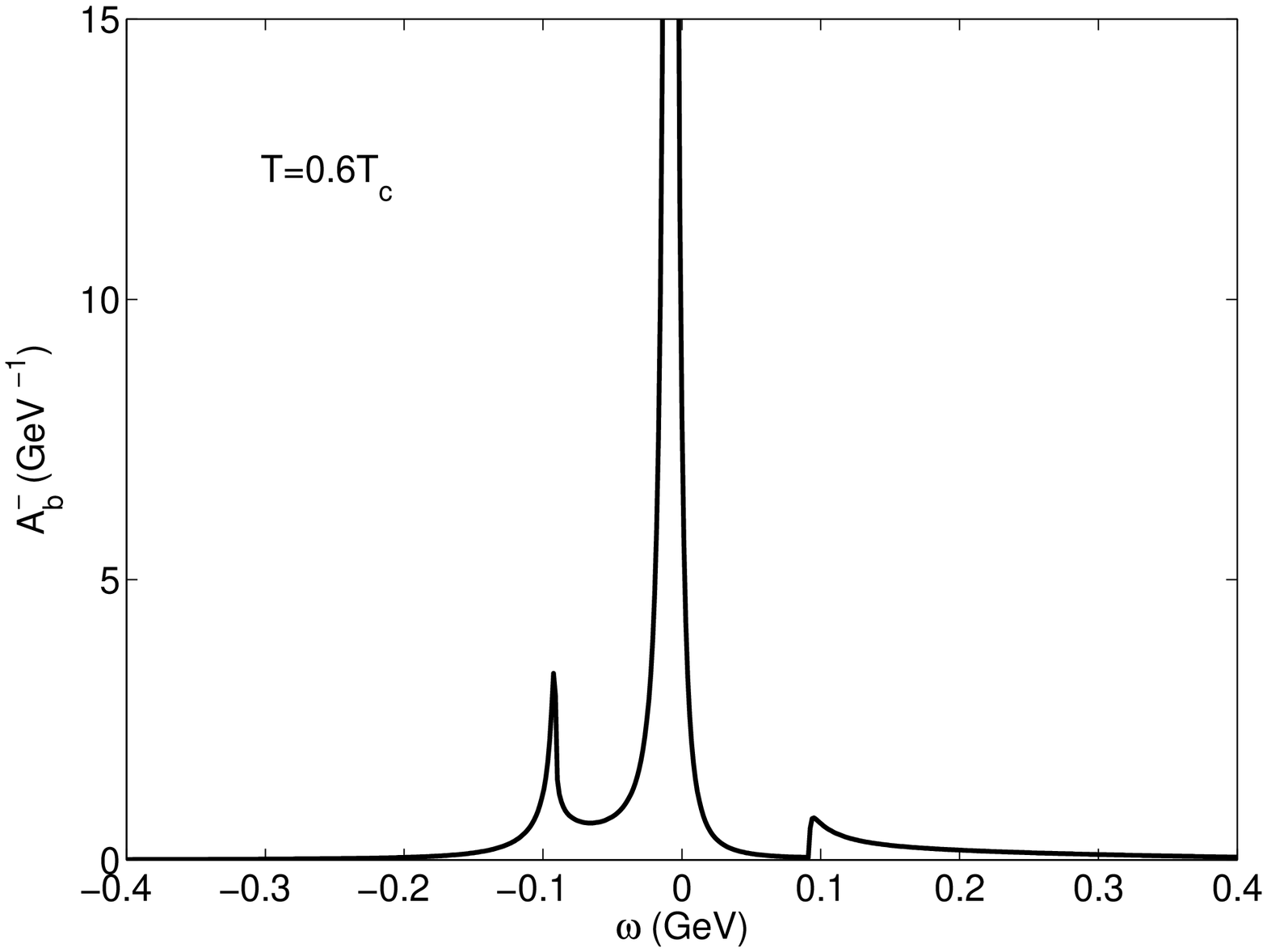}
\includegraphics[width=8cm]{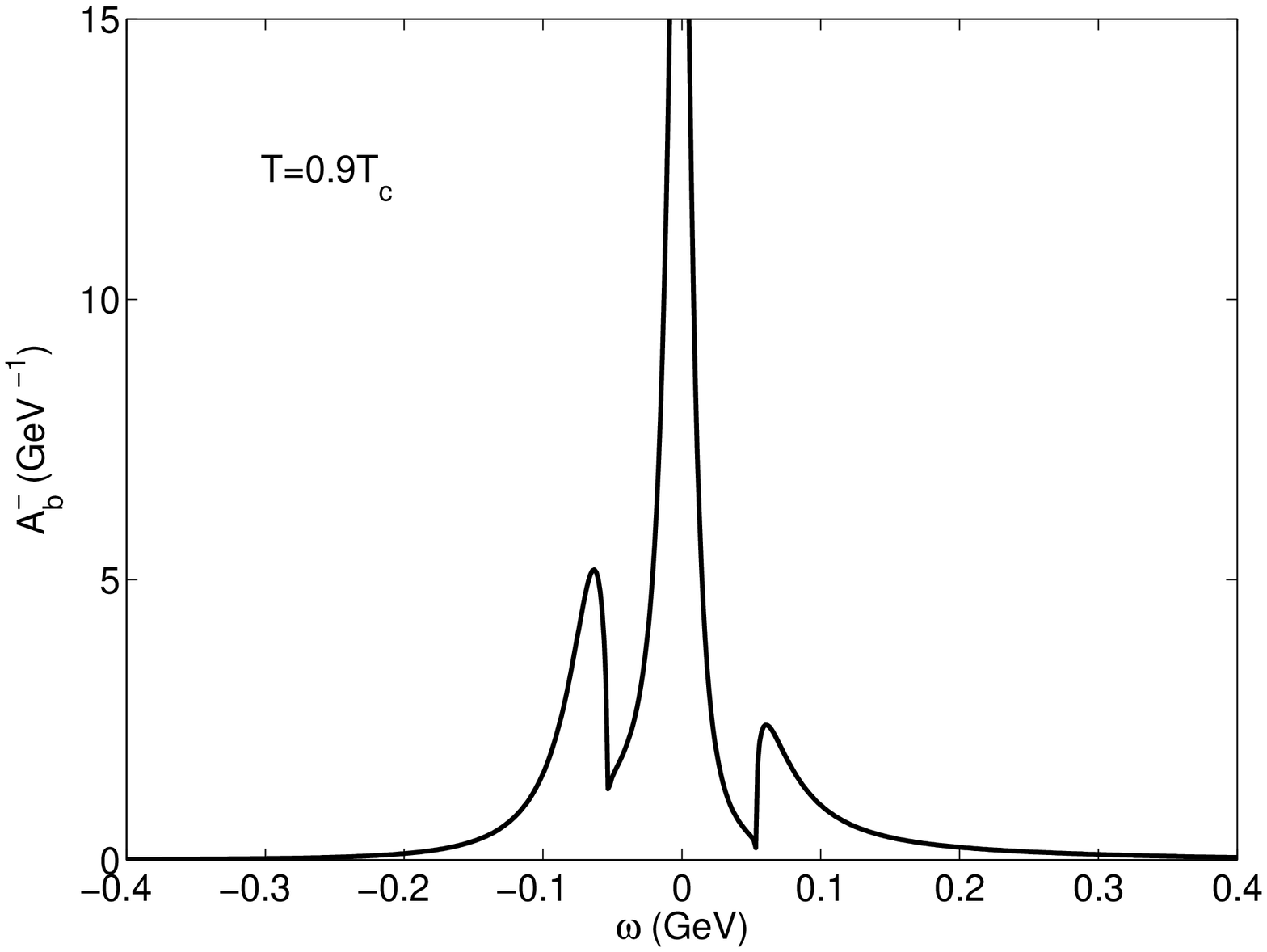}
\includegraphics[width=8cm]{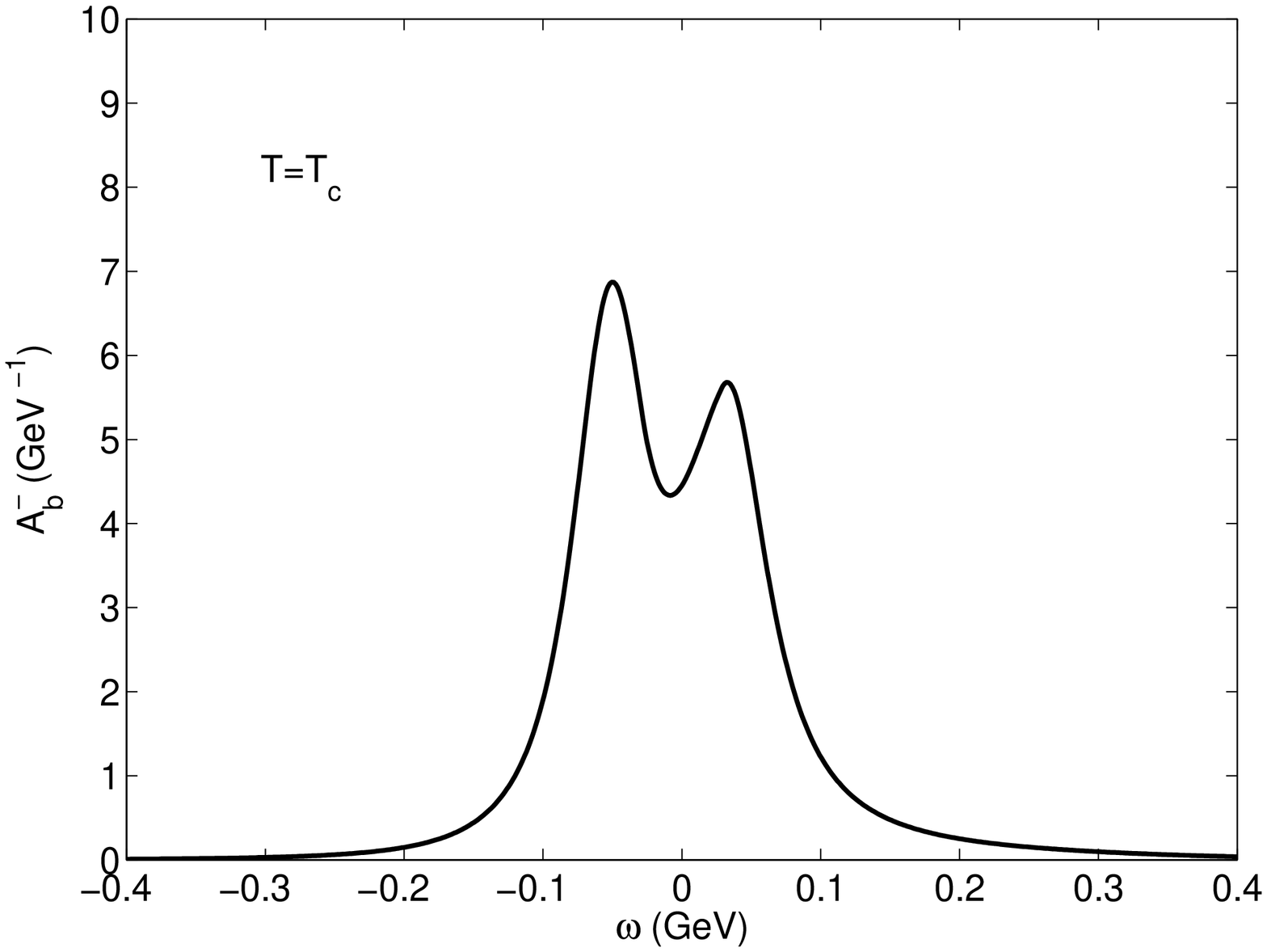}
\caption{The spectral density of the unpaired blue quark for $k=\mu=400$MeV at various temperatures $T\leq T_c$.
The zero temperature pairing gap is set to be $\Delta_0=100$MeV, corresponding to $G/4=3.59$GeV$^{-2}$. \label{fig7}}
\end{center}
\end{figure*}

Then the diagonal component of the self-energy $\Sigma^L(K)$ can be evaluated as
\begin{eqnarray}
\Sigma_{11}^L(K)=\Sigma_{\text{rg}}(K)\tau_0\lambda_{\text{rg}}+\Sigma_{\text{b}}(K)\tau_0\lambda_{\text{b}},
\end{eqnarray}
where $\Sigma_{\text{rg}}(K)$ and $\Sigma_{\text{b}}(K)$ correspond to the self-energies for paired and unpaired colors, respectively.
They take the same form as Eq. (\ref{self1}), i.e.,
\begin{eqnarray}
\Sigma_{\text{rg}}(K)&=&-\sum_Q[{\cal D}_{11}^{2}(Q){\cal G}_{\Delta}(Q-K)+{\cal D}_{11}^{5}(Q){\cal G}_{0}(Q-K)],\nonumber\\
\Sigma_{\text{b}}(K)&=&-2\sum_Q{\cal D}_{11}^{5}(Q){\cal G}_{\Delta}(Q-K).
\end{eqnarray}
The off-diagonal component reads
\begin{eqnarray}
\Sigma_{12}^L(K)=\sum_Q{\cal D}_{12}^{2}(Q){\cal F}_{\Delta}(Q-K)\tau_2\lambda_2.
\end{eqnarray}

Above the critical temperature, $T\geq T_c$, the off-diagonal component $\Sigma_{12}^L(K)$ vanishes and the self-energies for paired and unpaired
colors become degenerate, i.e., $\Sigma_{\text{rg}}(K)=\Sigma_{\text{b}}(K)$. This reflects the fact that the color SU(3) symmetry gets restored above the transition temperature. We have $\Sigma_{11}^L(K)=\Sigma_0(K)\tau_0\lambda_0$, where $\lambda_0$ is the identity matrix in the color
space and
\begin{eqnarray}
\Sigma_0(K)=-2\sum_Q{\cal D}_0(Q){\cal G}_0(Q-K).
\end{eqnarray}
Here ${\cal D}_0(Q)={\cal D}_{11}^{2,5,7}(Q)$ is the common diquark propagator in the normal phase. The above choice of the quark self-energy
$\Sigma_0(K)$ is identical to that adopted by Kitazawa, Koide, Kunihiro, and Nemoto~\cite{kitazawa} in studying the precursor and pseudogap
phenomena of color superconductivity. Therefore, our $T$-matrix approximation for the color superconducting phase is really a generalization of
the $T$-matrix approach for $T>T_c$ used in ~\cite{kitazawa}.

\begin{figure*}
\begin{center}
\includegraphics[width=8cm]{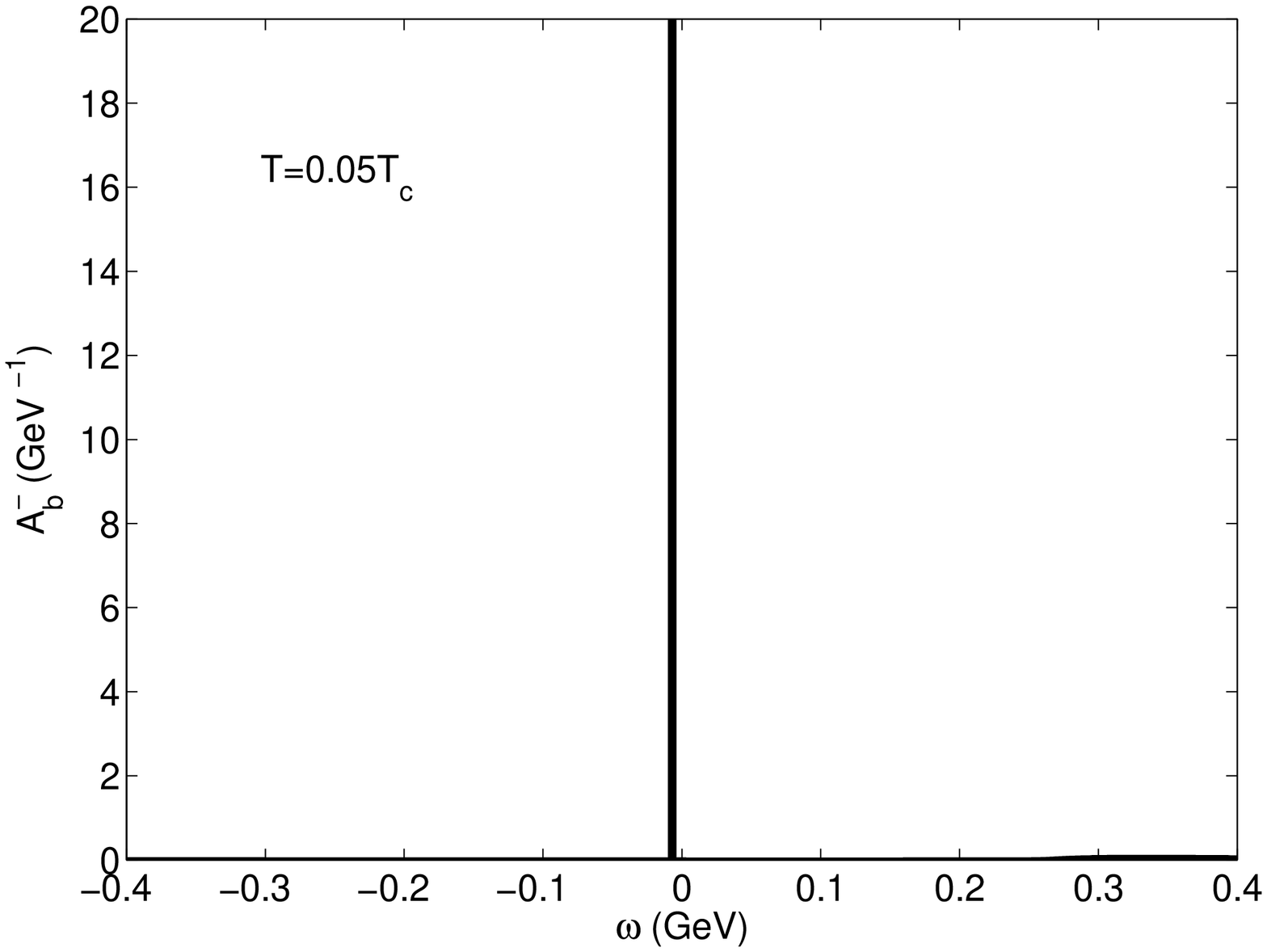}
\includegraphics[width=8cm]{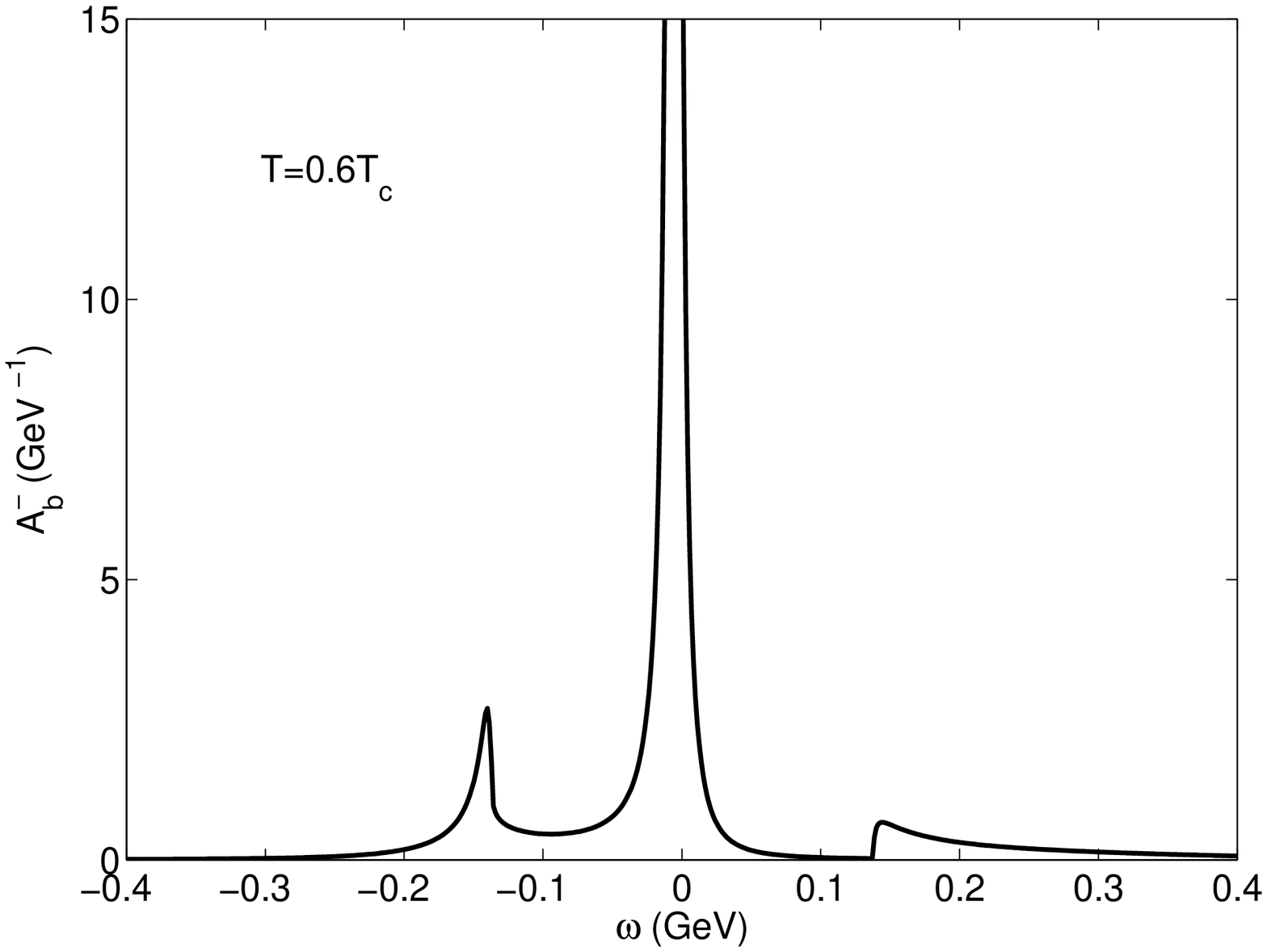}
\includegraphics[width=8cm]{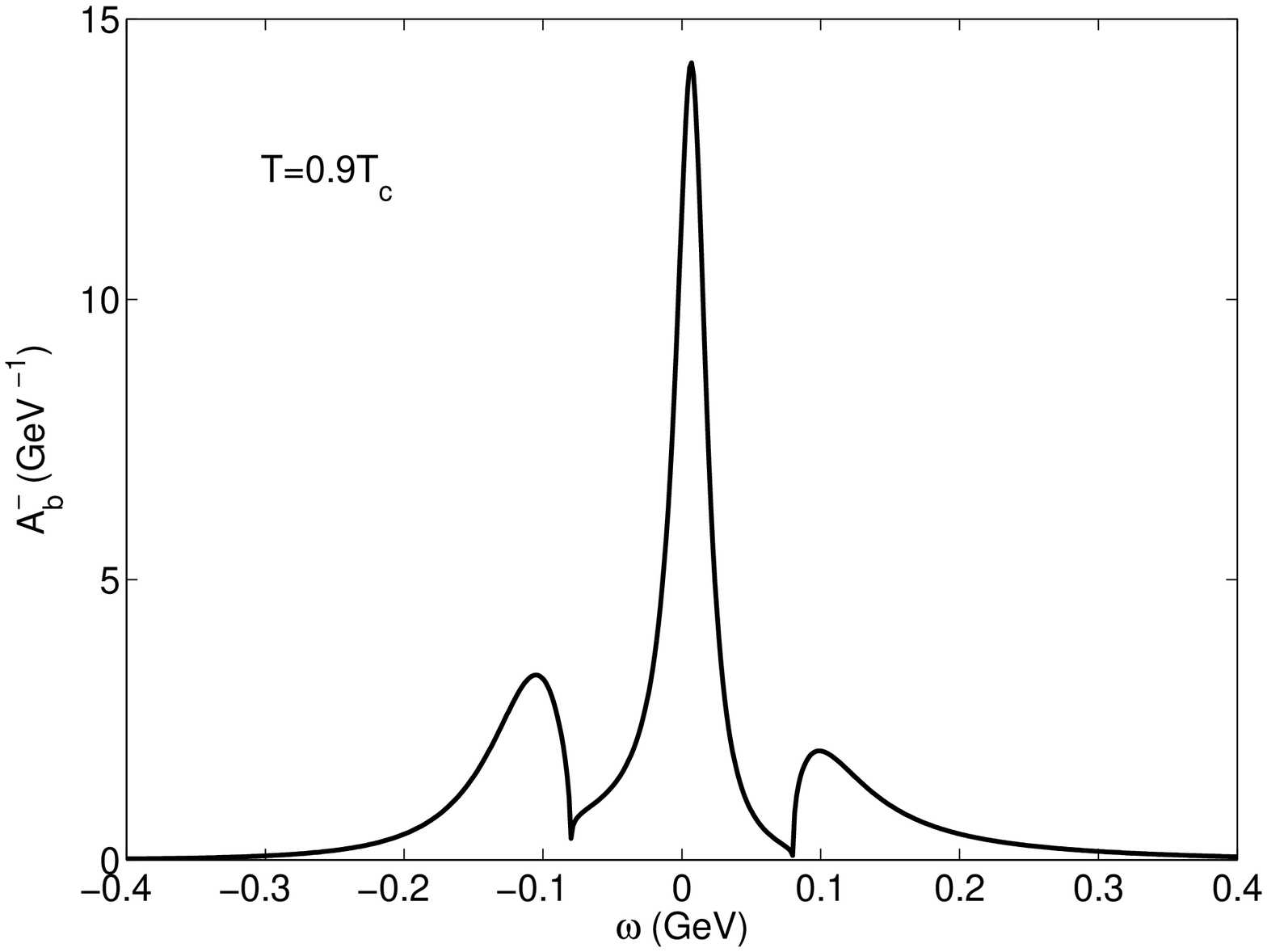}
\includegraphics[width=8cm]{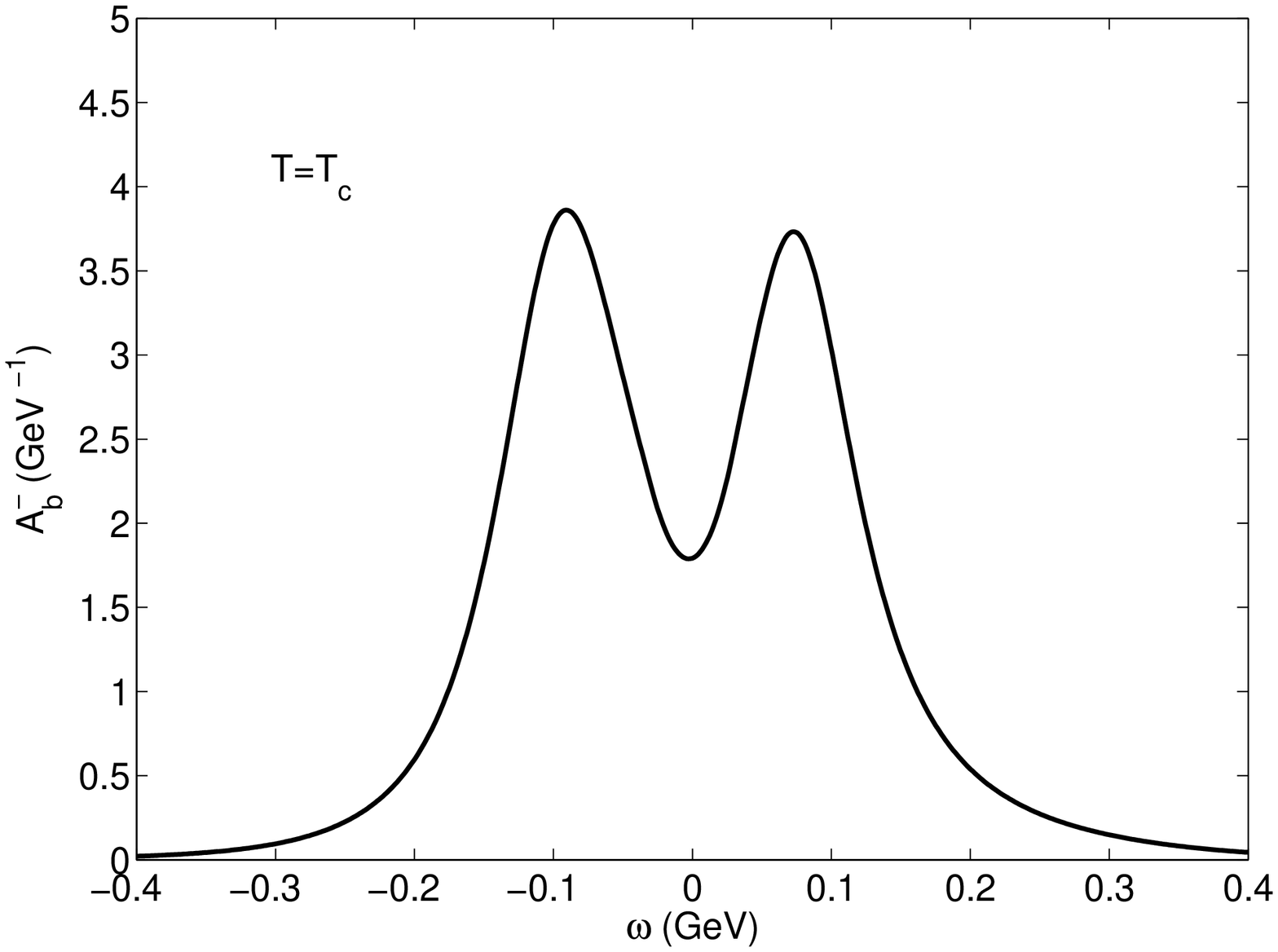}
\caption{ The spectral density of the unpaired blue quark for $k=\mu=400$MeV at various temperatures $T\leq T_c$.
The zero temperature pairing gap is set to be $\Delta_0=150$MeV, corresponding to $G/4=4.39$GeV$^{-2}$.\label{fig8}}
\end{center}
\end{figure*}

\subsection{Spectral density of blue quarks}
The dressed quark propagator ${\cal S}_{11}(K)$ including the pairing fluctuation effects can be expressed as
\begin{eqnarray}
{\cal S}_{11}(K)&=&{\cal G}_{\text{rg}}(K)\tau_0\lambda_{\text{rg}}+{\cal G}_{\text{b}}(K)\tau_0\lambda_{\text{b}},
\end{eqnarray}
where ${\cal G}_{\text{rg}}(K)$ and ${\cal G}_{\text{b}}(K)$ are the propagators for paired and unpaired colors, respectively. We focus on the excitation spectrum of the blue quark. Its spectral density function ${\cal A}_{\rm b}(\omega,{\bf k})$ is defined as
\begin{eqnarray}
{\cal A}_{\rm b}(\omega,{\bf k}) =-\frac{1}{4\pi}\text{Im}{\rm Tr}\left[\gamma^0{\cal G}_{\rm b}^{\rm R}(\omega,{\bf k})\right],
\end{eqnarray}
where the retarded Green's function ${\cal G}_{\rm b}^{\rm R}(\omega,{\bf k})$ is given by
\begin{eqnarray}
{\cal G}_{\rm b}^{\rm R}(\omega,{\bf k})=\frac{1}{[{\cal G}_0^{\rm R}(\omega,{\bf k})]^{-1}-\Sigma^{\rm R}_{\rm b}(\omega,{\bf k})}.
\end{eqnarray}

\begin{figure*}
\begin{center}
\includegraphics[width=8cm]{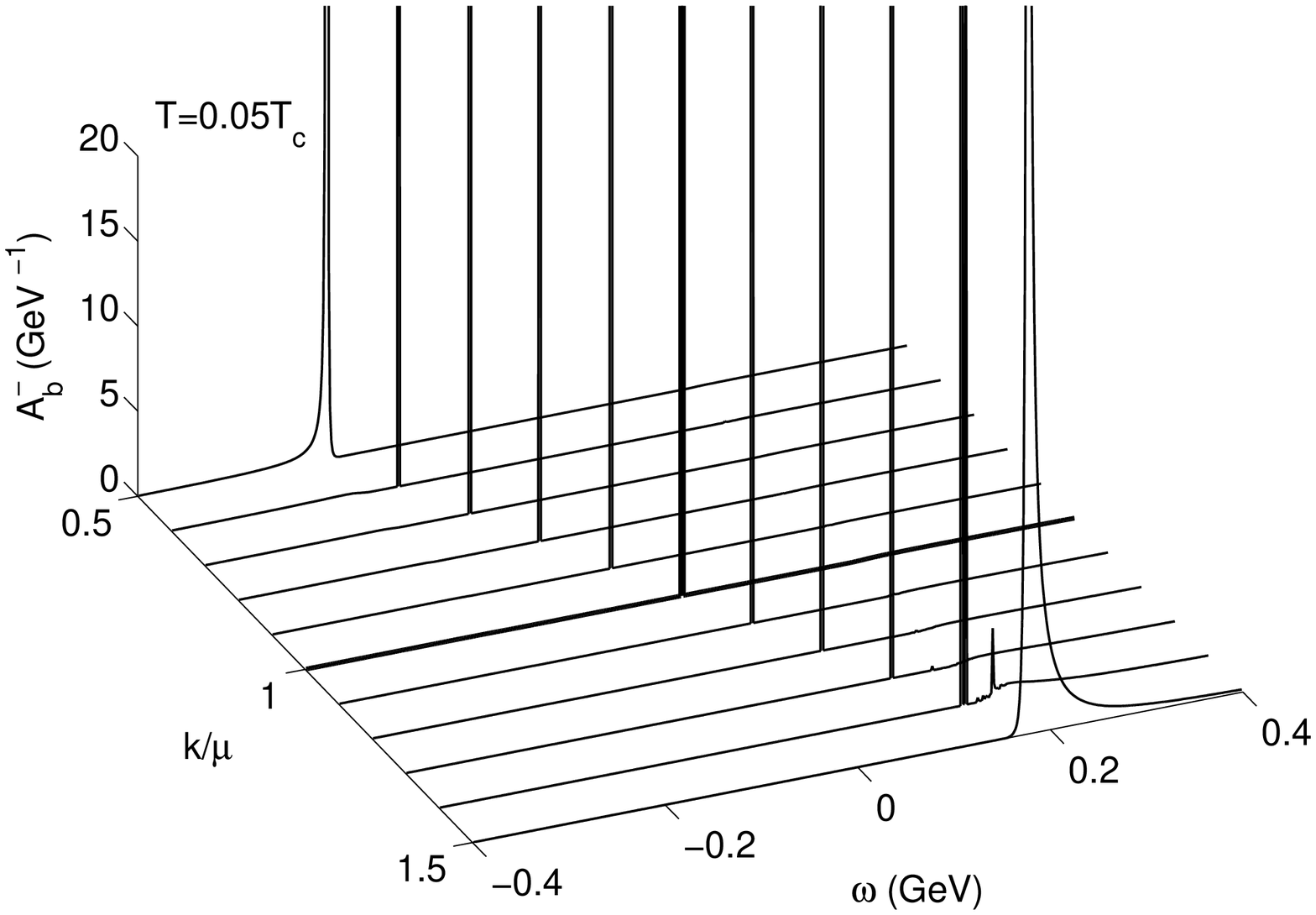}
\includegraphics[width=8cm]{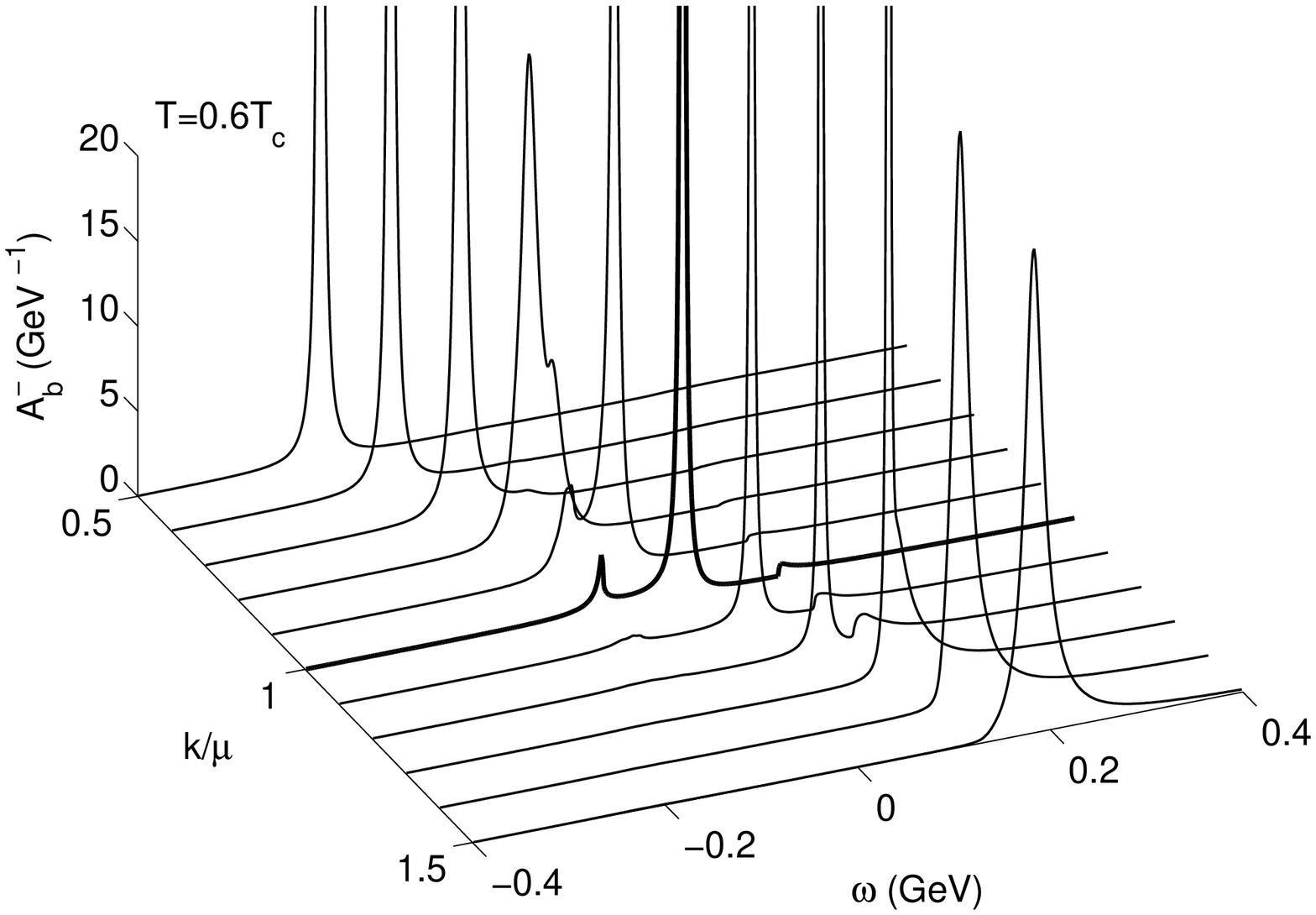}
\includegraphics[width=8cm]{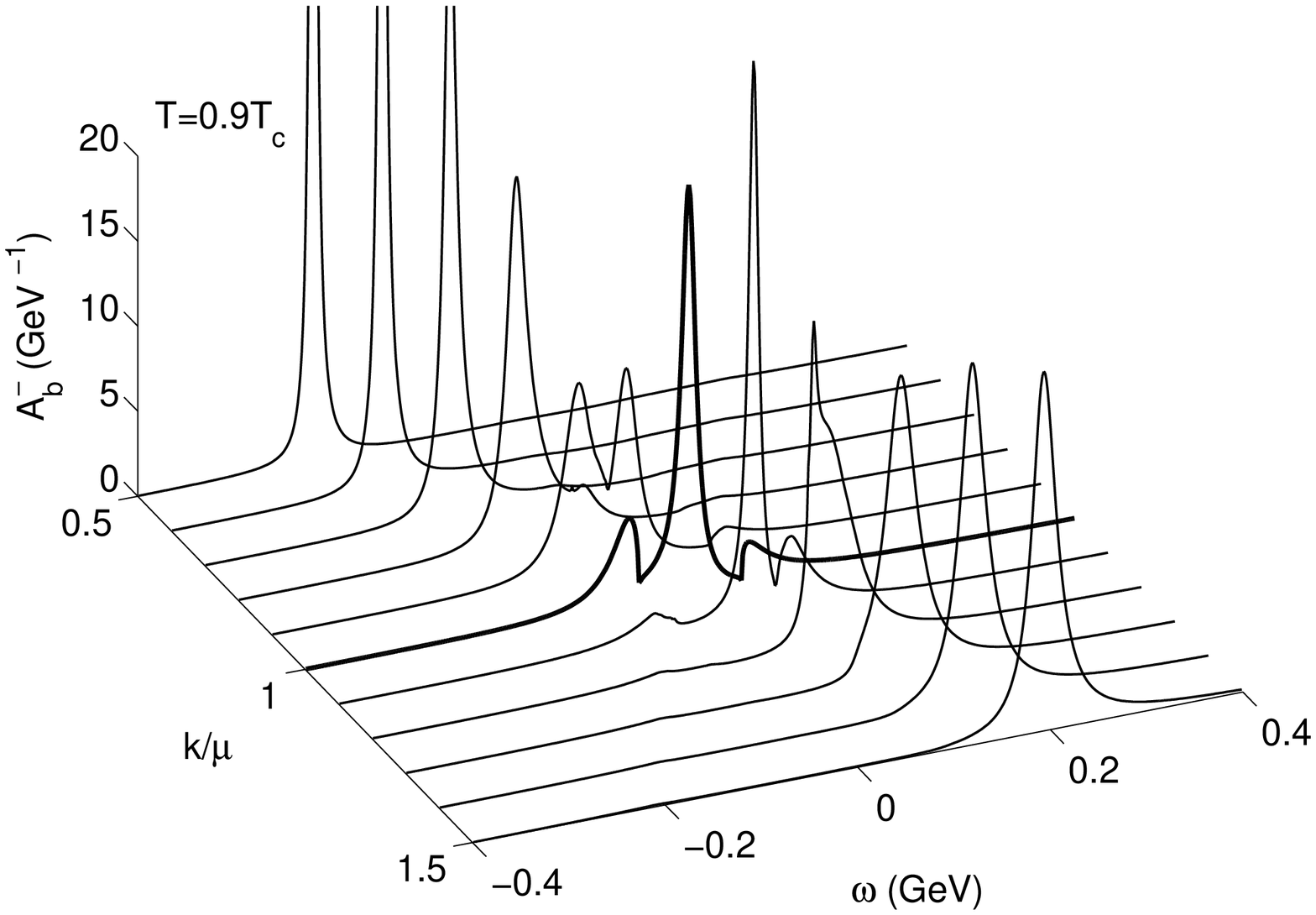}
\includegraphics[width=8cm]{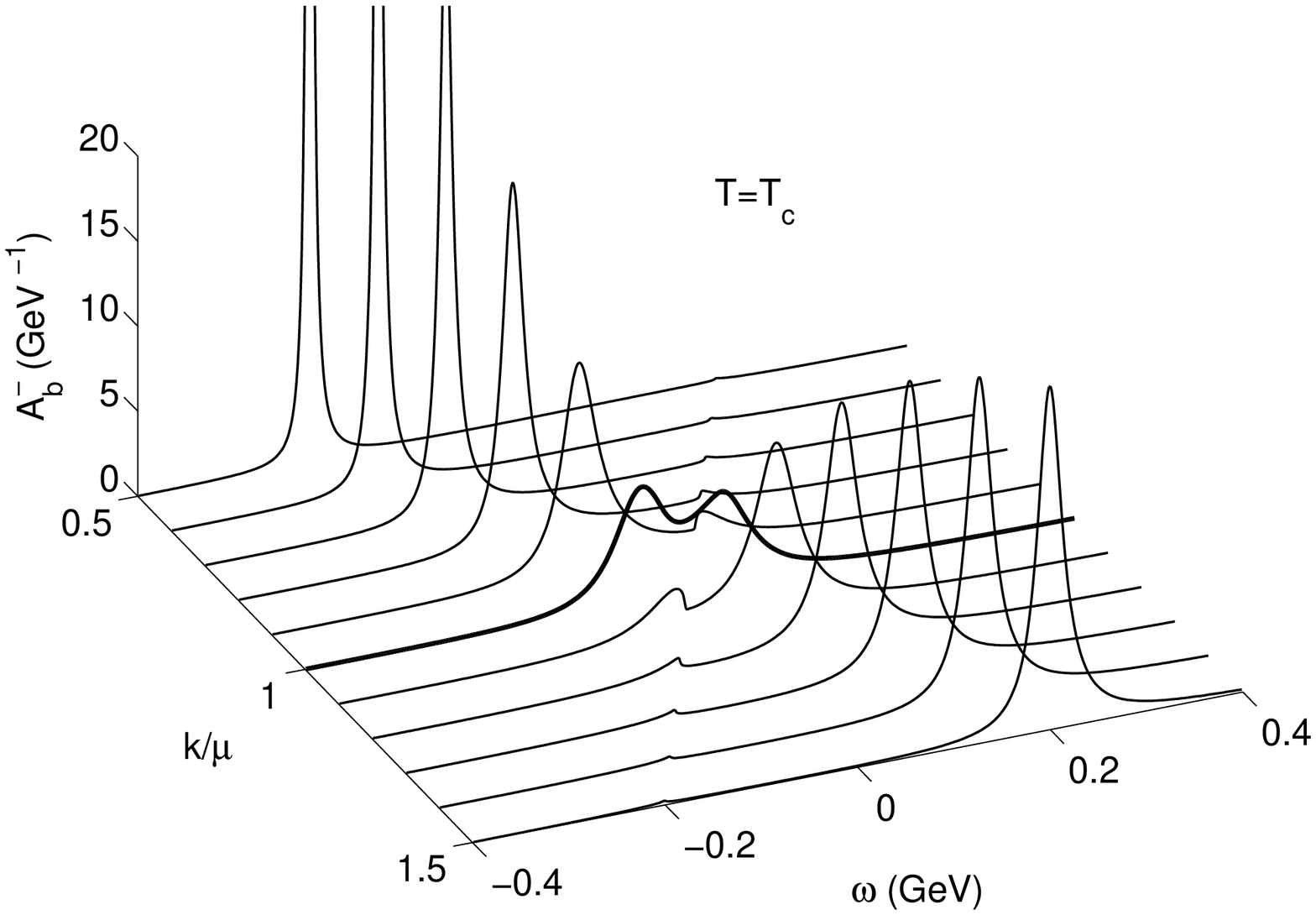}
\caption{The spectral density of the unpaired blue quark for different momenta $k$ at various temperatures $T\leq T_c$.
The zero temperature pairing gap is set to be $\Delta_0=100$MeV. \label{fig9}}
\end{center}
\end{figure*}

\begin{figure*}
\begin{center}
\includegraphics[width=8cm]{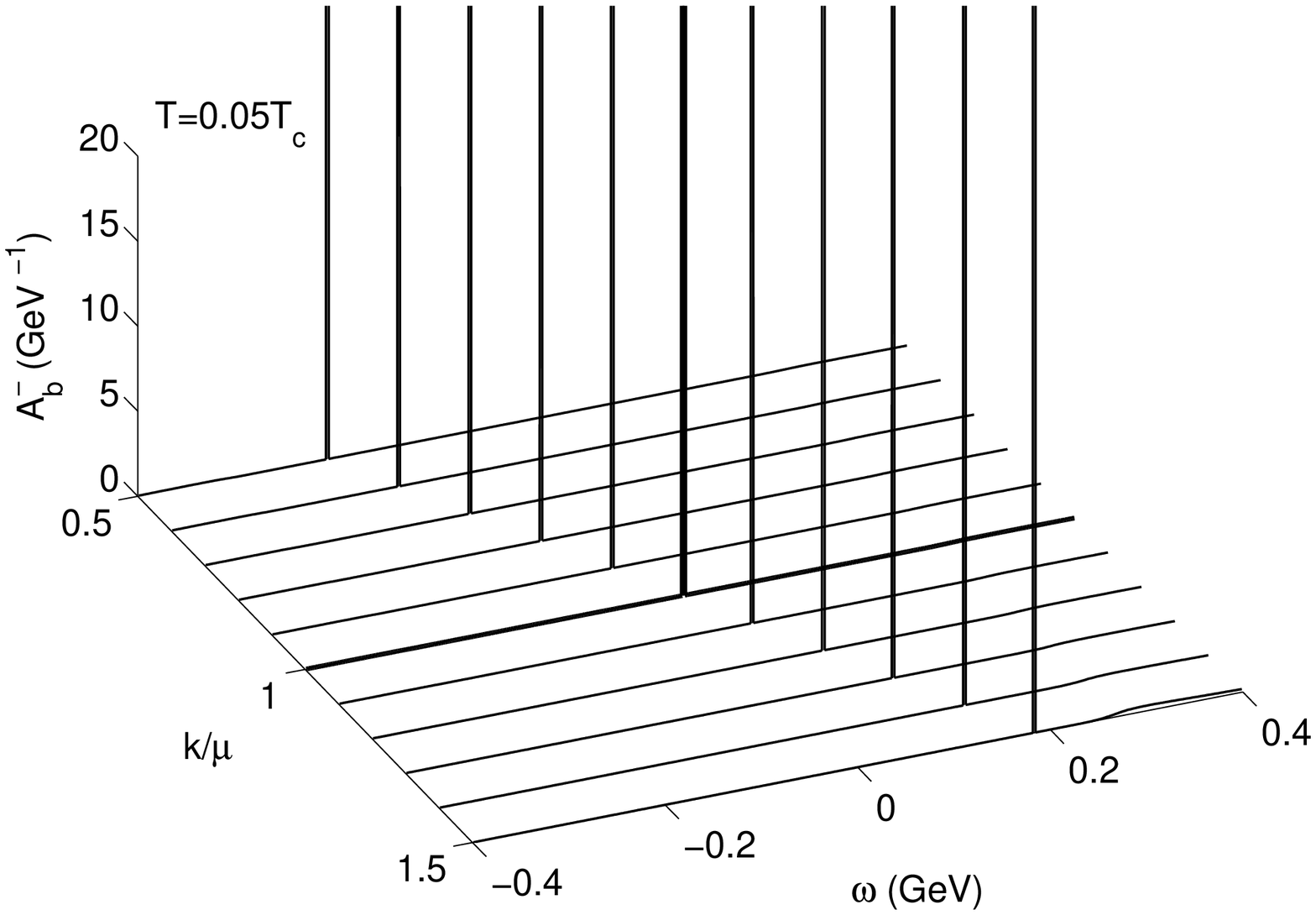}
\includegraphics[width=8cm]{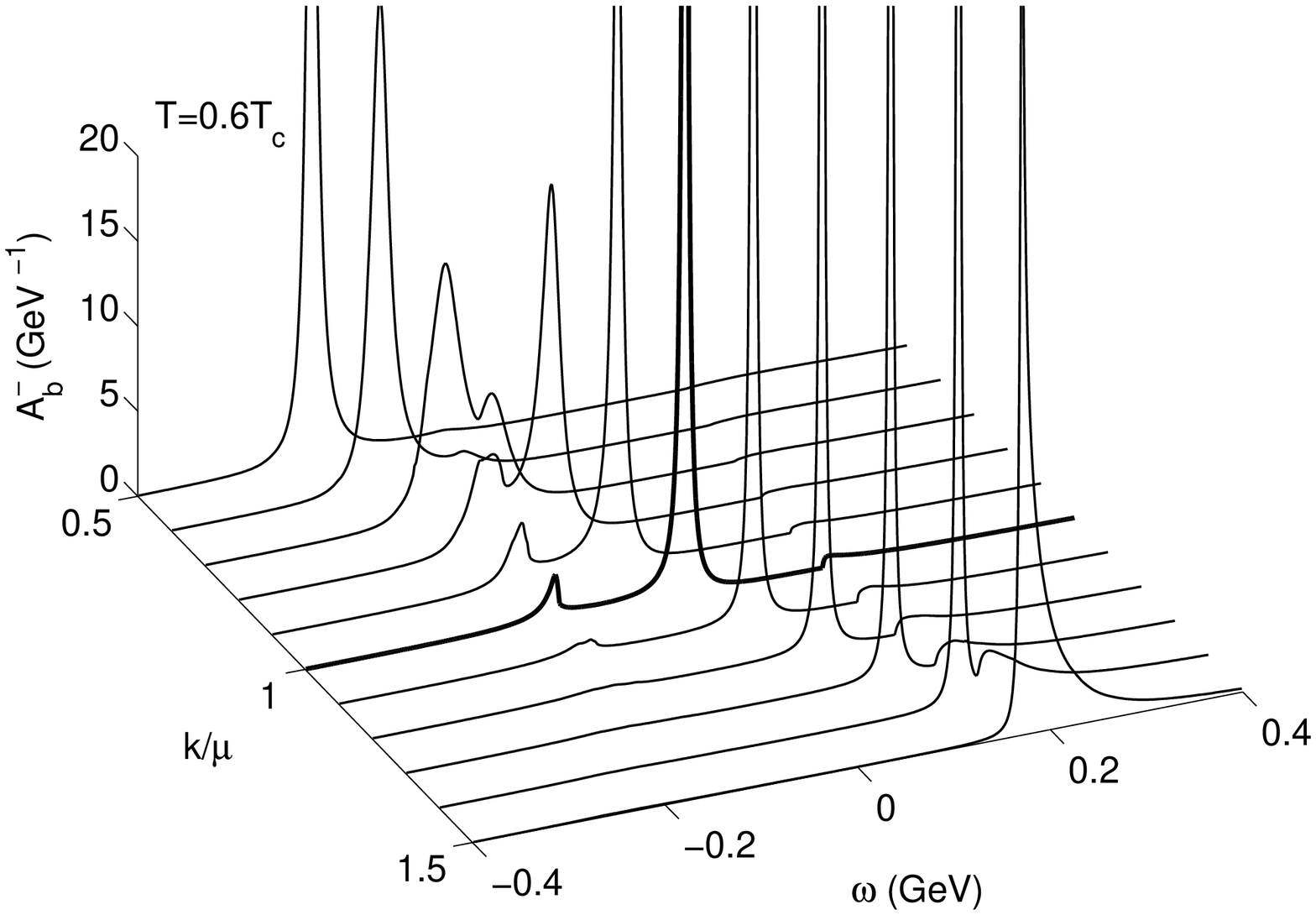}
\includegraphics[width=8cm]{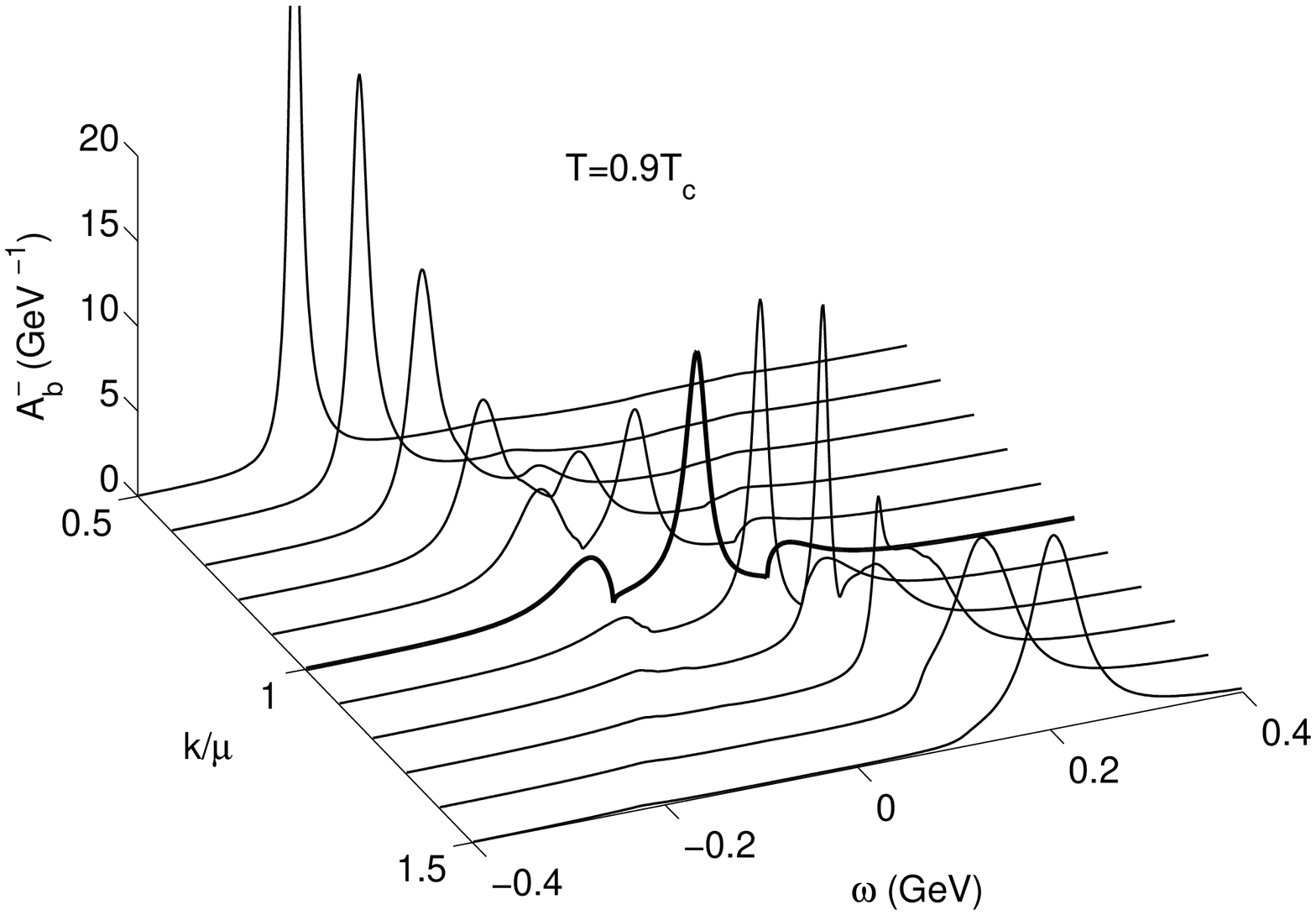}
\includegraphics[width=8cm]{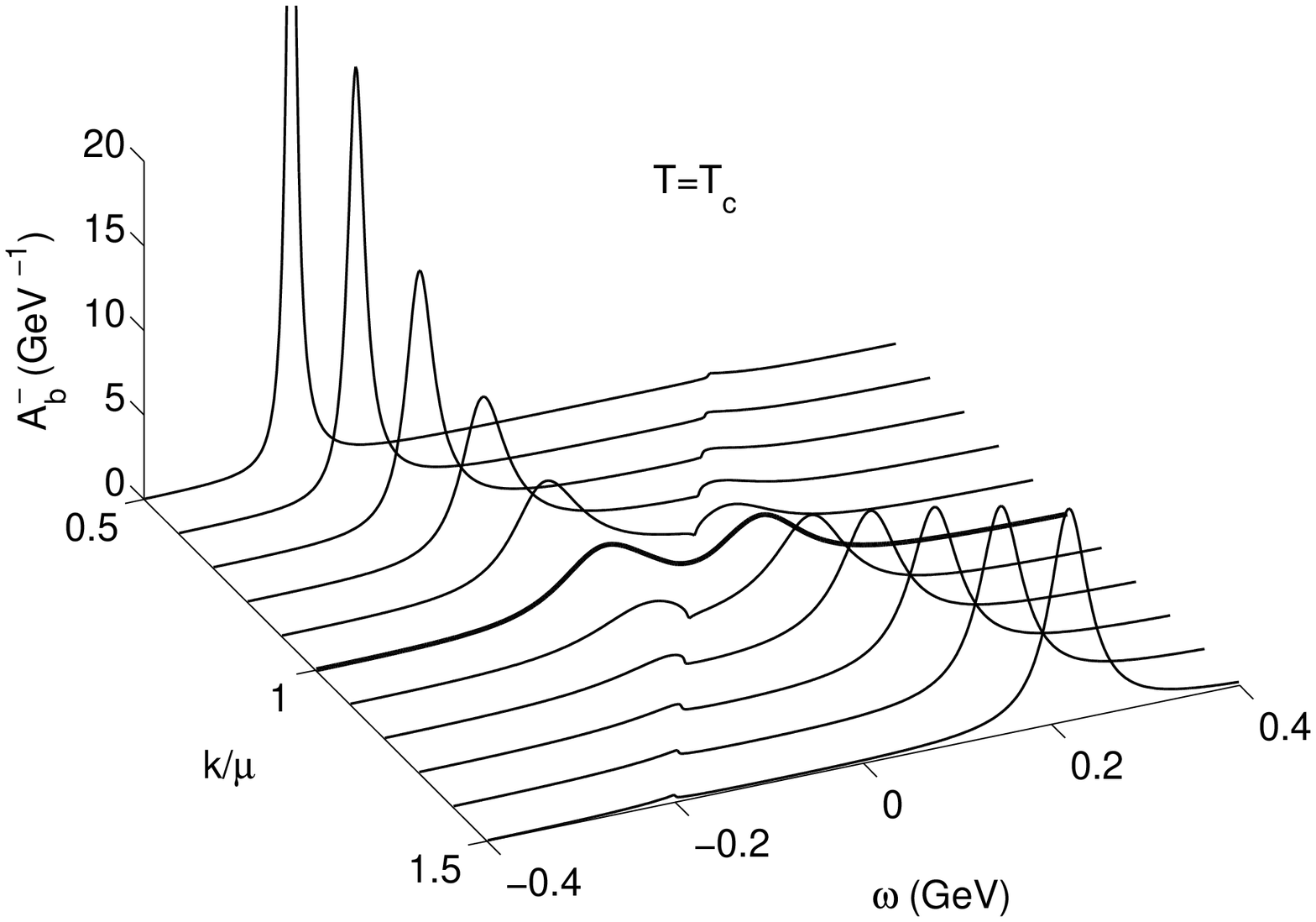}
\caption{The spectral density of the unpaired blue quark for different momenta $k$ at various temperatures $T\leq T_c$.
The zero temperature pairing gap is set to be $\Delta_0=150$MeV.\label{fig10}}
\end{center}
\end{figure*}

To obtain the explicit form of ${\cal A}_{\rm b}(\omega,{\bf k})$, we note that the retarded self-energy $\Sigma^{\rm R}_{\rm b}(\omega,{\bf k})$
has the following Dirac matrix structure
\begin{eqnarray}
\Sigma^{\rm R}_{\rm b}(\omega,{\bf k})=\Sigma^0_{\rm b}(\omega,{\bf k})\gamma^0
-\Sigma^{\rm v}_{\rm b}(\omega,{\bf k})\mbox{\boldmath{$\gamma$}}\cdot\hat{{\bf k}}
\end{eqnarray}
due to the fact that the quarks are massless. The imaginary part of the self-energy can be evaluated as
\begin{eqnarray}
&&{\rm Im}\Sigma^0_{\rm b}(\omega,{\bf k})=\sum_{\bf q}\left[H^-({\bf k},{\bf q})+H^+({\bf k},{\bf q})\right],\nonumber\\
&&{\rm Im}\Sigma^{\rm v}_{\rm b}(\omega,{\bf k})=\sum_{\bf q}\left[H^-({\bf k},{\bf q})-H^+({\bf k},{\bf q})\right]
(\hat{\bf q}-\hat{\bf k})\cdot\hat{\bf k},
\end{eqnarray}
where the functions $H^\pm({\bf k},{\bf q})$ are defined as
\begin{eqnarray}
H^s({\bf k},{\bf q})&=&\frac{E_{{\bf q}-{\bf k}}^s-s\xi_{{\bf q}-{\bf k}}^s}{2E_{{\bf q}-{\bf k}}^s}
\text{Im}{\cal D}_{11}^{5\text{R}}(\omega+E_{{\bf q}-{\bf k}}^s,{\bf q})\nonumber\\
&\times&\left[b(\omega+E_{{\bf q}-{\bf k}}^s)+f(E_{{\bf q}-{\bf k}}^s)\right]\nonumber\\
&+&\frac{E_{{\bf q}-{\bf k}}^s+s\xi_{{\bf q}-{\bf k}}^s}{2E_{{\bf q}-{\bf k}}^s}
\text{Im}{\cal D}_{11}^{5\text{R}}(\omega-E_{{\bf q}-{\bf k}}^s,{\bf q})\nonumber\\
&\times&\left[b(\omega-E_{{\bf q}-{\bf k}}^s)+f(-E_{{\bf q}-{\bf k}}^s)\right].
\end{eqnarray}
The imaginary part of ${\cal D}_{11}^{5\text{R}}(\omega,{\bf q})$ is given by
\begin{eqnarray}
\text{Im}{\cal D}_{11}^{5\text{R}}(\omega,{\bf q})=-\frac{\text{Im}\chi_{11}^{5\rm R}(\omega,{\bf q})}
{[\text{Re}\chi_{11}^{5\rm R}(\omega,{\bf q})]^2+[\text{Im}\chi_{11}^{5\rm R}(\omega,{\bf q})]^2}.
\end{eqnarray}
Here the explicit form of the pair susceptibility $\chi_{11}^5(\omega,{\bf q})$ reads
\begin{eqnarray}
&&\chi_{11}^5(\omega,{\bf q})\nonumber\\
&=&\frac{1}{G}-\sum_{\bf k}\left[J_1({\bf k},{\bf q})+J_1({\bf q}-{\bf k},{\bf q})\right][1-(\hat{\bf q}-\hat{\bf k})\cdot\hat{\bf k}]\nonumber\\
&+&\sum_{\bf k}\left[J_2({\bf k},{\bf q})+J_2({\bf q}-{\bf k},{\bf q})\right][1+(\hat{\bf q}-\hat{\bf k})\cdot\hat{\bf k}],
\end{eqnarray}
where the functions $J_1({\bf k},{\bf q})$ and $J_2({\bf k},{\bf q})$ are defined as
\begin{eqnarray}
J_1({\bf k},{\bf q})&=&\sum_{s=\pm}\frac{1-f(E_{\bf k}^s)-f(\xi_{{\bf q}-{\bf k}}^s)}
{E_{\bf k}^s+\xi_{{\bf q}-{\bf k}}^s+s\omega} \frac{E_{\bf k}^s+\xi_{\bf k}^s}{2E_{\bf k}^s}\nonumber\\
&-&\sum_{s=\pm}\frac{f(E_{\bf k}^s)-f(\xi_{{\bf q}-{\bf k}}^s)}{E_{\bf k}^s-\xi_{{\bf q}-{\bf k}}^s-s\omega}
\frac{E_{\bf k}^s-\xi_{\bf k}^s}{2E_{\bf k}^s}
\end{eqnarray}
and
\begin{eqnarray}
J_2({\bf k},{\bf q})&=&\sum_{s=\pm}\frac{1-f(E_{\bf k}^s)-f(\xi_{{\bf q}-{\bf k}}^{-s})}{E_{\bf k}^s+\xi_{{\bf q}-{\bf k}}^{-s}-s\omega}
\frac{E_{\bf k}^s-\xi_{\bf k}^s}{2E_{\bf k}^s}\nonumber\\
&-&\sum_{s=\pm}\frac{f(E_{\bf k}^s)-f(\xi_{{\bf q}-{\bf k}}^{-s})}{E_{\bf k}^s-\xi_{{\bf q}-{\bf k}}^{-s}+s\omega}
\frac{E_{\bf k}^s+\xi_{\bf k}^s}{2E_{\bf k}^s}.
\end{eqnarray}

At the quark chemical potential $\mu\sim 400$MeV, the antiquark degree of freedom is much less important than the quark degree of freedom.
Therefor we decompose the self-energy $\Sigma^{\rm R}_{\rm b}(\omega,{\bf k})$ as
\begin{eqnarray}
\Sigma^{\rm R}_{\rm b}(\omega,{\bf k})=\Sigma^-_{\rm b}(\omega,{\bf k})\Lambda_{\bf k}^-
+\Sigma^+_{\rm b}(\omega,{\bf k})\Lambda_{\bf k}^+.
\end{eqnarray}
Here the minus and plus signs correspond to quark and antiquark degrees of freedom, respectively. The self-energies
$\Sigma^\pm_{\rm b}(\omega,{\bf k})$ read
\begin{eqnarray}
\Sigma_{\rm b}^\pm(\omega,{\bf k})=\Sigma_{\rm b}^0(\omega,{\bf k})\pm\Sigma_{\rm b}^{\rm v}(\omega,{\bf k}).
\end{eqnarray}
Using these results, the total spectral density ${\cal A}_{\rm b}(\omega,{\bf k})$ can be expressed as a sum of the spectral densities
for quark and antiquark,
\begin{eqnarray}
{\cal A}_{\rm{b}}(\omega,{\bf k})={\cal A}_{\rm b}^-(\omega,{\bf k})+{\cal A}_{\rm b}^+(\omega,{\bf k}),
\end{eqnarray}
where
\begin{eqnarray}
{\cal A}_{\rm{b}}^s(\omega,{\bf k})=-\frac{1}{2\pi}\frac{\rm{Im}\Sigma_{\rm{b}}^s(\omega,{\bf k})}
{[\omega-\xi_{\bf k}^s-\rm{Re}\Sigma_{\rm{b}}^s(\omega,{\bf k})]^2+[\rm{Im}\Sigma_{\rm{b}}^s(\omega,{\bf k})]^2}.
\end{eqnarray}
In the following, we focus on the spectral density ${\cal A}_{\rm b}^-(\omega,{\bf k})$ for quark excitation.

In the numerical calculations, we fix the coupling constant $G$ by using the zero temperature pairing gap $\Delta_0$ through the
BCS gap equation (\ref{gapq}). The momentum cutoff $\Lambda$ is chosen as $\Lambda=650$ MeV. It is generally thought that the pairing
gap $\Delta$ at $\mu=400$ MeV is of order of 100 MeV \cite{CSCbegin,CSCreview}. Therefore, the ratio $\Delta/\mu$ reaches the order
$O(10^{-1})$, which is likely in the strong coupling regime.

In Fig. \ref{fig7} and Fig. \ref{fig8}, we show the numerical results for the spectral density ${\cal A}_{\rm b}^-(\omega,{\bf k})$ at
quark momentum $k=\mu$ and at quark chemical potential $\mu=400$MeV. The zero temperature gaps are set to be 100 MeV and 150 MeV in
Fig. \ref{fig7} and Fig. \ref{fig8}, respectively. They corresponds to the values of the coupling constant $G/4=3.59$GeV$^{-2}$ and $G/4=4.39$GeV$^{-2}$, respectively. The qualitative behavior of the self-energy and the spectral density are very similar
to those for atomic color superfluidity. At low temperature, the Fermi liquid behavior of the blue quark persists and spectral weight of the
continuum part induced the pairing fluctuations is rather small. As the temperature becomes high enough, this continuum part becomes two visible
gaplike peaks, or the so-called pesudogap peaks, which coexist with the broadened Fermi-liquid peak. At and above the critical temperature
$T_c$, the Fermi-liquid peak disappears completely and the spectral density shows purely pseudogap behavior. The understanding of the decay process
of the unpaired blue fermion presented in Sec. II remains valid for the unpaired blue quark. On the other hand, for weak coupling or small $\Delta_0$, the pairing fluctuation effects are rather weak and the Fermi-liquid behavior persists for arbitrary temperature, as we expected.

The naive analytical argument by using the approximation like Eq. (\ref{pgapp}) in Sec. II also applies to the present ultrarelativistic quark system. Therefore, the coexistence of the Fermi-liquid behavior and the pseudogap behavior is quite generic for both atomic color superfluid studied in Sec. II and the quark color superconductor studied in this section.

The spectral density of the blue quark for momenta away from the Fermi surface $k=\mu$ is shown in Fig. \ref{fig9} and Fig. \ref{fig10}. We find that
the situation here is quite different from the case of atomic color superfluid with resonant interaction. For quark momenta away from the Fermi surface, we find that the spectral density generally shows a single-peak structure. Only for stronger couplings it shows a two-peak structure
for some special momentum and temperature. The observation here is similar to the result for $T>T_c$~\cite{kitazawa}, where it was found that the suppression of the spectral weight of the Fermi-liquid peak is most pronounced at the Fermi surface $k=\mu$. This means the attractive strength in this quark system is much weaker than the coupling strength in resonant Fermi gases. Therefore, for two-flavor color superconducting quark matter at moderate density ($\mu\sim400$ MeV), the pairing fluctuation effects are only pronounced near the Fermi surface.

\begin{figure*}
\begin{center}
\includegraphics[width=8cm]{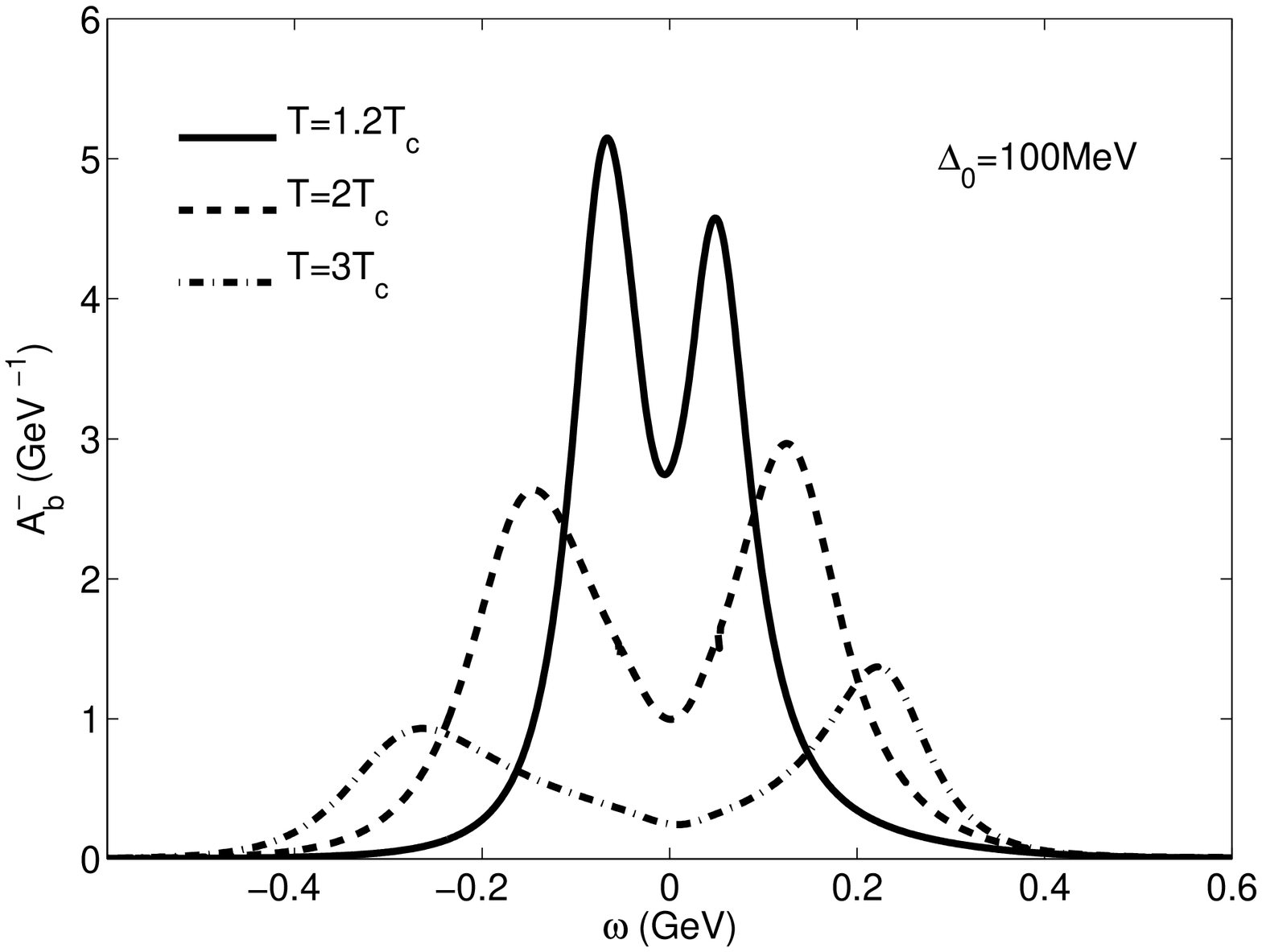}
\includegraphics[width=8.1cm]{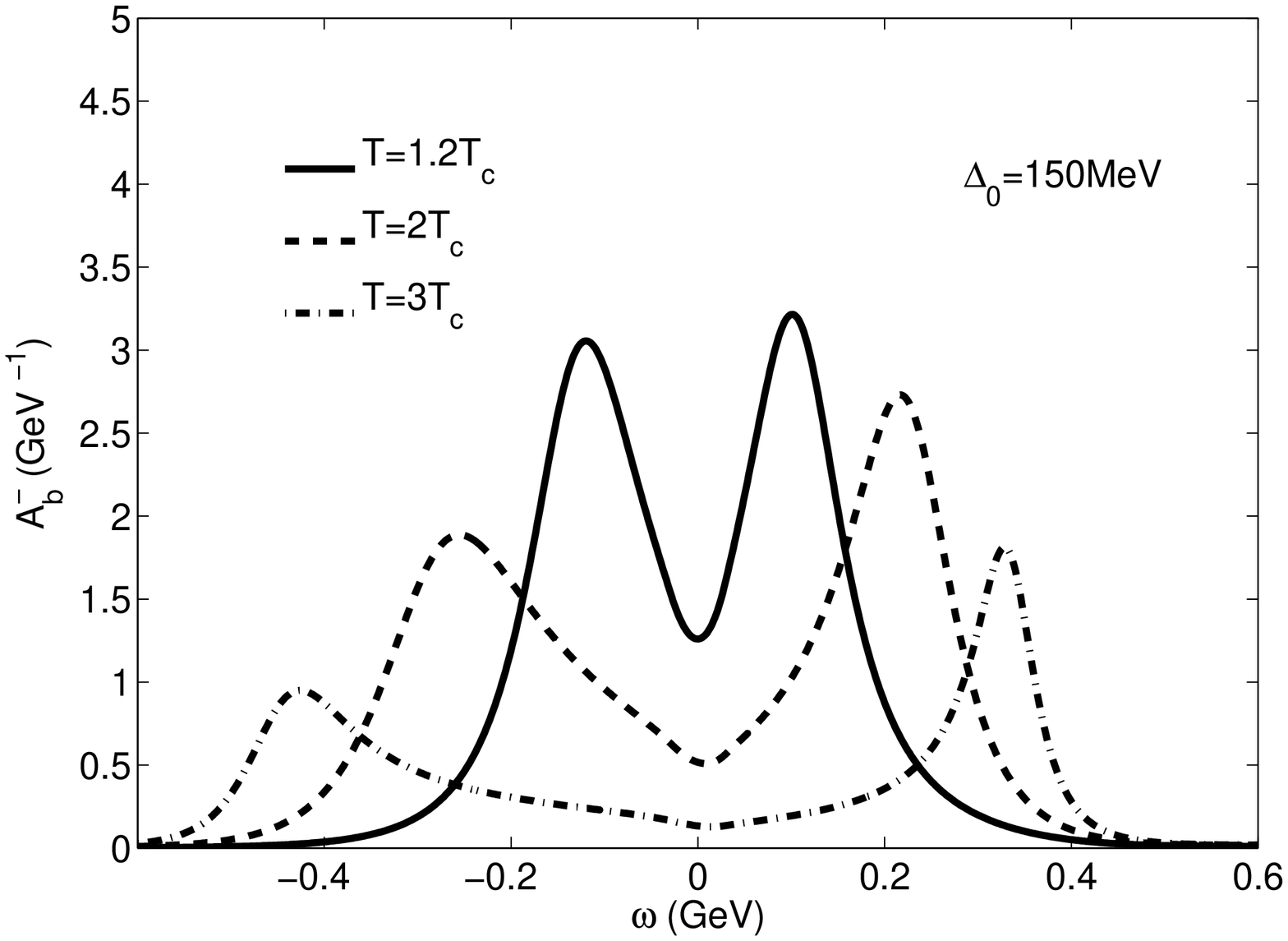}
\caption{The spectral density of the quarks for $k=\mu=400$MeV at various temperatures $T>T_c$. The left panel
is for $\Delta_0=100$MeV, while the right for $\Delta_0=150$MeV.
\label{fig11}}
\end{center}
\end{figure*}

In Fig. \ref{fig11}, we show the numerical results of the spectral density ${\cal A}_{\rm b}^-(\omega)$ at $k=\mu$ above the critical temperature $T_c$. For $T>T_c$, the color SU(3) symmetry is restored and hence the results valid for all three colors. We find that the pseudogap behavior
at the Fermi surface $k=\mu$ persists up to the temperature $T\sim (2-3)T_c$ for $\Delta_0\sim100$MeV and $\mu=400$MeV. Since we consider here only the spectral density for quarks, the sum rule $\int d\omega {\cal A}_{\rm b}^-(\omega,{\bf k})=1$ does not hold. Instead, we have $\int d\omega {\cal A}_{\rm b}(\omega,{\bf k})=1$. As the temperature increases, the two peaks become more asymmetric and their spectral weights become smaller, that is, the spectral density for antiquarks should be important at high temperatures. For $T\gg T_c$, the double-peak structure should disappear, as we expect. However, the validity of the four-fermion interaction model becomes questionable for $T\gg T_c$. Note that the quark spectral function and the pesudogap behavior have been studied by  Kitazawa, Koide, Kunihiro, and Nemoto \cite{kitazawa}. The two-peak structure of the quark spectral function shown here agrees with their results. The quantitative difference is due to the fact that we have used a different scheme to evaluate the quantities
${\rm Im}\chi_{11}^{5\rm R}(\omega,{\bf q})$ and ${\rm Re}\chi_{11}^{5\rm R}(\omega,{\bf q})$. In \cite{kitazawa}, since the case $T>T_c$ was focused, the imaginary part ${\rm Im}\chi_{11}^{5\rm R}(\omega,{\bf q})$ was analytically carried out and the real part
${\rm Re}\chi_{11}^{5\rm R}(\omega,{\bf q})$
was obtained by the dispersion relation with a cutoff associated with the frequency $\omega$. In this work, we consider mainly the case $T<T_c$,
the imaginary part ${\rm Im}\chi_{11}^{5\rm R}(\omega,{\bf q})$ cannot be carried out analytically. Therefore, we have to evaluate the imaginary and real
parts of $\chi_{11}^{5\rm R}(\omega,{\bf q})$ numerically. The cutoff $\Lambda$ is always associated with the integration over the quark momentum ${\bf k}$.

We have only studied the ideal case that the $u$ and $d$ quarks have a common chemical potential. For dense quark matter that may exists in the core of compact stars, the $u$ and $d$ quarks possess different chemical potentials due to the charge neutrality and beta equilibrium constraints. The presence of a chemical potential imbalance will generally suppress the pairing fluctuation effects. Therefore, the pairing fluctuation effects become less important if the charge neutrality and beta equilibrium constraints are imposed. Further, it is generally believed that the dense quark matter under compact star constraints should be in some exotic phases such as the Larkin-Ovhinnikov-Fulde-Ferrell phase~\cite{LOFF1,LOFF2}. Since the temperature in compact stars is much smaller than the critical temperature here, which is of order $50$ MeV, our studies here are likely irrelevant to the dense quark matter in compact stars. However, our study may be relevant to the systems where the charge neutrality and beta equilibrium constraints are not important, such as the hot and dense matter created in supernova explosions and heavy ion collisions (such as GSI-FAIR). For $T>T_c$, it was found that the dilepton production rate at low energy would be enhanced due to the pairing fluctuation effects~\cite{dil}. It will be interesting to explore in the future the phenomenological consequences of the pairing fluctuation effects, especially for the color superconducting phase studied in this paper.

\section{Summary}
In summary, we have investigated the pairing fluctuation effects in atomic color superfluids and two-flavor quark superconductors.
The common feature of these systems is that the pairing occurs among three colors of fermions. Because of the SU$(3)$ symmetry of the
Hamiltonian, one color does not participate in pairing, which is unique in the three-color systems. This branch of fermionic excitation
is gapless in the naive BCS mean-field description. In this paper, we have generalized the pairing fluctuation theory for color superfluidity/superconductivity to the low temperature domain (below the critical temperature). Especially, the theory has been used
to study how the pairing fluctuation effects influence the excitation spectrum of the unpaired color in these systems. The main
conclusions can be summarized as follows:
\\ (i) At low temperature, the Fermi-liquid behavior of the unpaired color persists even though the pairing fluctuations
are taken into account. The reason is that, at low temperature, a large pairing gap for the paired colors forms which suppresses the
pairing fluctuation effects for the unpaired color.
\\ (ii) As the temperature is increased, the continuum part of the spectral density, of which the spectral weight is rather small at
low temperature, evolves to two pseudogap peaks. Meanwhile, the Fermi-liquid broadens and its spectral weight gets suppressed. At and
above the critical temperature, the Fermi-liquid peak disappears completely and the all three colors exhibit pseudogap-like spectra.
\\ (iii) The coexistence of Fermi-liquid behavior and pseudogap behavior is generic for both atomic color superfluids and quark color
superconductors. The reason is likely due to the same symmetry structures of these systems.

{\bf Acknowledgement:} We thank Professor Qun Wang for a reading of the manuscript and helpful comments. J. P. and J. W. are supported by the National Natural Science Foundation of China under Grant No. 11125524. L. H. acknowledges the support from the Helmholtz International Center for FAIR within the framework of the LOEWE program.

\end{document}